\documentclass[applsci,article,accept,moreauthors,pdftex,10pt,a4paper]{mdpi} 
\usepackage{graphics}
\usepackage{epsfig}
\usepackage{graphics}  
\usepackage{color}     
\usepackage{amsmath,amsfonts,amssymb,amsthm}
\usepackage{pdfpages}
\usepackage{epsfig}
\usepackage{float}
\usepackage[labelformat=simple]{subfig}

\usepackage{natbib}
\graphicspath{{./figures/}}
\usepackage{booktabs} 
\usepackage{multirow}
\usepackage{soul} 
\usepackage{microtype}
\usepackage{upgreek}

\firstpage{1} 
\articlenumber{x}
\doinum{10.3390/------}
\pubvolume{7}
\pubyear{2017}
\copyrightyear{2017}
\externaleditor{Academic Editor: Christophe Finot} % It is okay. We confirm this.
\history{Received: 25 April 2017; Accepted: 13 June 2017; Published: date}

\Title{Modulational Instability in Linearly %please hyphenate
Coupled Asymmetric Dual-Core Fibers}

\Author{Arjunan Govindarajan $^{1,\dagger}$, Boris A. Malomed $^{2, 3}$, Arumugam Mahalingam $^{4}$ and \mbox{Ambikapathy Uthayakumar} $^{1,}$*} %We have now corrected the accuracy of names and affiliations.

\AuthorNames{Arjunan Govindarajan, Boris A. Malomed, Arumugam Mahalingam and Ambikapathy Uthayakumar}

\address{%
	$^{1}$ \quad Department of Physics, Presidency College, Chennai 600 005, Tamilnadu, India; govind@cnld.bdu.ac.in\\
	$^{2}$ \quad Department of Physical Electronics, School of Electrical Engineering, Faculty of Engineering, Tel Aviv University, Tel Aviv 69978, Israel; malomed@post.tau.ac.il\\% We have now separated the affiliations
	$^{3}$ \quad Laboratory of Nonlinear-Optical Informatics, ITMO University, \mbox{St. Petersburg 197101}, Russia\\
	$^{4}$ \quad Department of Physics, Anna University, Chennai 600 025, Tamilnadu, India; mahabs22@gmail.com}

\corres{Correspondence: uthayk@yahoo.com}

\firstnote{Current address: Centre for Nonlinear Dynamics, School of Physics, Bharathidasan University, Tiruchirappalli 620 024, Tamilnadu, India.}

\abstract{We investigate modulational instability (MI) in asymmetric dual-core nonlinear directional couplers incorporating the effects of the differences in effective mode areas and group velocity dispersions, as well as phase- and group-velocity mismatches. Using coupled-mode equations for this system, we identify MI conditions from the linearization with respect to small perturbations. First, we compare the MI spectra of the asymmetric system and its symmetric counterpart in the case of the anomalous group-velocity dispersion (GVD). In particular, it is demonstrated that the increase of the inter-core linear-coupling coefficient leads to a reduction of the MI gain spectrum in the asymmetric coupler. The analysis is extended for the asymmetric system in the normal-GVD regime, where the coupling induces and controls the MI, as well as for the system with opposite GVD signs in the two cores. Following the analytical consideration of the MI, numerical simulations are carried out to explore nonlinear development of the MI, revealing the generation of periodic chains of localized peaks with growing amplitudes, which may transform into arrays of solitons.}


\begin{document}

\section{Introduction}

\label{sec:1}

The modulational instability (MI)\ is a ubiquitous phenomenon originating
from the interplay of linear dispersion or diffraction and the nonlinear
self-interaction of wave fields. This effect was first theoretically
identified by Benjamin and Feir in 1967 for waves on deep water \cite{benjamin1967}; hence, MI is often called the Benjamin--Feir instability.
\textls[-18]{Studies of the MI draw steadily growing interest in nonlinear optics~\cite{hasegawa1984,tai1986,agarwal1}, fluid dynamics \cite%
{zakharov2006freak,melville1982instability}, Bose--Einstein condensates \cite%
{konotop2002,li2005,SOC}, plasma physics \cite%
{taniuti1968self,galeev1975nonlinear} and other fields.}

In its standard form, the MI applies to continuous waves (CWs) or quasi-CW
states in media featuring cubic (Kerr) self-focusing nonlinearity and
anomalous group-velocity dispersion (GVD), giving rise to the instability
against infinitesimal perturbations in the form of amplitude and phase
modulations, which eventually generates trains of soliton-like pulses \cite%
{zakharov2009}. MI can also~be observed in the normal-GVD regime in systems
incorporating additional ingredients, such as the cross-phase modulation
interaction between two components \cite{boggio2001}, in the case of the
co-propagation of optical fields and other effects, in particular the
loss dispersion \cite{tanemura2004} or fourth-order GVD \cite{hook1993}. In
all of these cases, destabilizing perturbation may originate from quantum
noise or from an additional weak frequency-shifted wave \cite%
{agrawal2006nonlinear}. Based on the nature of the underlying optical
propagation, the MI is classified as the temporal (longitudinal) instability
\cite{hasegawa1973transmission,rehberg1988}, if the CW is subject to the
GVD in fibers, or spatial (transverse) instability \cite{malendevich2001},
if the CW state experiences the action of diffraction in a planar waveguide.
More general spatio-temporal MI occurs in bulk optical media when both the
GVD and diffraction are essential \cite{liou1992}.

The MI has found many important applications, including the creation of
pulses with ultra-high repetition rates \cite{greer1989,agrawal2001applications}, the expansion of the bandwidth of Raman fiber
amplifiers \cite{ellingham2005}, the generation of optical supercontinuum \cite%
{dudley2009} and all-optical switching \cite{trillo1988soliton}. In the
context of nonlinear fiber optics, MI can also drive the four-wave mixing
initiated by the interaction of a signal wave with random noise \cite%
{boggio2001}. MI is also often regarded as a precursor to soliton formation,
since the same nonlinear Schr\"{o}dinger equation, which governs the MI,
gives rise to stable solitary pulses. Indeed, the breakup of the original CW
into soliton arrays may be an eventual outcome of the development of the MI
\cite{agrawal2006nonlinear}.

Starting from the theoretical analysis by Jensen \cite{jensen1982nonlinear},
followed by the experimental verification~\cite{Maier1982}, nonlinear
directional couplers (NLDC), which are built as dual-core fibers, have been
one of the promising elements of integrated photonic circuits for the
realization of ultrafast all-optical switches, as well as a subject of
intensive fundamental studies \cite%
{trillo1988soliton,kivshar1993,friberg1988,malomed1996symmetric,chiang1995intermodal,Chen1992,Govin2}%
. The operation of the NLDC is governed by the interplay of the Kerr
self-focusing, which induces a change in the refractive index in each core,
intra-core linear GVD, and linear coupling between the cores. The
linear-coupling coefficient determines the critical value of the power, which
gives rise to the spontaneous breaking of the symmetry between the two cores
\cite{Snyder}. Based on such power-dependent transmission characteristics,
many applications of the NLDC have been proposed, such as all-optical
switching and power splitting~\cite{trillo1988soliton}, logic operations
\cite{yang1991,yang1992}, pulse compression \cite{kitayama1983} and
bistability \cite{thirstrup1995}.

The MI\ dynamics in NLDC models was investigated in many works. In
particular, in \cite{trillo1989parametric}, Trillo {et al.}, who
first studied soliton switching in NLDC \cite{trillo1988soliton}, also
investigated the MI, considering different combinations of linear and
nonlinear effects in a saturable nonlinear medium. In~\cite{tasgal1999}%
, the MI was investigated for antisymmetric and asymmetric CW states in the
dual-core fibers, demonstrating that they are subject to the MI even in the
normal-GVD regime. In \cite{li1}, MI was explored by considering
the effects of intermodal dispersion, along with higher-order effects, such as
the third-order dispersion (TOD) and self-steepening, leading to the
conclusion that the intermodal dispersion does not affect the MI growth rate
of symmetric or antisymmetric CW states, but can drastically modify the MI
of asymmetric CW configurations. Moreover, TOD, as usual, has no influence
on the MI gain spectrum in NLDC, while self-steepening can significantly
shift the dominant MI band at a sufficiently high input power level. In
\cite{li2}, Li {et al.} extended the MI to birefringent fiber
couplers by including the cross-phase modulation, polarization mode
dispersion, and polarization-dependent coupling. Furthermore, in \cite%
{nithya}, MI was studied under the combined effects of the intermodal
dispersion and saturable nonlinear response. In \cite{porseziancgl},
Porsezian {et al.} carried out analytical and numerical investigation
of MI for asymmetric CW states in a dissipative NLDC model, based on
cubic-quintic complex Ginzburg--Landau equations. In a similar way, in
\cite{ganapathy}, MI was investigated for asymmetric dissipative fiber
couplers, which are used in fiber lasers. In that work, the system was
asymmetric, as the bar channel was an active (amplified) one, while the
cross channel was a passive lossy core (the same setting was investigated as
a nonlinear amplifier \cite{Peng}).

In all of the works dealing with the MI, except for \cite{ganapathy}, it
was assumed that the NLDC is completely symmetric with respect to the two
cores. Extension of the analysis to asymmetric nonlinear couplers is a
subject of obvious interest, as new degrees of freedom introduced by the
asymmetry may enhance the functionality of NLDC-based devices \cite%
{Kaup98,Govin3}. In a simple way, an asymmetric NLDC (ANLDC) can be
manufactured using the difference in diameters of the cores, which tends to
produce not only the phase-velocity mismatch between them, but also a change
in nonlinearity coefficients. Further, the asymmetry can be imposed by
deforming transverse shapes of the cores, while maintaining their areas
equal. In such birefringent couplers, one can induce a phase-velocity
mismatch without a change in the nonlinearity coefficients. Furthermore, to
attain the asymmetry, cores with different GVD coefficients may be used as well.
A number of works addressed the switching dynamics \cite%
{Li2012,Govin1,Chen1992,He2011,shum2002,Nobrega}, stability of solitons \cite%
{Kaup98,Kaup97,atai200355,atai2002spatial,Chaoszaf}, logic operations \cite%
{yang1991,yang1992}, etc., to elucidate possible advantages of the ANLDC
over the symmetric couplers.

In particular, switching of bright solitons has been studied \cite{Li2012}
in the model taking into regard the group- and phase-velocity mismatch and
differences in the GVD coefficients and effective mode areas of the two
cores. Recently, switching dynamics of dark solitons and interaction
dynamics of bright solitons have been investigated in \cite%
{Govin3,Govin4}. However, systematic investigation of the MI dynamics and
ensuing generation of pulse arrays in ANLDC has not been reported, as of yet.
This is the subject of the present work.

\textls[-20]{The remainder of the paper is structured as follows. Section \ref{sec:2}
introduces the coupled-mode system for the propagation of electromagnetic
fields in the asymmetric coupler. Section \ref{sec:3} presents the
linear-stability analysis for the MI induced by small perturbations,
followed by further analysis in Section~\ref{sec:4}.} Section \ref{sec:5}
reports direct simulations of the nonlinear development of the MI. Section \ref{sec:6} concludes the paper.

\section{Coupled-Mode Equations}

\label{sec:2}

The propagation of optical waves in asymmetric nonlinear couplers is
governed by a pair of linearly-coupled nonlinear Schr\"{o}dinger equations
\cite{agrawal2001applications,Govin3};
\begin{gather}
i\frac{\partial q_{1}}{\partial z}+i\beta _{11}\frac{\partial q_{1}}{%
\partial t}-\frac{\beta _{21}}{2}\frac{\partial ^{2}q_{1}}{\partial t^{2}}%
+\gamma _{1}|q_{1}|^{2}q_{1}+cq_{2}+\delta _{a}q_{1}=0, \label{eqn:1} \\
i\frac{\partial q_{2}}{\partial z}+i\beta _{12}\frac{\partial q_{2}}{%
\partial t}-\frac{\beta _{22}}{2}\frac{\partial ^{2}q_{2}}{\partial t^{2}}%
+\gamma _{2}|q_{2}|^{2}q_{2}+cq_{1}-\delta _{a}q_{2}=0, \label{eqn:2}
\end{gather}%
where $q_{1}$, $q_{2}$ and $\gamma _{1}$, $\gamma _{2}$ are amplitudes of
slowly varying envelopes and nonlinearity coefficients in the two cores of
the ANLDC, while $\delta _{a}\ $ accounts for the phase-velocity difference
between the cores. Further, $\beta _{1j}\equiv 1/v_{gj}$ and $\beta _{2j}$ ($%
j=1,2$) are the group-velocity and GVD parameters in the $j$-th core, and $c$
is the coefficient of the linear coupling between the cores.

To derive normalized coupled equations, we perform rescaling,
\begin{equation}
q_{j}\equiv (\gamma _{1}L_{D})^{1/2}u_{j},\tau \equiv t-\beta
_{11}z/T_{0},\xi \equiv z/L_{D}, \label{eqn:3}
\end{equation}%
where $L_{D}=T_{0}^{2}/\left\vert \beta _{21}\right\vert $ is the dispersion
length corresponding to a characteristic pulse width $T_{0}$, the result
being:
\begin{gather}
i\frac{\partial u_{1}}{\partial \xi }+\frac{\sigma _{1}}{2}\frac{\partial
^{2}u_{1}}{\partial \tau ^{2}}+\left\vert u_{1}\right\vert ^{2}u_{1}+\kappa
u_{2}+\chi u_{1}=0, \label{eqn:4} \\
i\frac{\partial u_{2}}{\partial \xi }+i\rho \frac{\partial u_{2}}{\partial
\tau }+\alpha \frac{\sigma _{1}}{2}\frac{\partial ^{2}u_{2}}{\partial \tau
^{2}}+\Gamma \left\vert u_{2}\right\vert ^{2}u_{2}+\kappa u_{1}-\chi u_{2}=0.
\label{eqn:5}
\end{gather}%

Here, the normalized coupling coefficient is $\kappa \equiv cL_D,$ $\sigma
_{1}=+1$ and $-1$ correspond to the anomalous and normal GVD in the first
core, \textls[-15]{while the normalized phase- and group-velocity mismatches and
differences in the GVD and effective mode areas are represented,
respectively, by:}
\begin{equation}
\chi =\delta _{a}L_{D},\,\rho =(\beta _{12}-\beta _{11})L_{D}/T_{0},\,\alpha
=\beta _{22}/\beta _{21},\,\Gamma =\gamma _{2}/\gamma _{1}. \label{eqn:6}
\end{equation}%

To design such asymmetric fiber couplers and to calculate the asymmetry
coefficients, we adopt physical parameters for the first core, corresponding
to standard nonlinear directional couplers, as follows: $\beta_{21}=0.02$
ps$^2$/m, $\gamma_1=10$ kW$^{-1}$/m, $T_D=50$ fs % We agree with the units not to be italic and we have changed the kW.m format
at wavelength $\lambda=1.5\ \upmu \rm{m}$.
Physical parameters for the second core are then determined by normalized
coefficients, according to the design outlined above. Furthermore, in terms of this
normalized system, we will call ``bar'' and ``cross'' the cores corresponding
to Equations (\ref{eqn:4}) and (\ref{eqn:5}), respectively.

\section{The Linear-Stability Approach}

\label{sec:3}

Steady-state CW\ solutions with common propagation constant $Q$ are looked
for as:
\begin{equation}
u_{1}=A_{1}\exp (iQ\xi ),\quad u_{2}=A_{2}\exp (iQ\xi ), \label{eqn:7}
\end{equation}%
where $A_{1},A_{2}$ are real amplitudes, which determine the total intensity
and asymmetry ratio:
\begin{equation}
P=A_{1}^{2}+A_{2}^{2},~\eta =A_{1}/A_{2}. \label{P}
\end{equation}%

The substitution of {Ansatz} \eqref{eqn:7} in Equations \eqref{eqn:4} and %
\eqref{eqn:5} yields an expression for propagation constant $Q$ and a
relation between $\eta $ and the phase velocity mismatch, $\chi $:
\begin{gather}
Q=\frac{P(\Gamma +\eta ^{2})}{2(1+\eta ^{2})}+\frac{\kappa (\eta ^{2}+1)}{%
2\eta }, \label{eqn:8} \\
\chi =\frac{P(\Gamma -\eta ^{2})}{2(1+\eta ^{2})}+\frac{\kappa (\eta ^{2}-1)%
}{2\eta }, \label{eqn:9}
\end{gather}

Next, we add infinitesimal perturbations $a_{j}$ to the CW solutions, as:
\begin{equation}
\begin{split}
u_{1}& =[A_{1}+a_{1}]\exp (iQ\xi ), \\
u_{2}& =[A_{2}+a_{2}]\exp (iQ\xi ).
\end{split}
\label{eqn:10}
\end{equation}%

Substituting Expression \eqref{eqn:10} into Equations \eqref{eqn:4} and \eqref{eqn:5}%
, we arrive at linearized equations for the complex perturbations:
\begin{gather}
i\frac{\partial a_{1}}{\partial \xi }+\frac{\sigma _{1}}{2}\frac{\partial
^{2}a_{1}}{\partial \tau ^{2}}+\eta ^{2}\frac{P}{1+\eta ^{2}}%
(a_{1}+a_{1}^{\ast })+\kappa a_{2}-\kappa \eta ^{-1}a_{1}=0, \label{eqn:11}
\\
i\frac{\partial a_{2}}{\partial \xi }+i\rho \frac{\partial a_{2}}{\partial
\tau }+\alpha \frac{\sigma _{1}}{2}\frac{\partial ^{2}a_{2}}{\partial \tau
^{2}}+\Gamma \frac{P}{1+\eta ^{2}}(a_{2}+a_{2}^{\ast })+\kappa a_{1}-\kappa
\eta a_{2}=0. \label{eqn:12}
\end{gather}%

Solutions to Equations (\ref{eqn:11}) and (\ref{eqn:12}) are looked for, in the
usual form, as:
\begin{gather}
a_{1}=F_{1}e^{i(K\xi -\Omega \tau )}+G_{1}e^{-i(K\xi -\Omega \tau )},
\label{eqn:13} \\
a_{2}=F_{2}e^{i(K\xi -\Omega \tau )}+G_{2}e^{-i(K\xi -\Omega \tau )},
\label{eqn:14}
\end{gather}%
where $K$ and $\Omega $ are a (generally, complex) wave number and an
arbitrary frequency of the perturbation. A set of linear coupled equations
for perturbation amplitudes $F_{j}$ and $G_{j}$ are derived by substituting
Expressions \eqref{eqn:13} and \eqref{eqn:14} in Equations \eqref{eqn:11} and %
\eqref{eqn:12}:
\begin{equation}
\mathbf{M}\times (F_{1},F_{2},G_{1},G_{2})^{T}=0, \label{eqn:15}
\end{equation}%
where $\mathbf{M}$ is a $4\times 4$ matrix, whose elements are written in
Appendix \ref{Appendix A}. % We confirm this change.
 A nontrivial solution exists under condition $\det \mathbf{M}%
=0 $. Straightforward algebraic manipulations transform the latter condition
into a dispersion relation, in the form of a quartic equation for $K$ as a
function of $\Omega $:
\begin{equation}
K^{4}-aK^{3}+bK^{2}+cK+d=0. \label{eqn:16}
\end{equation}%

Rather cumbersome expressions for coefficients $\left( a,b,c,d\right) $ are
also given in Appendix \ref{Appendix A}. The~MI growth rate (gain), defined here for the
amplitude of the waves (rather than for the power), is determined by the
largest absolute value of the imaginary part of the wave number:%
\begin{equation}
G=\left\{ |\text{Im}(K)|\right\} _{\max }. \label{eqn:17}
\end{equation}%

\section{Analysis of the Modulational Instability}
\label{sec:4}
\vspace{-6pt}
\subsection{The Anomalous-Dispersion Regime}

We start by considering the case of the anomalous GVD in both cores, i.e., $%
\sigma _{1}=1$ in \mbox{Equations (\ref{eqn:4}) and (\ref{eqn:5})}, as in this case the
MI is well known to occur in nonlinear optical fibers. First, in Figure \ref%
{fig:1}, the red line shows the MI gain in the conventional symmetric NLDC
(``SNLDC''), with $\alpha =\Gamma =1$ and $\rho =\chi =0$. In the same
figure, the solid blue line shows the gain for the asymmetric NLDC
(``ANLDC'') with a particular choice of asymmetry parameters (the reason for
choosing these values is explained below), such that the effective mode area
of the second core is twice that of the first core, and the GVD of the bar
channel is ten-times higher than in the cross one. The figure makes it
evident that the MI gain increases by a factor $>$2 in the ANLDC, and the MI
bandwidth is wider by a factor $\simeq$4. \textls[-15]{The enhancement of the MI is a
new result in the context of the nonlinear directional coupler (similar
enhancement was earlier found in the single-core decreasing-GVD fibers with
a tapered core \cite{xu2001}.}
\begin{figure}[H]
\centering
\includegraphics[width=0.35\textwidth]{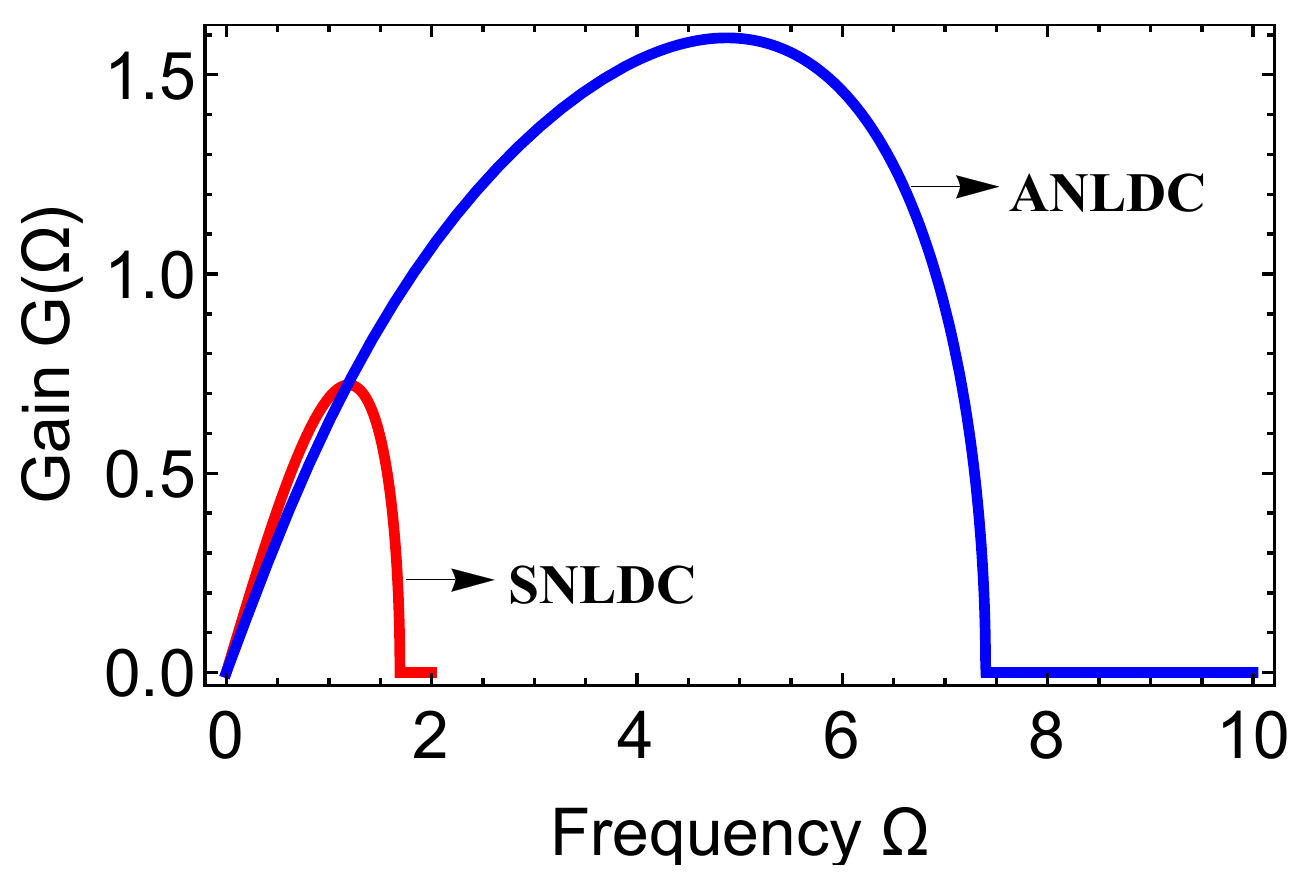}
\caption{Modulational instability (MI) gain spectra for symmetric nonlinear directional couplers (SNLDC) and asymmetric (ANLDC)
couplers in the anomalous group-velocity dispersion (GVD) regime ($\protect\sigma _{1}=1$). Parameters
of the symmetric system are $P=\protect\eta =\protect\alpha =\Gamma =\protect%
\kappa =1$ and $\protect\rho =\protect\chi =0$. For the asymmetric ones, the
parameters are the same, except for $\protect\alpha =0.1,\Gamma =2$, $%
\protect\chi =0.66$ and $\protect\rho =0.1$. }
\label{fig:1}
\end{figure}

\subsubsection{The Effect of the Input Power on the Instability Spectrum}

\begin{figure}[H]
\centering
\subfloat[]{\includegraphics[width=0.35\linewidth]{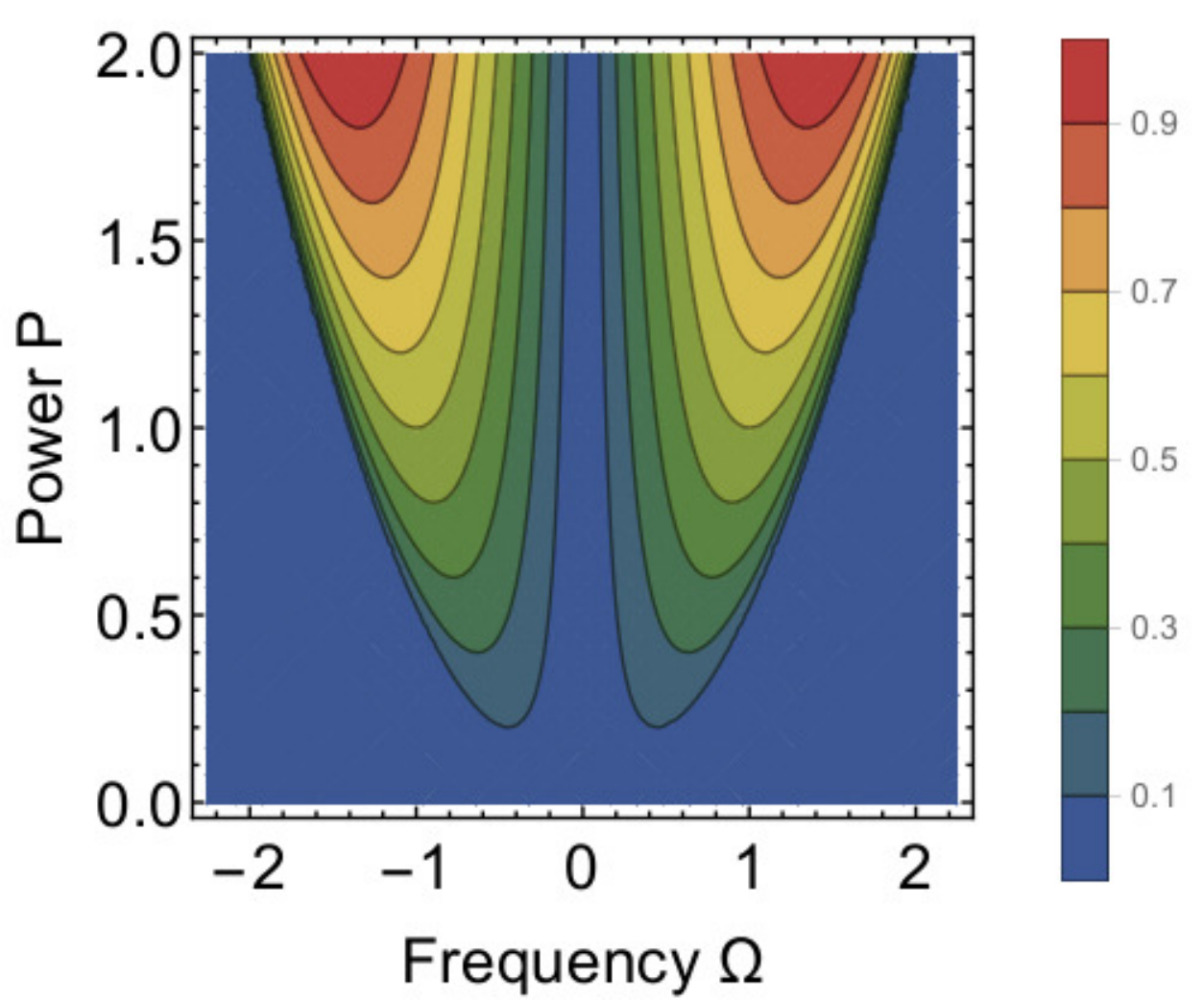}} %
\subfloat[]{\includegraphics[width=0.35\linewidth]{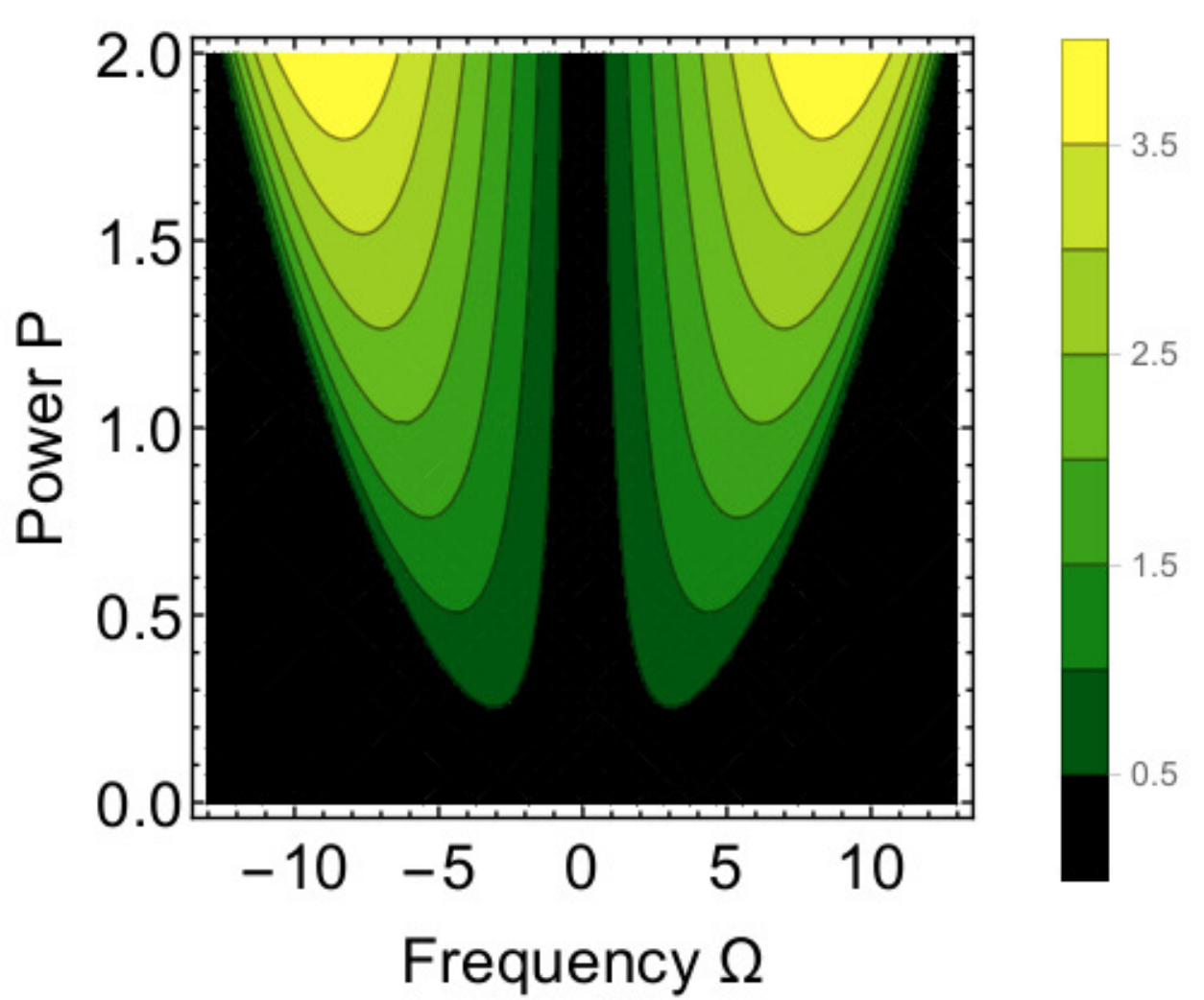}}
\caption{Contour plots showing the dependence of the MI gain on the continuous wave (CW) power, $P$, and perturbation frequency, $\Omega $, for symmetric and
asymmetric couplers in the anomalous-GVD regime ($\protect\sigma _{1}=1$).
Parameters of symmetric system (\textbf{a}) are $\protect\eta =\protect\alpha =\Gamma
=1$, $\protect\kappa =2$ and $\protect\rho =\protect\chi =0$. For the
asymmetric system (\textbf{b}), $\protect\eta =\protect\alpha =0.1$, $\Gamma =2$, $%
\protect\kappa =1$, $\protect\rho =0.1$, and the phase-velocity mismatch is
defined in terms of $P$, in order to produce the largest gain: $\protect\chi %
=-4.95+0.985P$. Note the difference in horizontal scales between ({a}) and
({b}). }
\label{fig:2}
\end{figure}

To elucidate the role of individual effects in the dramatic expansion of the
MI region in the asymmetric coupler, we first examine the variation of the MI gain
spectrum as a function of the CW power, in both the symmetric and asymmetric
systems. Figure \ref{fig:2}a clearly demonstrates that the MI gain in the
former case increases as in the case of the usual MI \cite%
{agrawal2006nonlinear}, i.e., linearly with the power. For the asymmetric
system, Figure \ref{fig:2}b shows not only the growth of the MI gain with
the increase of the power, but also strong expansion of the MI\ bandwidth.

\subsubsection{The Role of the Coupling Coefficient}

Figure \ref{fig:3}a shows the MI spectrum as a function of the normalized
coupling coefficient in the ANLDC, i.e., $\kappa $ in Equations (\ref{eqn:11})
and (\ref{eqn:12}). The limit case of zero coupling, i.e., the system with
decoupled cores, is included too. It is seen that the dependence of the
largest gain and MI bandwidth on $\kappa $ is very weak.

\subsubsection{The Impact of Asymmetry Parameters}
\begin{figure}[H]
\centering
\subfloat[\label{fig:3a}]{\includegraphics[width=0.32\linewidth]{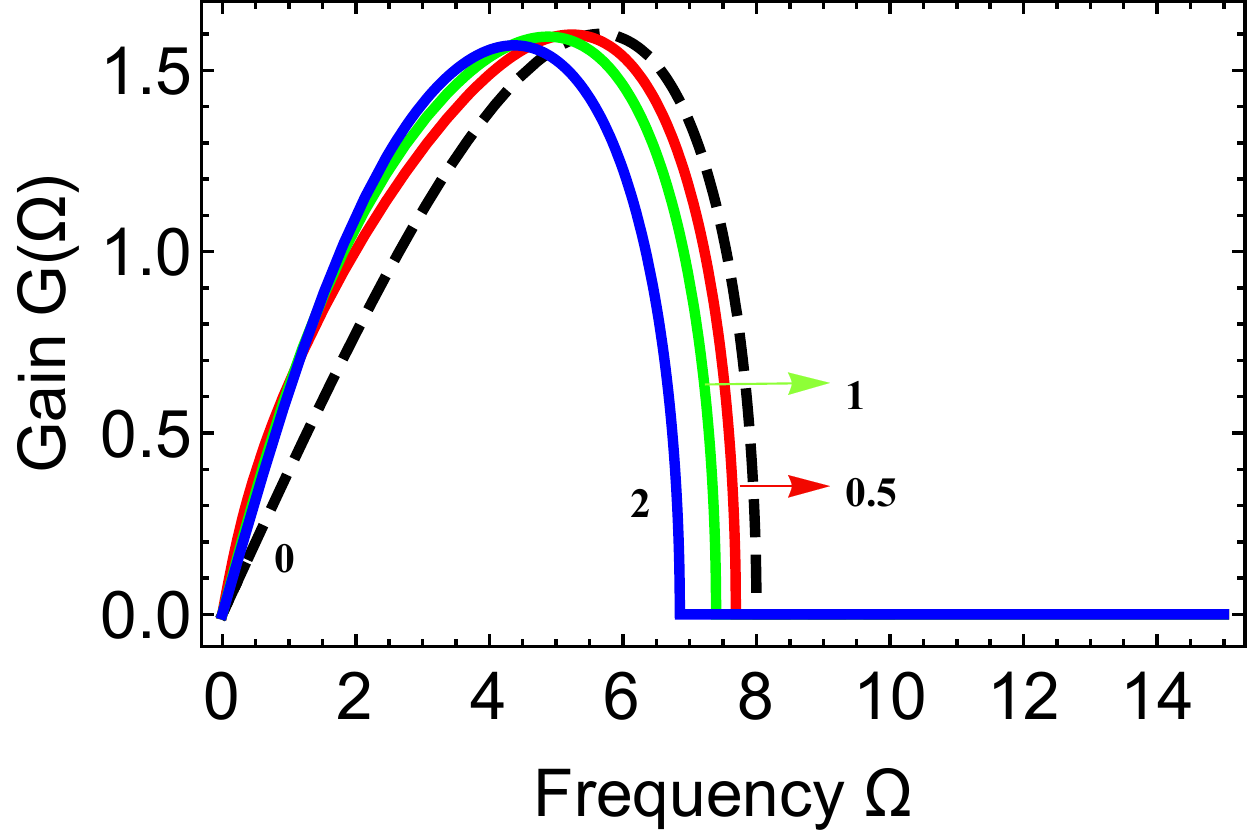}} %
\subfloat[\label{fig:4a}]{\includegraphics[width=0.32\linewidth]{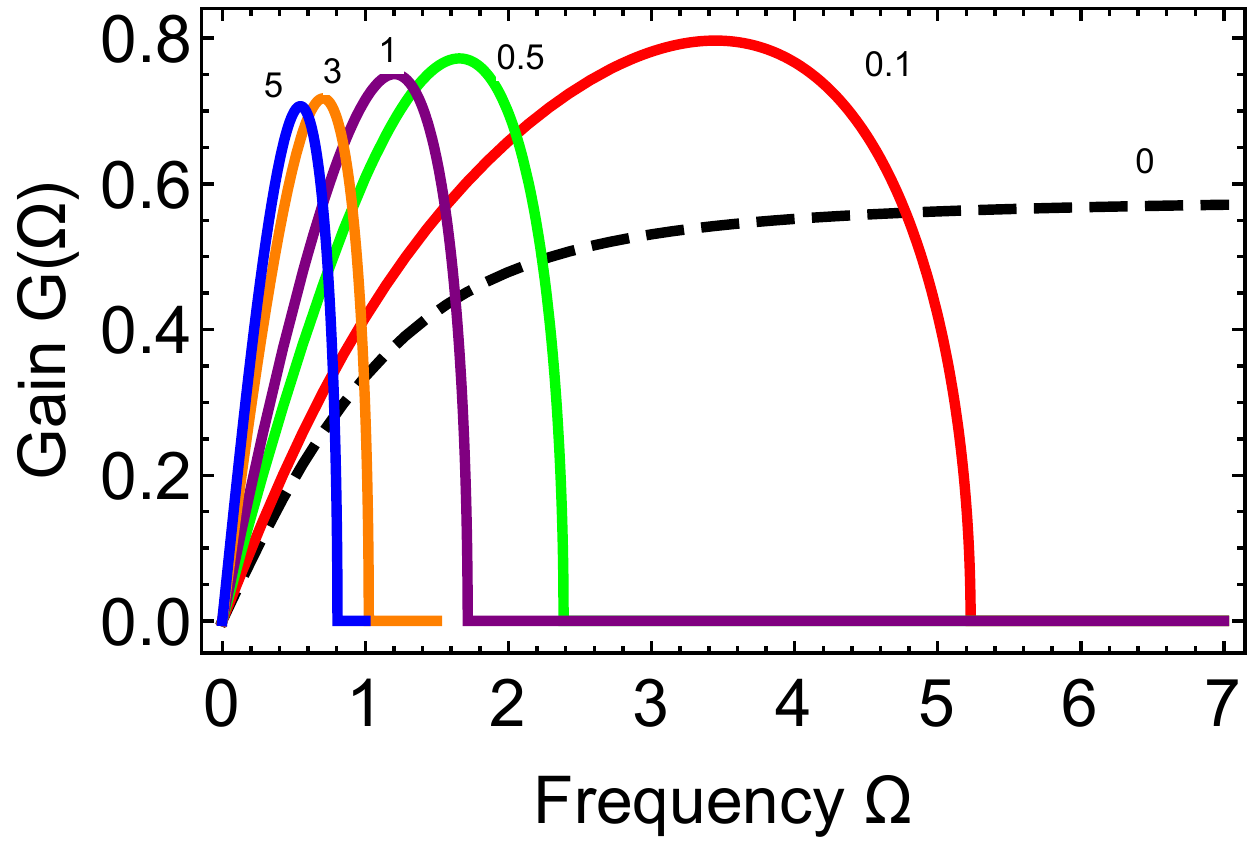}} %
\subfloat[\label{fig:5a}]{\includegraphics[width=0.32\linewidth]{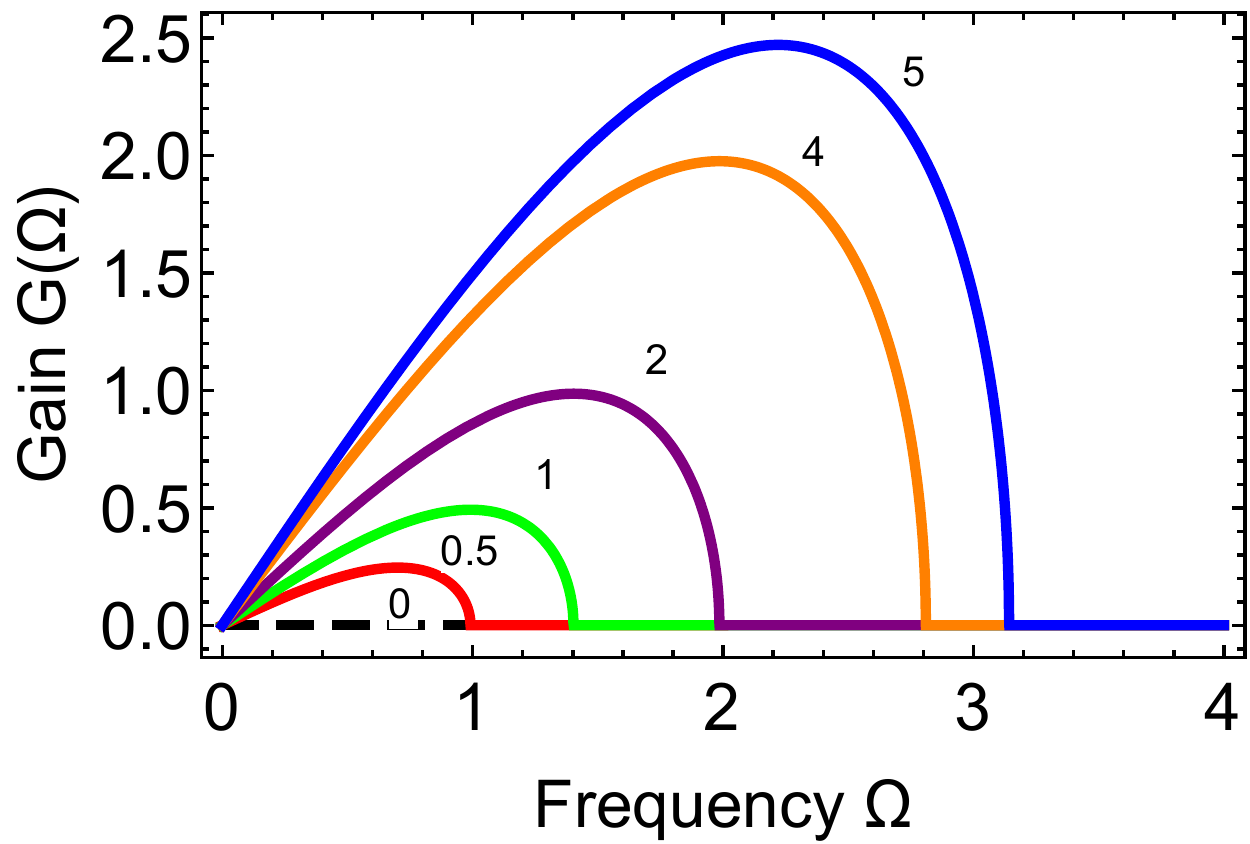}}
\caption{The MI gain spectra in the asymmetric coupler with anomalous GVD ($%
\protect\sigma _{1}=1$). (\textbf{a})~The~results for different values of the
normalized coupling coefficient, $\protect\kappa $. Parameters of the system
are \mbox{$P=2,\protect\eta =0.5,\protect\alpha =0.1,\Gamma =2$ }and $\protect\rho %
=0.01$. (\textbf{b}) For different values of the ratio of the GVD coefficients in the
cross and bar channels, $\protect\alpha $, indicated near each curve with $%
P=1,\protect\eta =\protect\kappa =0.5,\Gamma =1$ and $\protect\rho =0.01$.
(\textbf{c})~For different values of the ratio of the nonlinearity coefficients in
the two cores, $\Gamma $, which are indicated near the curves. Other
parameters are $P=0.5,\protect\eta =0.1,\protect\kappa =0.2,\protect\alpha %
=1 $ and $\protect\rho =0.1$.}
\label{fig:3}
\end{figure}

\textls[-15]{The influence of the GVD difference, $\alpha $, on the instability spectra
is presented in Figure \ref{fig:3}b. The~limit case of the coupler with zero
GVD in the cross channel, $\alpha =0$, is included as well. }As seen in the
figure, the MI bandwidth of MI is infinite in the limit case. Both the gain
and bandwidth of the MI monotonically decrease with the increase of $\alpha $%
, with the MI vanishing in the limit of $\alpha \rightarrow +\infty $. In~other words, relatively weak anomalous GVD in the cross channel strongly
affects MI bandwidth in the ANLDC.

The influence of the difference in effective mode areas of two cores ($%
\Gamma $) is illustrated by Figure \ref{fig:3}c. In this case too, we start
with the limit case of an extremely asymmetric coupler, in which the second
core is purely dispersive, with zero nonlinearity ($\Gamma =0$). In this
limit, the MI gain vanishes. The MI gain and bandwidth monotonically increase
with the growth of $\Gamma $. This dependence on $\Gamma $ is opposite to
that on $\alpha $, which is displayed in Figure \ref{fig:3}b. Thus, the MI can
be effectively controlled by means of the two asymmetry parameters, $\Gamma $
and $\alpha $.
\vspace{-12pt}
\vspace{-18pt}

\begin{figure}[H]
\centering
\subfloat[\label{fig:6a}]{\includegraphics[width=0.4\linewidth]{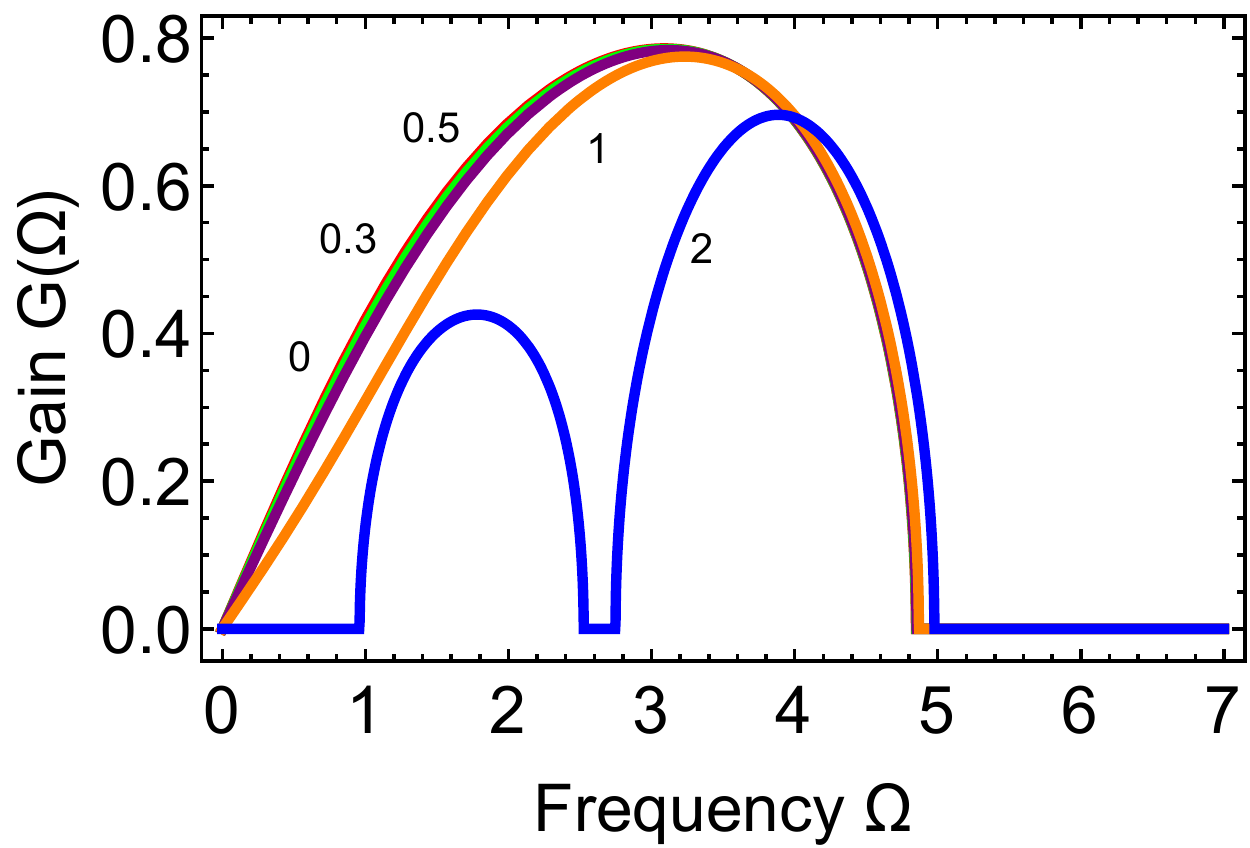}}
\hspace{0.5in} \subfloat[\label{fig:7a}]{\includegraphics[width=0.4%
\linewidth]{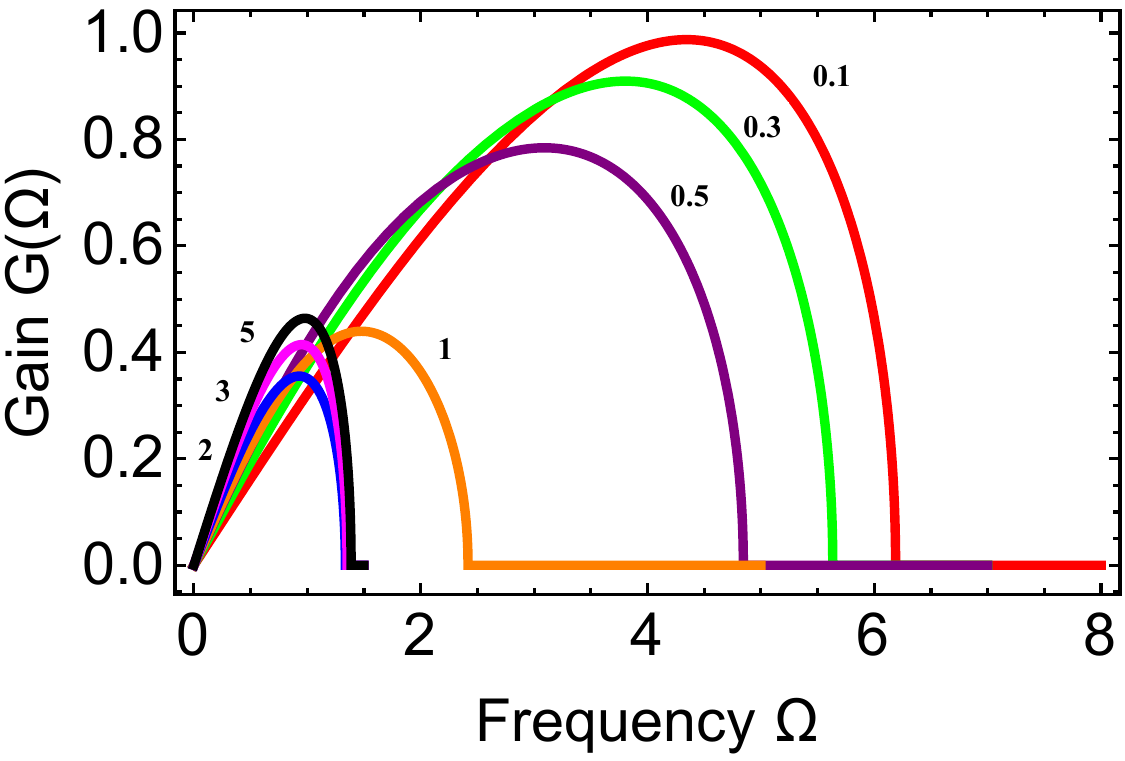}}
\caption{The MI gain spectra in the anomalous-GVD regime ($\protect\sigma %
_{1}=1$). (\textbf{a}) Results for different values of the group-velocity mismatch ($%
\protect\rho $ in Equation (\protect\ref{eqn:5})), which are indicated near the
curves. Other parameters are $P=\protect\kappa =\Gamma =1$, $\protect\eta %
=0.5$ and $\protect\alpha =0.1$ (\textbf{b}) Results for different values of
asymmetry ratio $\protect\eta $ of the CW state (see Equation (\protect\ref{P})),
which are indicated near the curves. Other parameters are $P=0.5,\protect%
\kappa =1$, $\protect\alpha =0.1,\Gamma =2$ and $\protect\rho =0.01$.}
\label{fig:6}
\end{figure}

Next, we study the effect of the group-velocity mismatch (walk-off % No, it should not be italic
between the cores), $\rho $. Figure~\ref{fig:6}a shows the impact of $\rho $
when the asymmetry is represented only by the GVD ratio, $\alpha =0.1$,
while the nonlinear coefficients in both cores are equal. The figure
demonstrates that the variation of $\rho $ in the range of $\rho \lesssim 1$
weakly affects the MI spectrum. The effect is much stronger at larger values
of the walk-off. In particular, the MI spectral band splits into two at $\rho
=2$. The latter effect seems interesting even if the value of $\rho =2$ may
be difficult to attain in real couplers. On the other hand, the analysis
demonstrates that the variation of $\rho $ produces almost no effect on the
MI gain in the case when the asymmetry is determined by the difference in
the nonlinearity coefficients ($\Gamma \neq 1$), while the GVD coefficients
are equal ($\alpha =1$). The latter result is not shown here in detail, as
it does not display noteworthy features.

It is obviously interesting as well to investigate the effect of the CW
asymmetry ratio, $\eta $ (see~Equation~(\ref{P})), on the MI. These results are
presented in Figure \ref{fig:6}b, which makes it obvious that the gain and
bandwidth of the MI quickly decrease with the increase of $\eta $ from small
values $\sim $$0.1$ to $\eta =2$. With the further increase of the asymmetry
ratio to values $\eta >2$, the largest MI gain slightly increases, while the
bandwidth remains practically constant.
\vspace{-6pt}
\subsection{The Normal-Dispersion Regime}
\begin{figure}[H]
\centering
\subfloat[\label{fig:N1a}]{\includegraphics[width=0.35\linewidth]{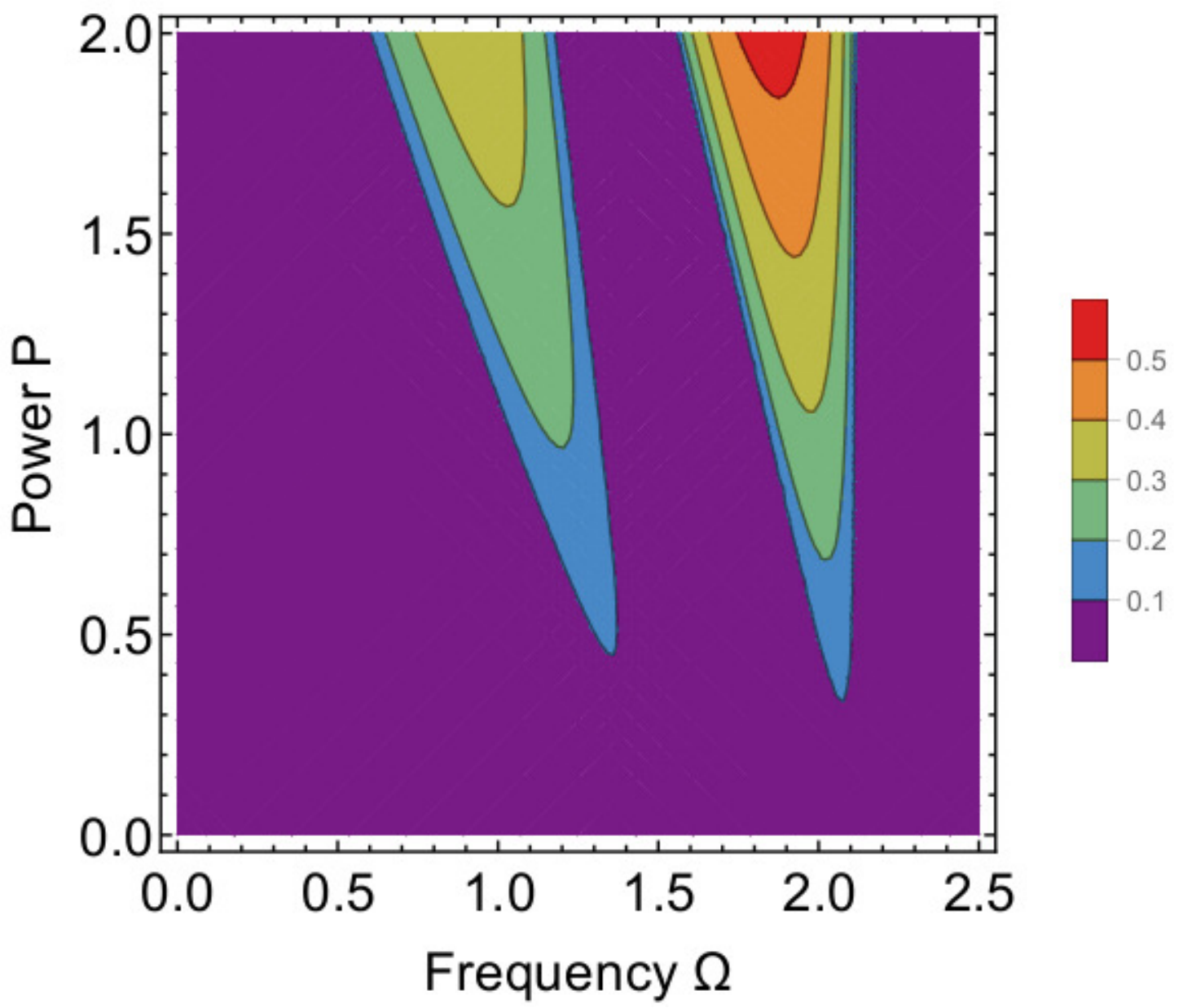}}
\hspace{0.5in} \subfloat[\label{fig:N1b}]{\includegraphics[width=0.35%
\linewidth]{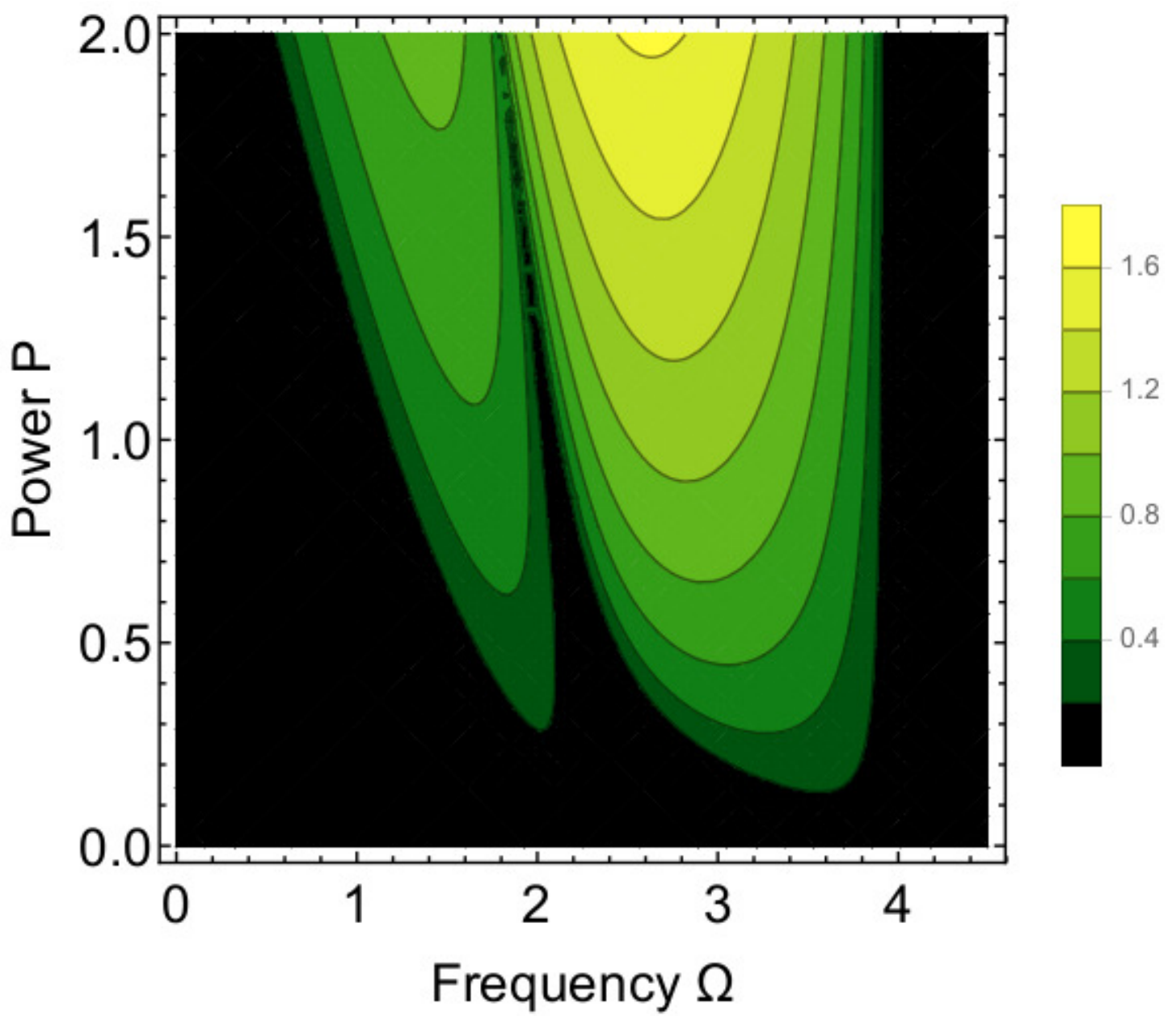}}
\caption{Contour plots showing the dependence of the MI gain on the CW
power, $P$, in the normal-GVD regime ($\protect\sigma =-1$) in the symmetric
and asymmetric systems. Parameters of the symmetric system (\textbf{a}) are $\protect%
\alpha =\Gamma =1$, $\protect\eta =2$, $\protect\kappa =0.9$ and $\protect%
\rho =\protect\chi =0$. For the asymmetric system (\textbf{b}), $\protect\eta =0.5,%
\protect\alpha =0.1,\Gamma =2,\protect\kappa =1.1$ and $\protect\rho =0.01$,
$\protect\chi =-0.825+0.7P$.}
\label{fig:N1}
\end{figure}

The combination of the self-focusing Kerr nonlinearity and normal GVD
usually supports stable CWs. However, as mentioned in the Introduction, MI
may occur under the normal GVD in more complex systems, including couplers.
Following the pattern of the MI investigation presented above for the
anomalous GVD, we first consider the effect of the CW power, $P$, on the MI
gain. We also compare the instability spectrum of the asymmetric system with
that of the symmetric one in \mbox{Figure \ref{fig:N1}a,b},
respectively. As seen in Figure \ref{fig:N1}, in both cases, two distinct MI
bands determine the instability, and (similar to the anomalous-GVD regime)
the MI gain of the asymmetric system linearly grows with $P$, featuring a
broad bandwidth.

\begin{figure}[H]
\centering
\subfloat[\label{fig:N2a}]{\includegraphics[width=0.35\linewidth]{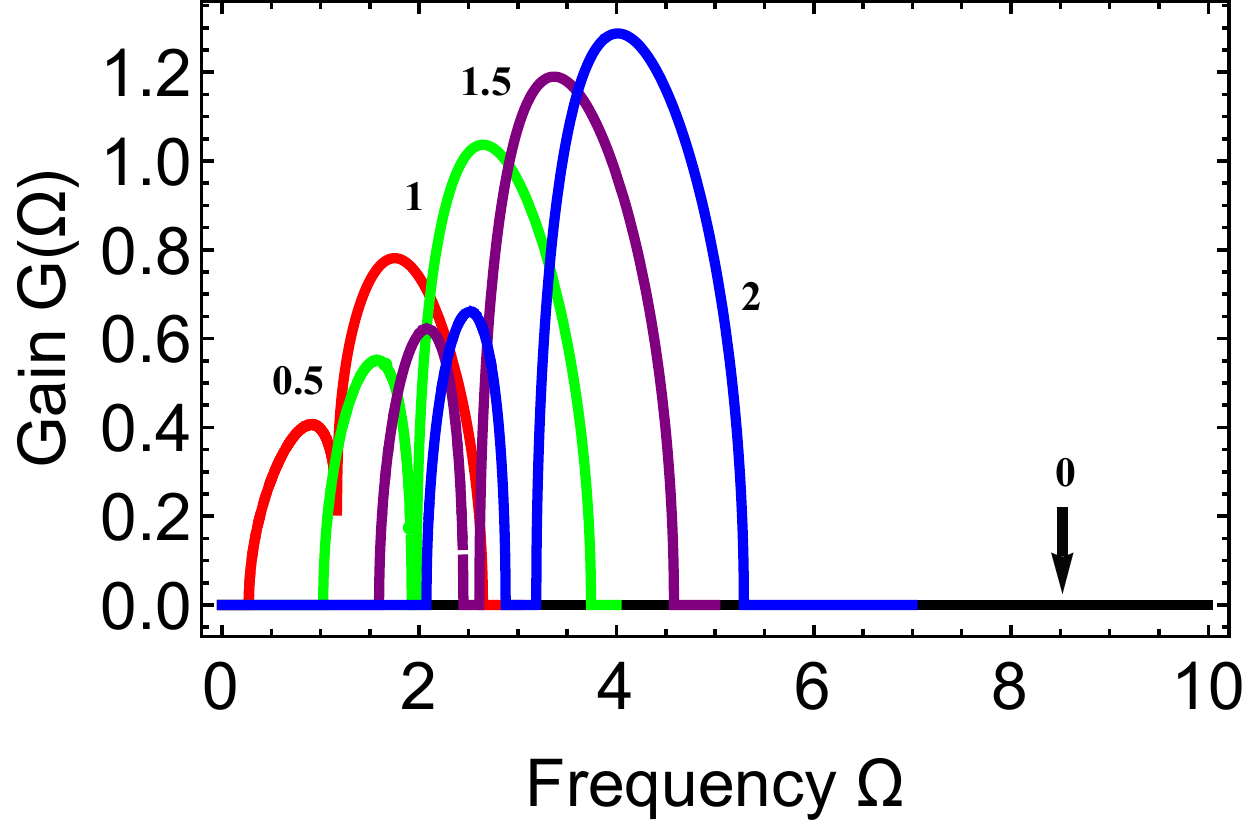}}
\hspace{0.5in} \subfloat[\label{fig:N3a}]{\includegraphics[width=0.35%
\linewidth]{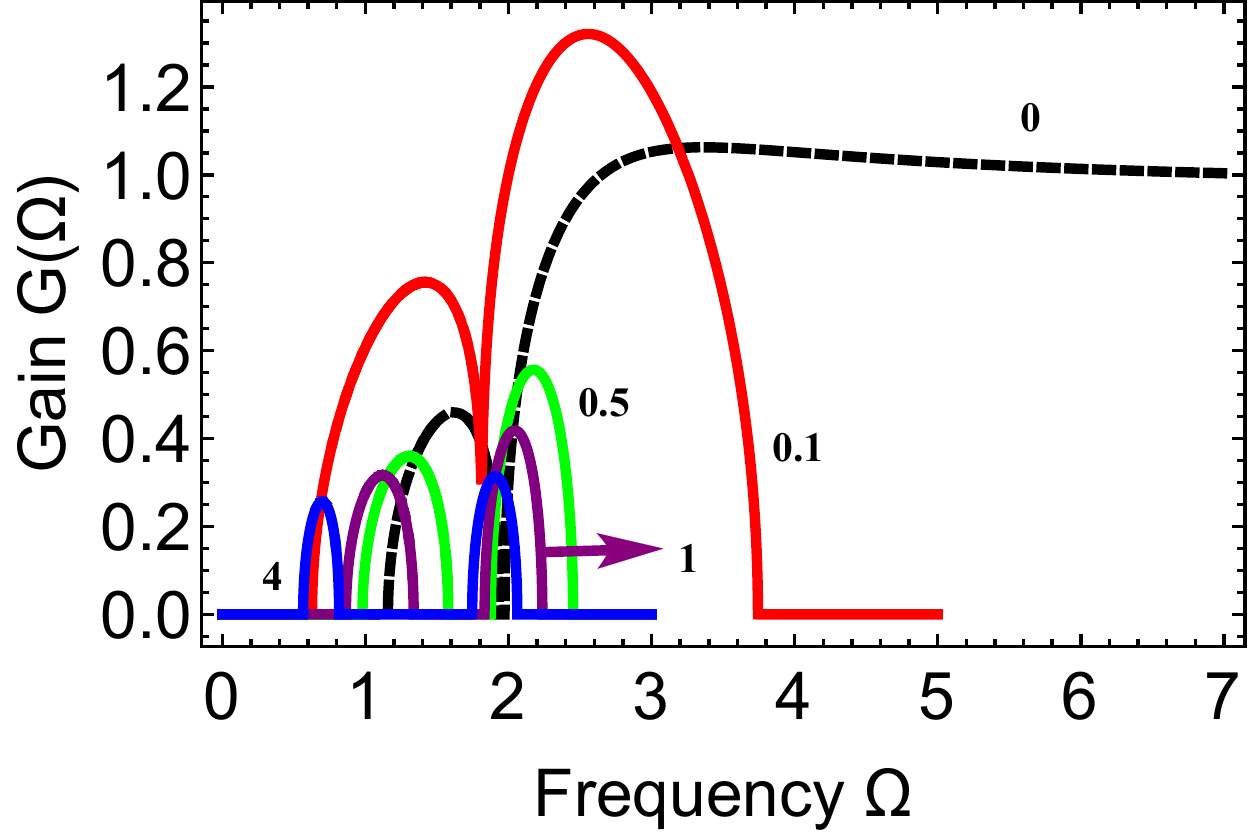}}
\caption{(\textbf{a}) The MI gain spectra in the normal-GVD regime ($\protect\sigma %
_{1}=-1$) for different values of (\textbf{a}) the coupling coefficient, $\protect%
\kappa $, indicated near the curves. Other parameters are $P=1,\protect\eta %
=0.5$, $\protect\alpha =0.1,\Gamma =2$ and $\protect\rho =0.01$. (\textbf{b}) The
change of values of the difference in normal-GVD coefficients ($\protect%
\sigma _{1}=-1$), $\protect\alpha $, in the two cores of the nonlinear
coupler (the values of $\protect\alpha $ are indicated near the curves).
Other parameters are $P=1.5,\protect\eta =0.5$, $\protect\kappa =\Gamma =1$
and $\protect\rho =0.1$.}
\label{fig:N2}
\end{figure}

To illustrate the essential effect of the coupling coefficient $\kappa $,
Figure \ref{fig:N2}a depicts the MI gain spectra for various values of $\kappa
$. Naturally, no MI takes place in the normal-GVD regime in the absence of
the coupling, $\kappa =0$. It is worthy to note the appearance of two
separated MI bands at $\kappa >1$, the MI gain increasing in both bands, along with their widths, with the growth of $\kappa $.

The effect of the relative difference in the magnitude of the normal GVD in
the two cores, $\alpha $, is shown in Figure \ref{fig:N2}b. Like in the
anomalous-GVD regime, here, as well, the MI bandwidth is infinite for $\alpha =0$
(it also contains a separate finite MI\ band). The MI spectrum features two
separate bands at $\alpha >0$ and the largest gain at $\alpha =0.1$. The
gain decreases with the subsequent increase of $\alpha $.

\begin{figure}[H]
\centering
\subfloat[\label{fig:N4a}]{\includegraphics[width=0.4\linewidth]{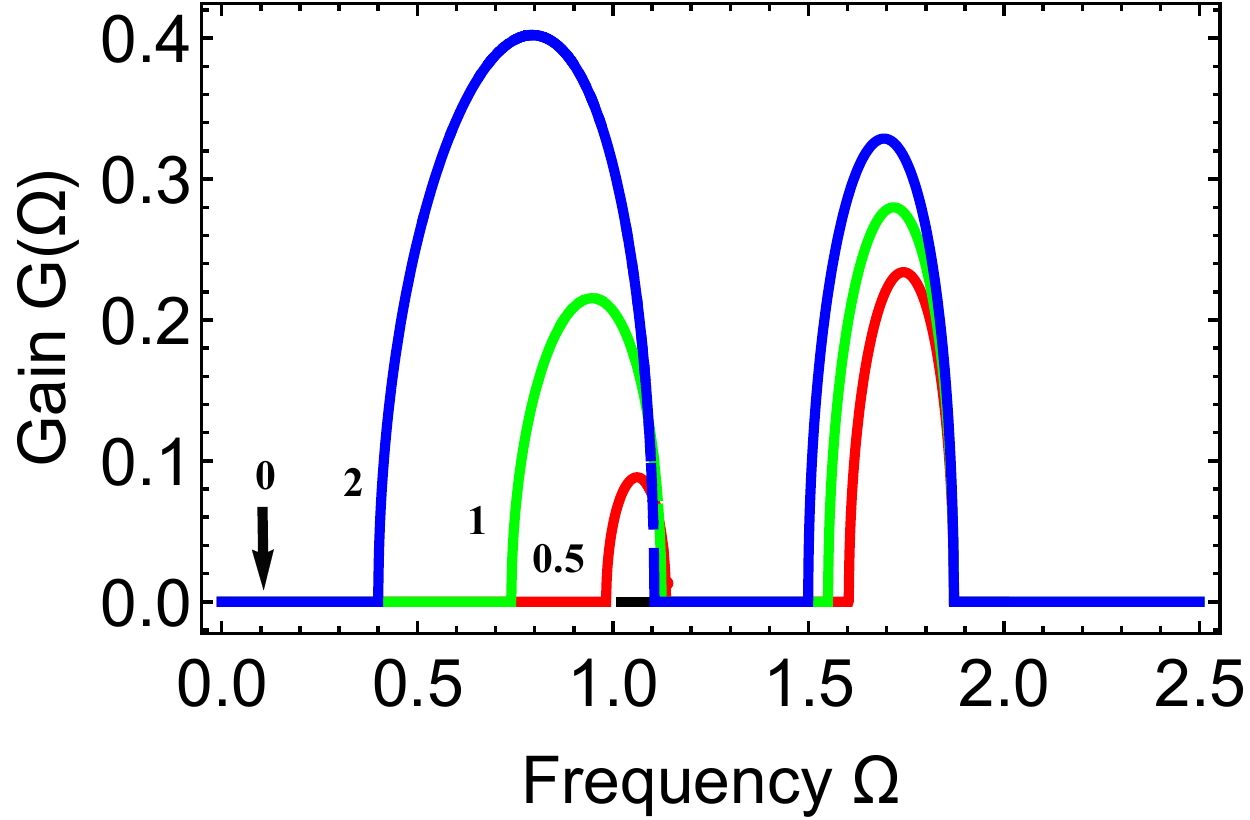}}
\hspace{0.5in} \subfloat[\label{fig:N5a}]{\includegraphics[width=0.4%
\linewidth]{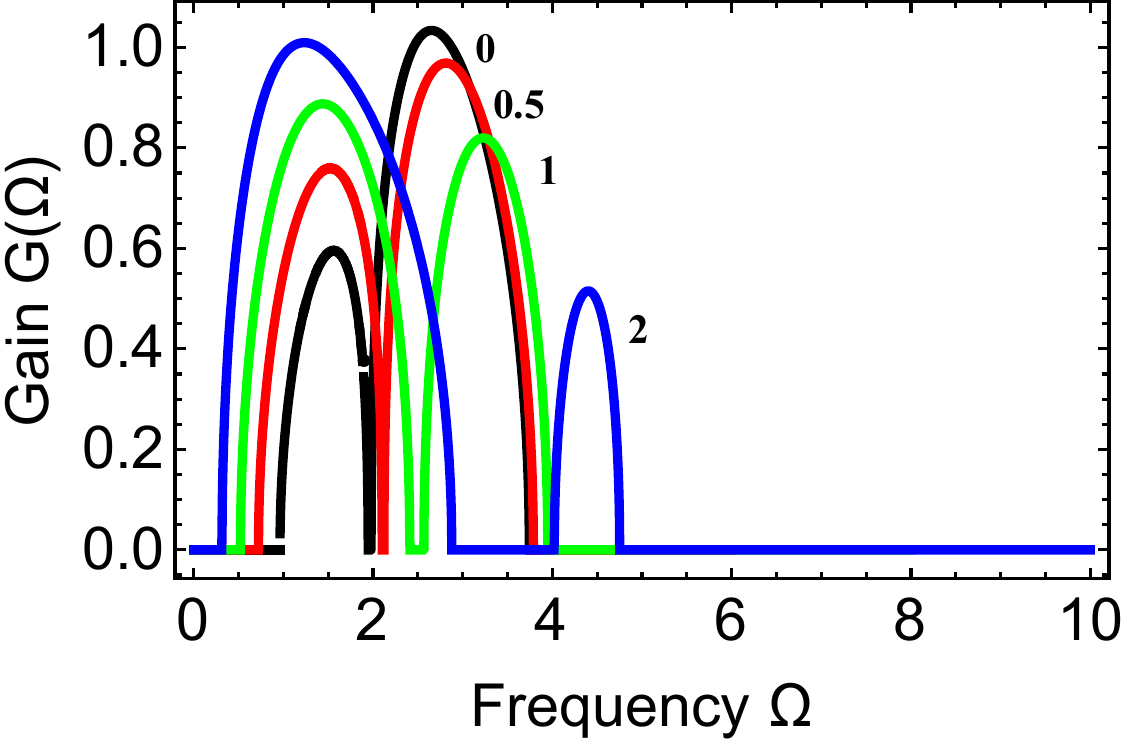}}
\caption{(\textbf{a}) The MI gain spectra in the normal-GVD regime ($\protect\sigma %
_{1}=-1$); (\textbf{a}) the results for different values of the ratio of the nonlinearity
coefficient in the two cores ($\Gamma $), indicated near the curves. Other
parameters are $P=1,\protect\eta =0.5,\protect\alpha =1,\protect\kappa =0.7$
and $\protect\rho =0.1$. (\textbf{b}) The results for different values of the
group-velocity mismatch ($\protect\rho $), indicated near the curves. Other
parameters are $P=1,\protect\eta =0.5$, $\protect\alpha =0.1,\Gamma =2,%
\protect\kappa =1$.}
\label{fig:N4}
\end{figure}

Figure \ref{fig:N4}a shows the effect of the relative difference in the
effective mode areas between the two channels, i.e., the ratio of the
nonlinearity coefficient, $\Gamma $. It is seen that no MI occurs when the
cross channel is linear ($\Gamma =0$), and two distinct MI bands emerge and
expand, featuring a growing largest value of the instability gain, with the
increase of $\Gamma $.

The influence of the group-velocity mismatch (walk-off between the cores), $%
\rho $, is depicted in Figure~\ref{fig:N4}b. Once again, the MI appears in the
form of two separated bands. The MI gain and bandwidth nontrivially depend
on $\rho $: at $\rho <1$ the low-frequency band is narrower, with smaller values
of the instability gain, while at $\rho \geq 1$, the situation is inverted.

We have also analyzed the effect of the CW's asymmetry $\eta $ (see Equation (\ref%
{P})) on the MI in the normal-GVD regime. No MI occurs for small values of $%
\eta $, {viz.} $\eta <0.2$. At $\eta >0.2$ (in particular, at $%
\eta=0.5 $), there again emerge two separate MI bands, as shown in Figure \ref%
{fig:N6}. The MI gain and bandwidth attain their maxima at $\eta =1$ (equal
amplitudes of the CW in the two cores), decreasing with the further increase
of $\eta $.

\vspace{-18pt}
\vspace{-18pt}

\begin{figure}[H]
\centering
\includegraphics[width=0.35\linewidth]{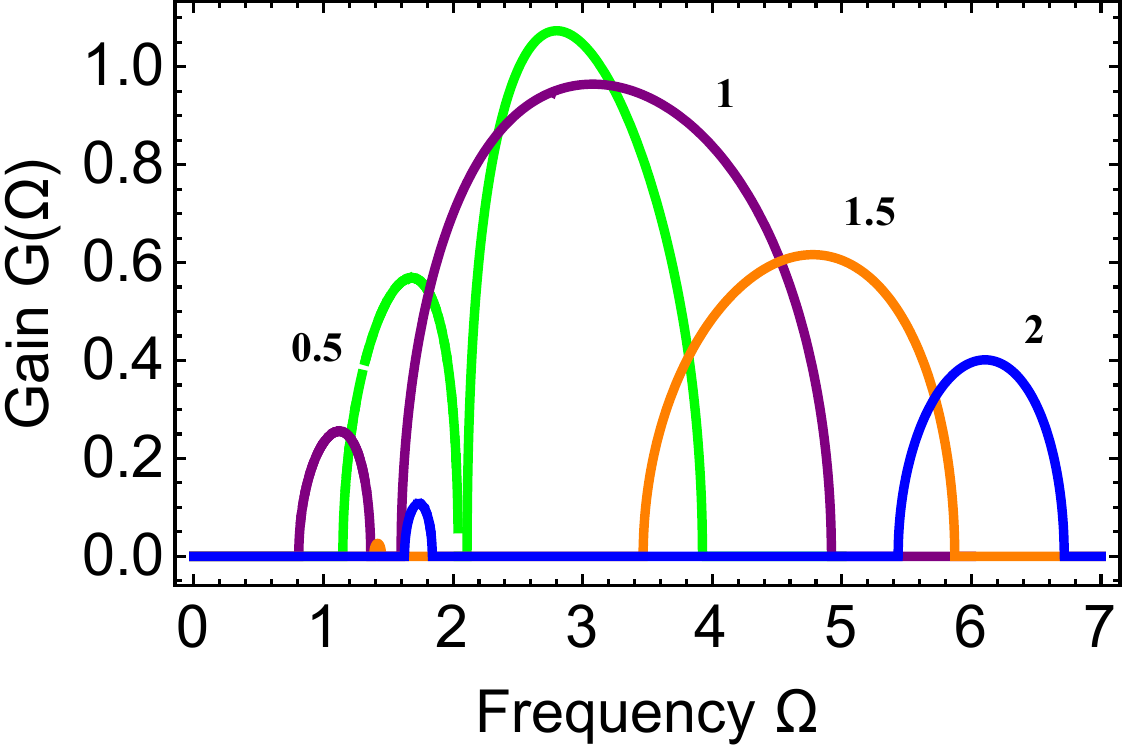}
\caption{\textls[-18]{The MI gain spectra in the normal-GVD regime ($\protect\sigma %
_{1}=-1$) at different values of the asymmetry ratio of the CW state ($%
\protect\eta $ in Equation (\protect\ref{P})). Other parameters are $P=1,\protect%
\kappa =1.1$, $\protect\alpha =0.1,\Gamma =2$ and $\protect\rho =0.01$.}}
\label{fig:N6}
\end{figure}

\subsection{The Coupler with Opposite Signs of the Dispersion in the Two
Cores}
The case of the opposite (``mixed'') GVD signs in the two cores of the
coupler, which corresponds to $\alpha <0$ in Equation (\ref{eqn:5}), is obviously
interesting, as well \cite{Kaup98}. For this purpose, we assume the anomalous and
normal GVD in the bar and cross channels, respectively.

\begin{figure}[H]
\centering
\includegraphics[width=0.4\linewidth]{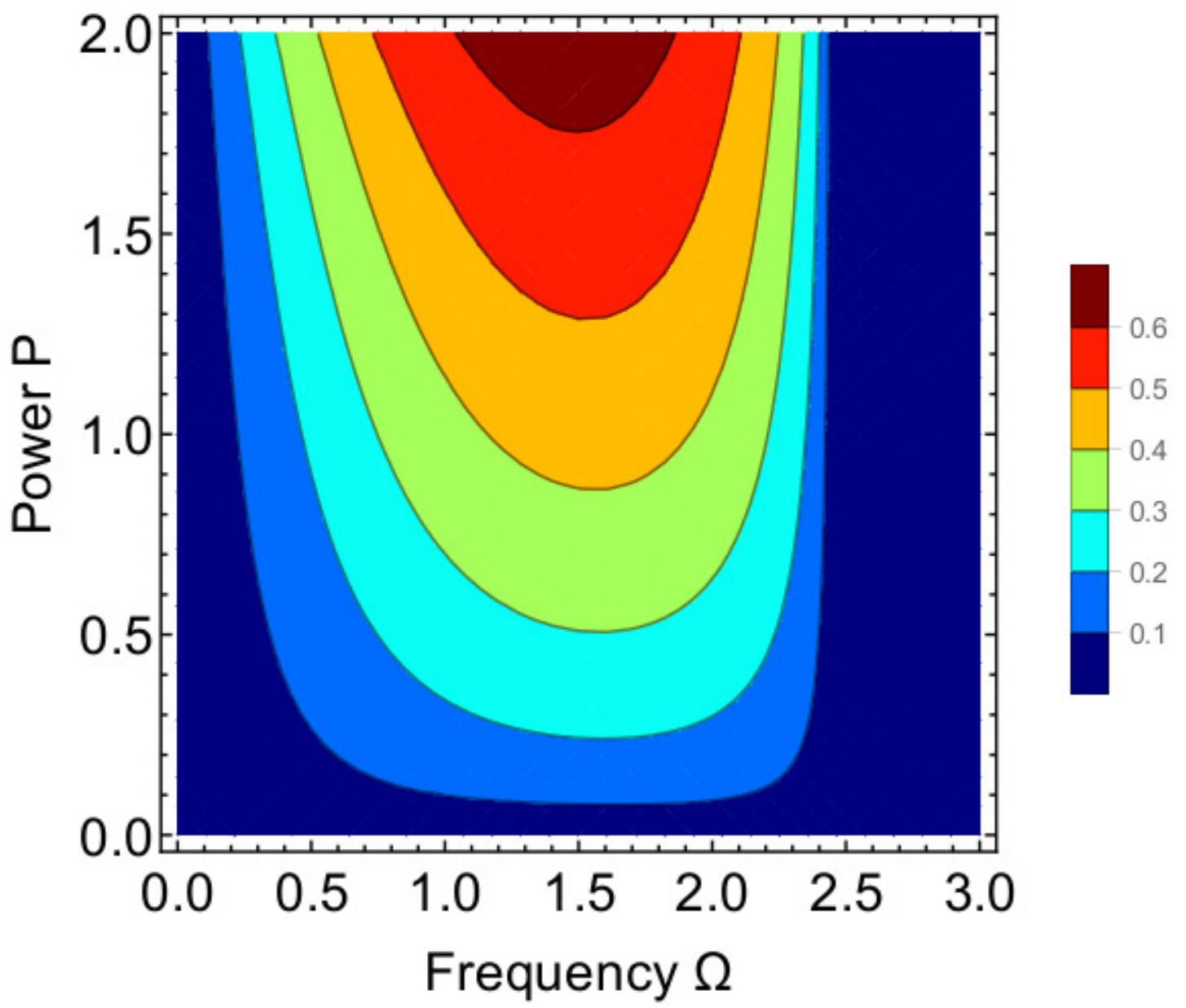}
\caption{The contour plot showing the dependence of MI gain on the CW total
power, $P$, in the mixed-GVD coupler ($\protect\alpha <0$). The parameters
are $\protect\eta =0.5,\protect\alpha =-0.1,\Gamma =2,\protect\kappa =1$,
and $\protect\rho =0.1$.}
\label{fig:AN1}
\end{figure}

\begin{figure}[H]
\centering\includegraphics[width=0.4\linewidth]{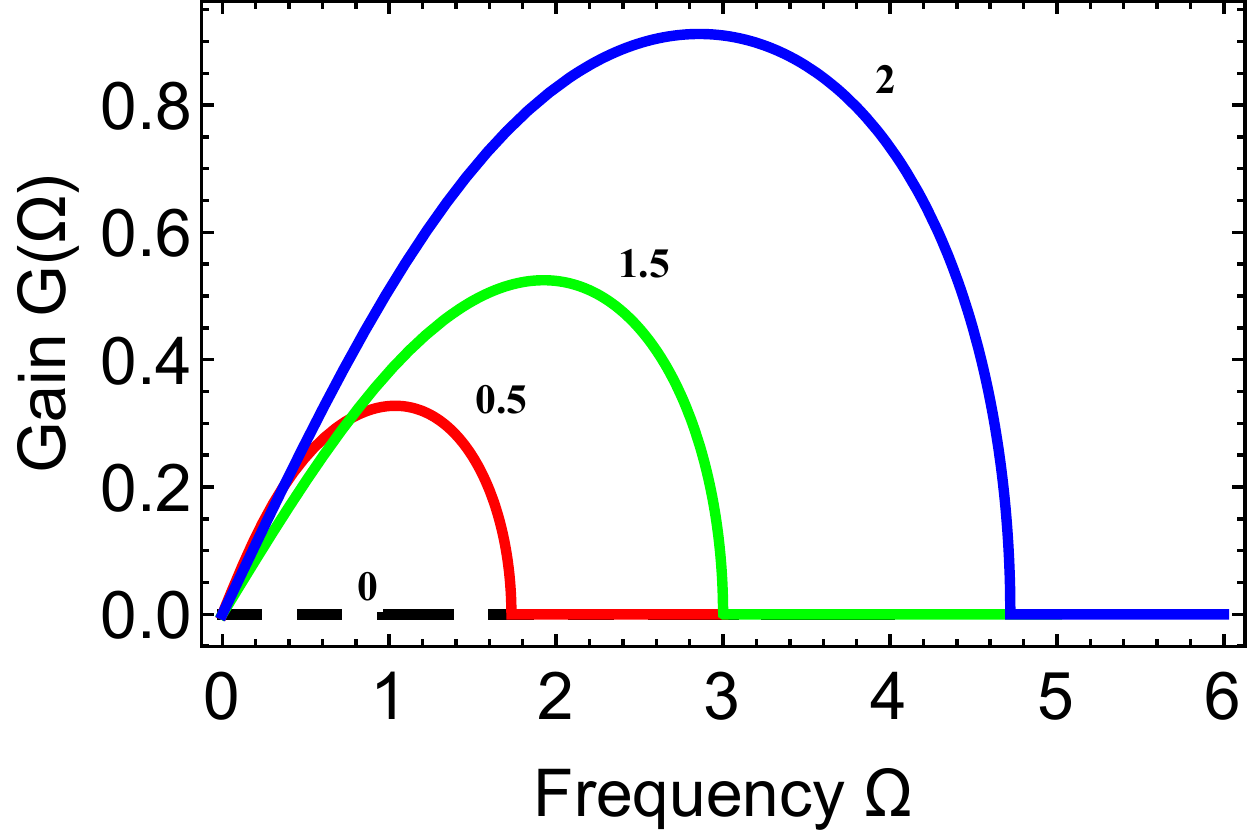}
\caption{The MI gain spectra for different values of the coupling
coefficient, $\protect\kappa $ (indicated near the curves), in the mixed-GVD
coupler ($\protect\alpha <0$) for $P=1,\protect\eta =0.7$, $\protect\alpha %
=-0.1,\Gamma =2$ with $\protect\rho =0.01$.}
\label{fig:AN2}
\end{figure}

\textls[-15]{Figure \ref{fig:AN1} shows the effect of the CW power, $P$, on the MI in the
mixed-GVD coupler. The figure demonstrates that the MI gain and bandwidth
monotonically increase with the growth of $P$.} It should be noted that the
spectra obtained for this case are somewhat different in comparison with the
conventional MI spectra, as the gain is stretched over a broad interval of
the perturbation frequency when the CW power is low ($P<1$). In the present
case, the effect of the coupling coefficient, $\kappa $, on the MI, which is
shown in Figure \ref{fig:AN2}, is essentially the same as demonstrated above
for the coupler with the normal GVD in both cores; see Figure \ref{fig:N2}a.
Namely, the MI\ gain and bandwidth increase with the growth of $\kappa $.

\begin{figure}[H]
\centering
\subfloat[\label{fig:AN3a}]{\includegraphics[width=0.4%
\linewidth]{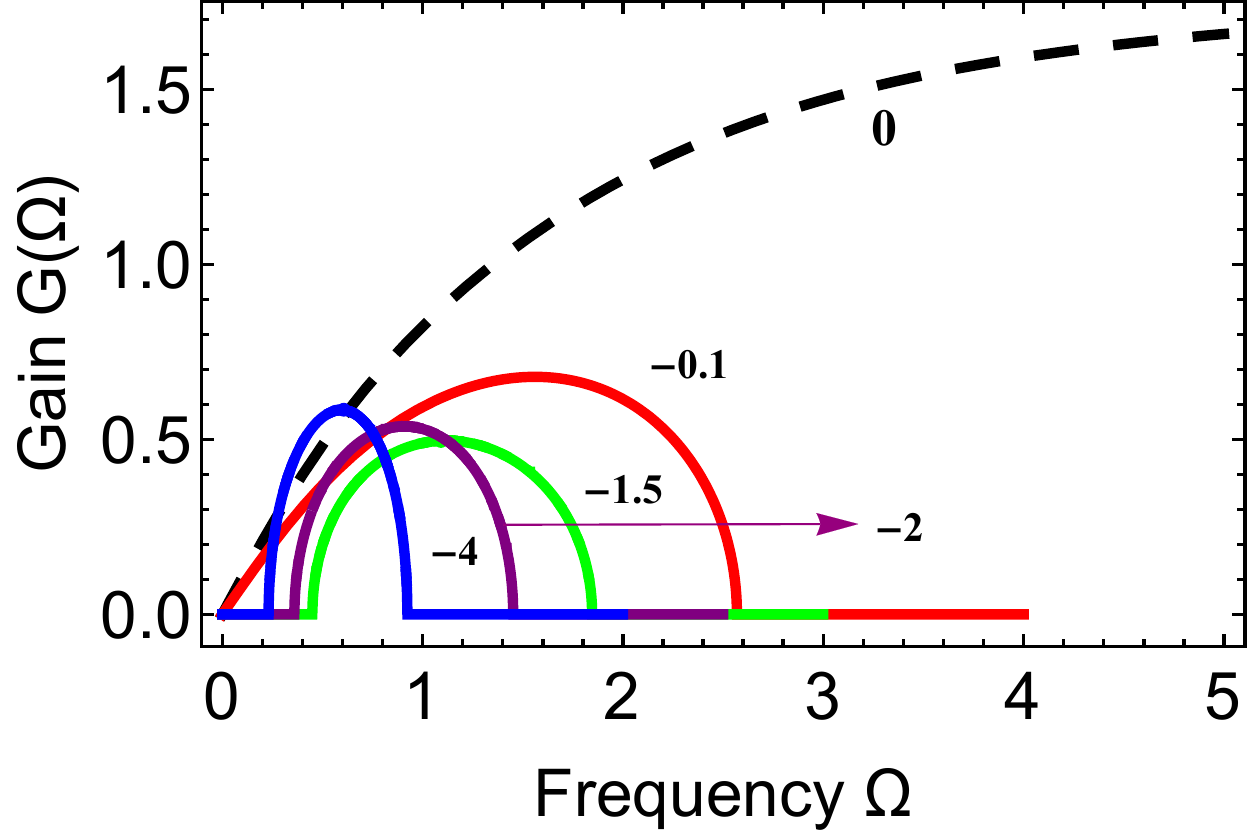}} \hspace{0.5in} \subfloat[\label{fig:AN4a}]{%
\includegraphics[width=0.4\linewidth]{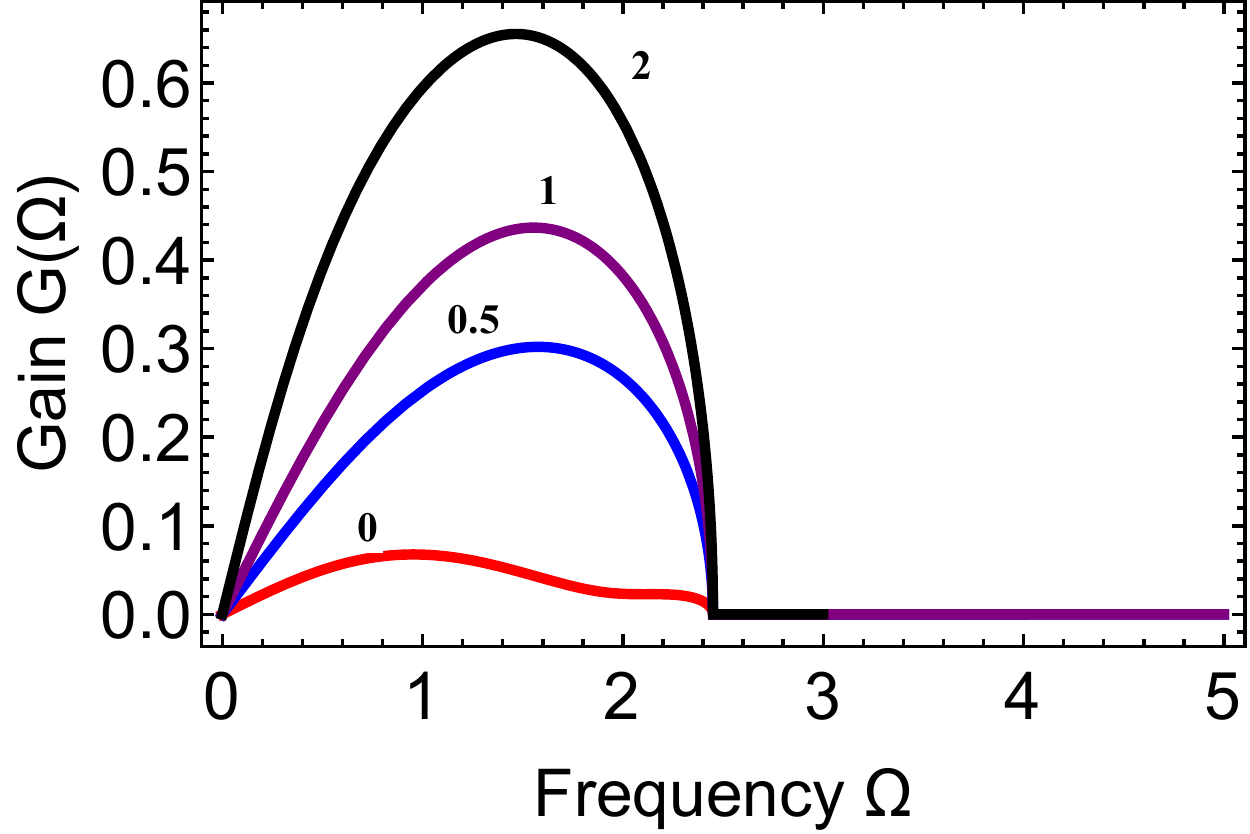}}
\caption{The MI gain spectra in the mixed-GVD coupler ($\protect\alpha <0$).
(\textbf{a}) The results for different negative values of the ratio of the GVD
coefficients in the two cores ($\protect\alpha $), which are indicated near
the curves for $P=2,\protect\eta =0.5,\protect\kappa =1.1,\Gamma =2$ with $%
\protect\rho =0.01$.
 (\textbf{b}) The results for different values of the ratio of
the nonlinearity coefficients in the two cores ($\Gamma $), which are
indicated near the curves. Parameters are same as in (a), except for $\protect\kappa =1$ and $\protect\alpha =-0.1$.}
\label{fig:AN3}
\end{figure}

The effects of the negative value of the ratio of the GVD coefficients, $%
\alpha<0$, and the ratio of the nonlinearity coefficients ($\Gamma $) in the
two cores are shown in Figure \ref{fig:AN3}. Similar to the coupler with the
anomalous GVD in each core, cf. Figure \ref{fig:3}b, the increase of $\alpha$
(see Figure \ref{fig:AN3}a) leads to shrinkage of the MI band. Like in the
coupler with the anomalous GVD in both cores, cf. Figure \ref{fig:3}c, the MI
gain increases with the growth of $\Gamma $, which is depicted in Figure \ref%
{fig:AN3}b; however, the difference is that, in the present case of the
mixed-GVD\ coupler, the bandwidth is not affected by the variation of $%
\Gamma $.

\begin{figure}[H]
\centering
\includegraphics[width=0.4\linewidth]{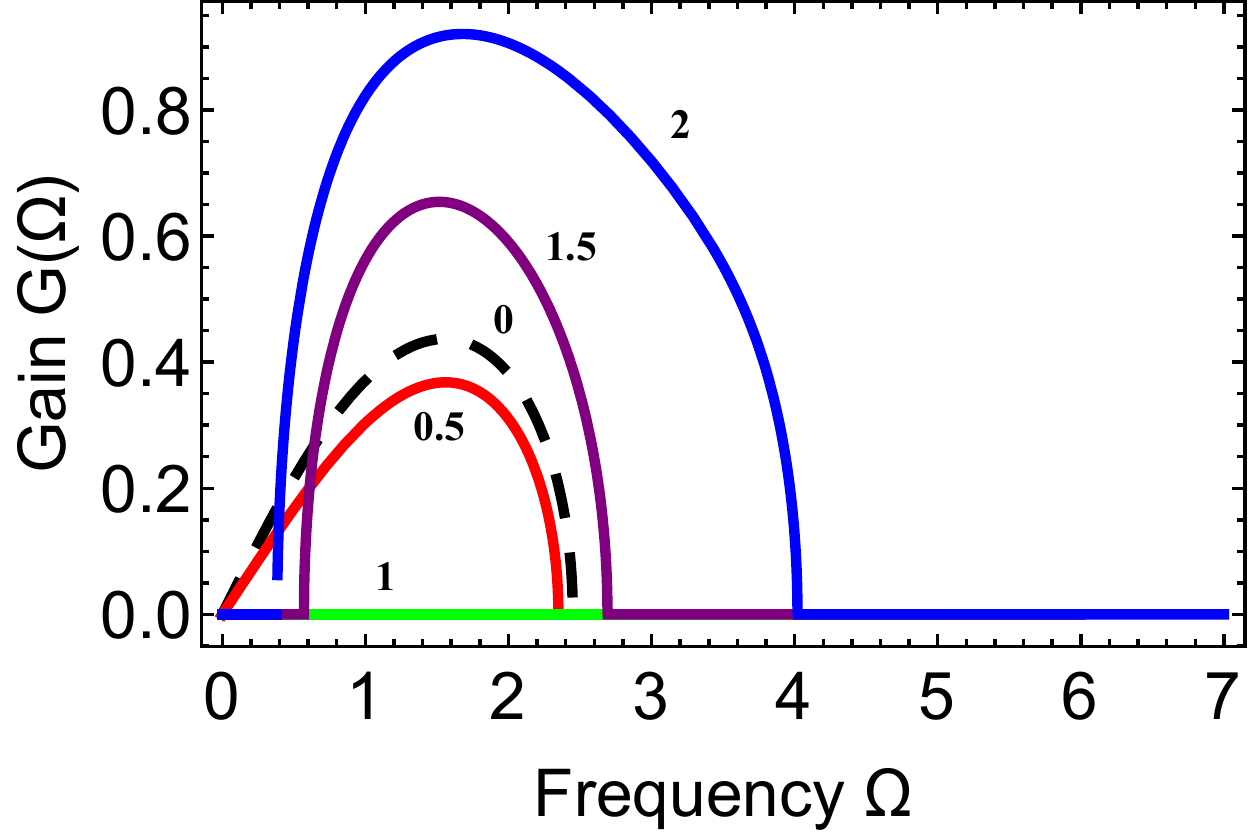}
\caption{The MI gain spectra for different values of the group-velocity
mismatch (walk-off) between the cores in the mixed-GVD coupler ($\protect%
\alpha <0$) for $P=\protect\kappa =1,\protect\eta =0.5,\protect\alpha %
=-0.1,\Gamma =2$.}
\label{fig:AN5}
\end{figure}

Figure \ref{fig:AN5} displays quite nontrivial evolution of the MI spectra
with the variation of the group-velocity mismatch (walk-off) between the cores, $%
\rho $ in Equation (\ref{eqn:5}). The evolution is very different from what is
demonstrated above for the coupler with the anomalous GVD in both cores, cf.
Figure~\ref{fig:6}a. Namely,~Figure~\ref{fig:AN5} shows that the increase of $%
\rho $ from zero to one suppresses the MI, which completely vanishes at $\rho
=1$. The system recovers the MI, which features monotonically increasing
gain and bandwidth, with the further increase of $\rho $ to values $\rho >1$%
.
\unskip
\vspace{-18pt}
\unskip

\begin{figure}[H]
\centering
\includegraphics[width=0.35\linewidth]{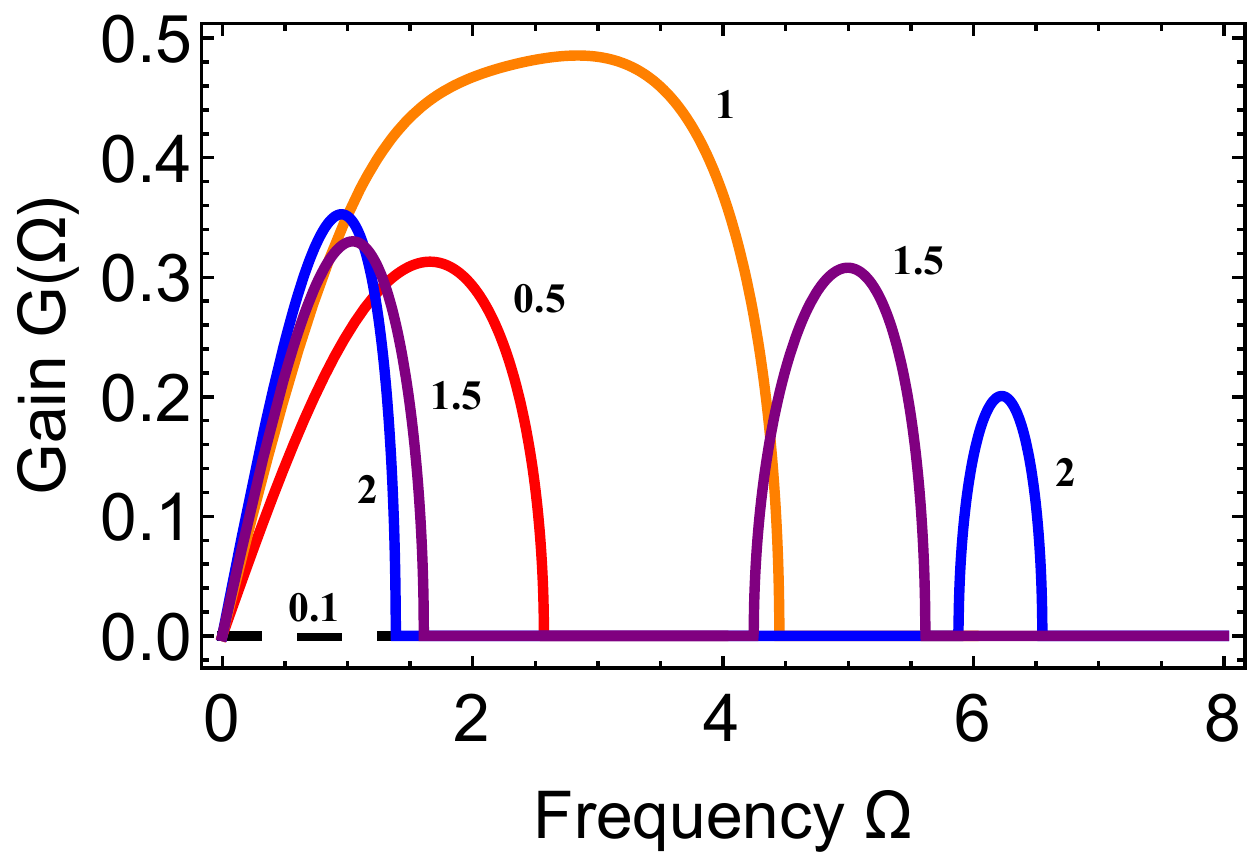}
\caption{The MI gain spectra for different values of the asymmetry ratio, $%
\protect\eta $, of the CW state (see Equation (\protect\ref{P})), which are
indicated near the figures, in the mixed-GVD coupler ($\protect\alpha <0$)
for \mbox{$P=0.5,\protect\kappa =1.1,\protect\alpha =-0.1,\Gamma =2$} with $\protect%
\rho =0.01$.}
\label{fig:AN6}
\end{figure}

Next, we consider the impact of the asymmetry parameter $\eta $ in the CW
state; see Equation (\ref{P}). As~show in Figure \ref{fig:AN6}, there is no MI at
small values of $\eta $, such as $\eta =0.1$. With the subsequent increase
of $\eta $ up to $\eta =1$, the MI gain and bandwidth increase, similar to
what was observed above in the coupler with anomalous GVD in both cores; see
Figure \ref{fig:6}b. However, the situation becomes completely different at $%
\eta >1$, when the CW amplitude is higher in the bar channel: the MI band
splits into two narrower ones, with smaller values of the gain.

\section{Direct Simulations}
\textls[-15]{The analytical results obtained above for the MI have been checked against
numerical calculations of the instability spectra. Numerical methods are
actually more relevant for direct simulations of the nonlinear evolution of
the MI, which was analyzed above in the linear approximation. The
simulations were carried out by dint of the well-known split-step Fourier
method \cite{Govin3} (using MATLAB).} Most results displayed below were
obtained using numerical meshes with $512$ Fourier points and periodic
boundary conditions with respect to variable $\tau $. Simulations performed
with denser meshes have produced virtually identical results. Furthermore,
results of the nonlinear development of the MI are not sensitive to details
of initial small perturbations, which initiate the onset of the MI.

The initial conditions were taken in the form of the CW to which a small
periodic perturbation was added:
\begin{equation}
u_{j}(0,\tau )=A_{j}+a_{0}\cos (\omega _{0}\tau ),\quad (j=1,2),
\label{eqn:18}
\end{equation}%
where $a_{0}$ is a small amplitude of the perturbation and $\omega _{0}$ is
its frequency.

\begin{figure}[H]
\centering
\subfloat[]{\includegraphics[width=0.4\linewidth]{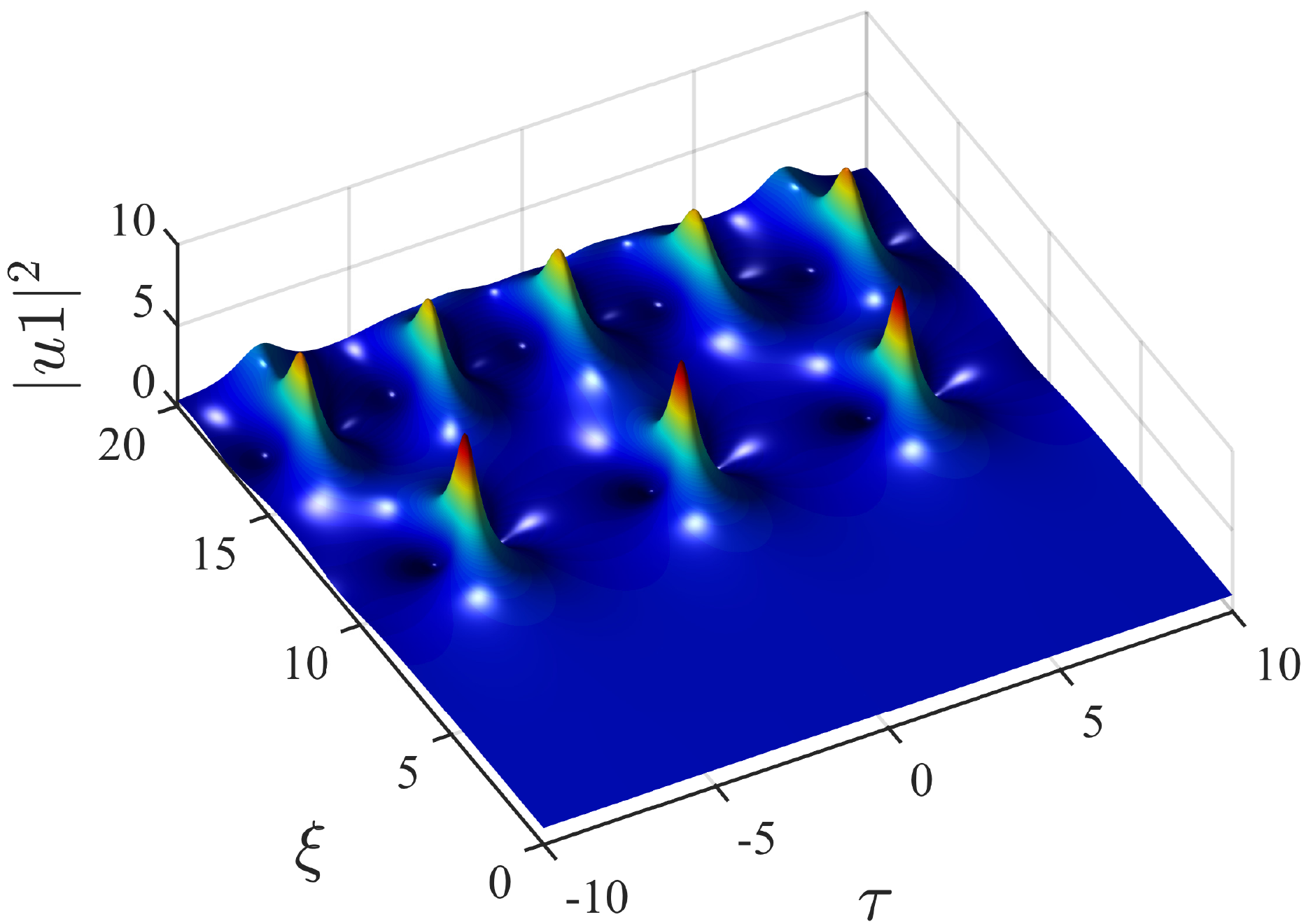}} %
\subfloat[]{\includegraphics[width=0.4\linewidth]{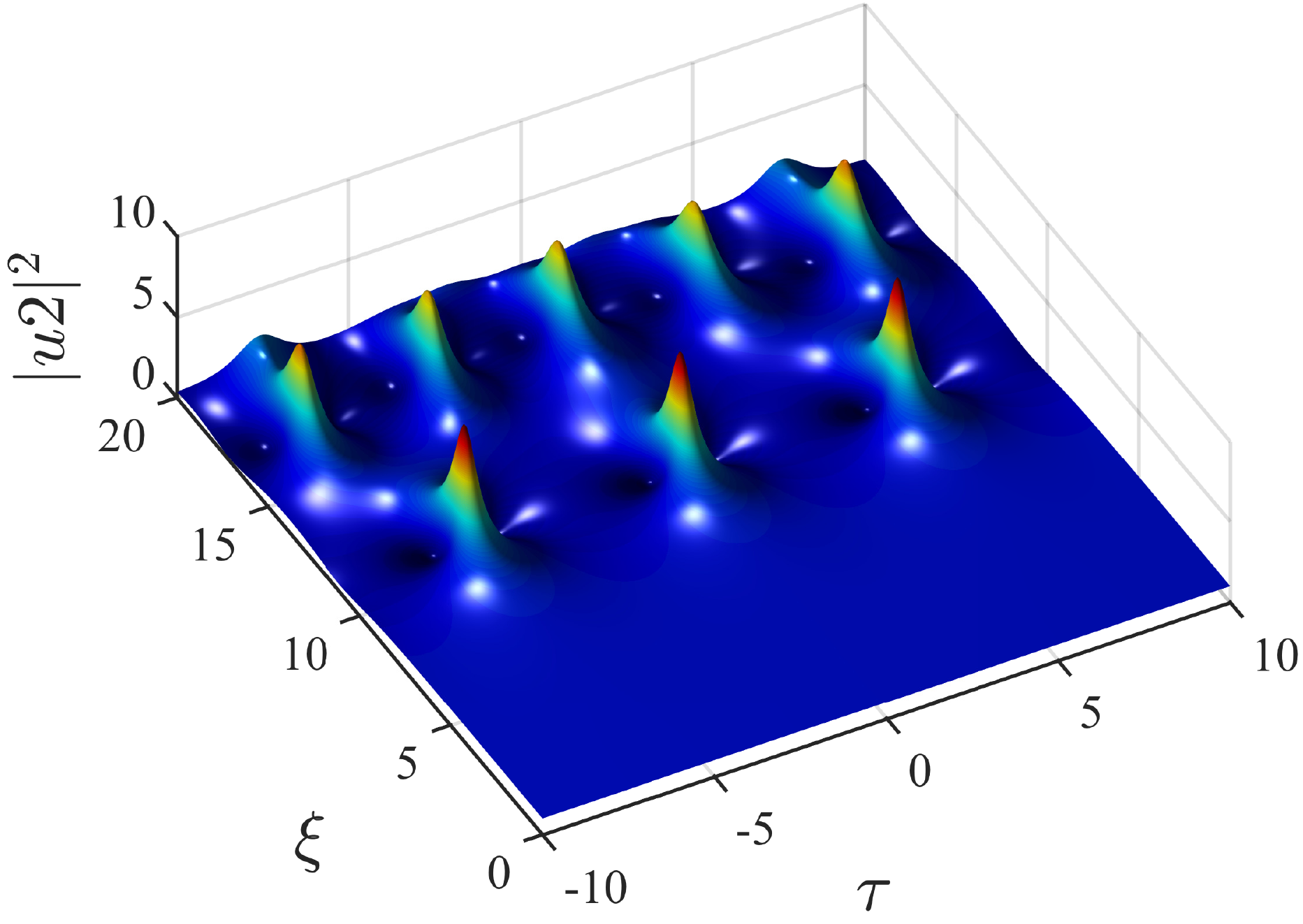}}
\caption{The evolution of the MI in the symmetric coupler with anomalous GVD
($\protect\sigma =1$) in the bar (\textbf{a}) and cross (\textbf{b}) channels for equal
amplitudes of the underlying CW state, $A_{1}=A_{2}=1$, with perturbation
parameters $a_{0}=0.0001$ and $\protect\omega _{0}=1$, in Equation (\protect\ref%
{eqn:18}). Other parameters are $\ \protect\alpha =\Gamma =\protect\kappa =1$
and $\protect\rho =\protect\chi =0$.}
\label{fig:S1}
\end{figure}

\begin{figure}[H]
\centering
\subfloat[]{\includegraphics[width=0.38\linewidth]{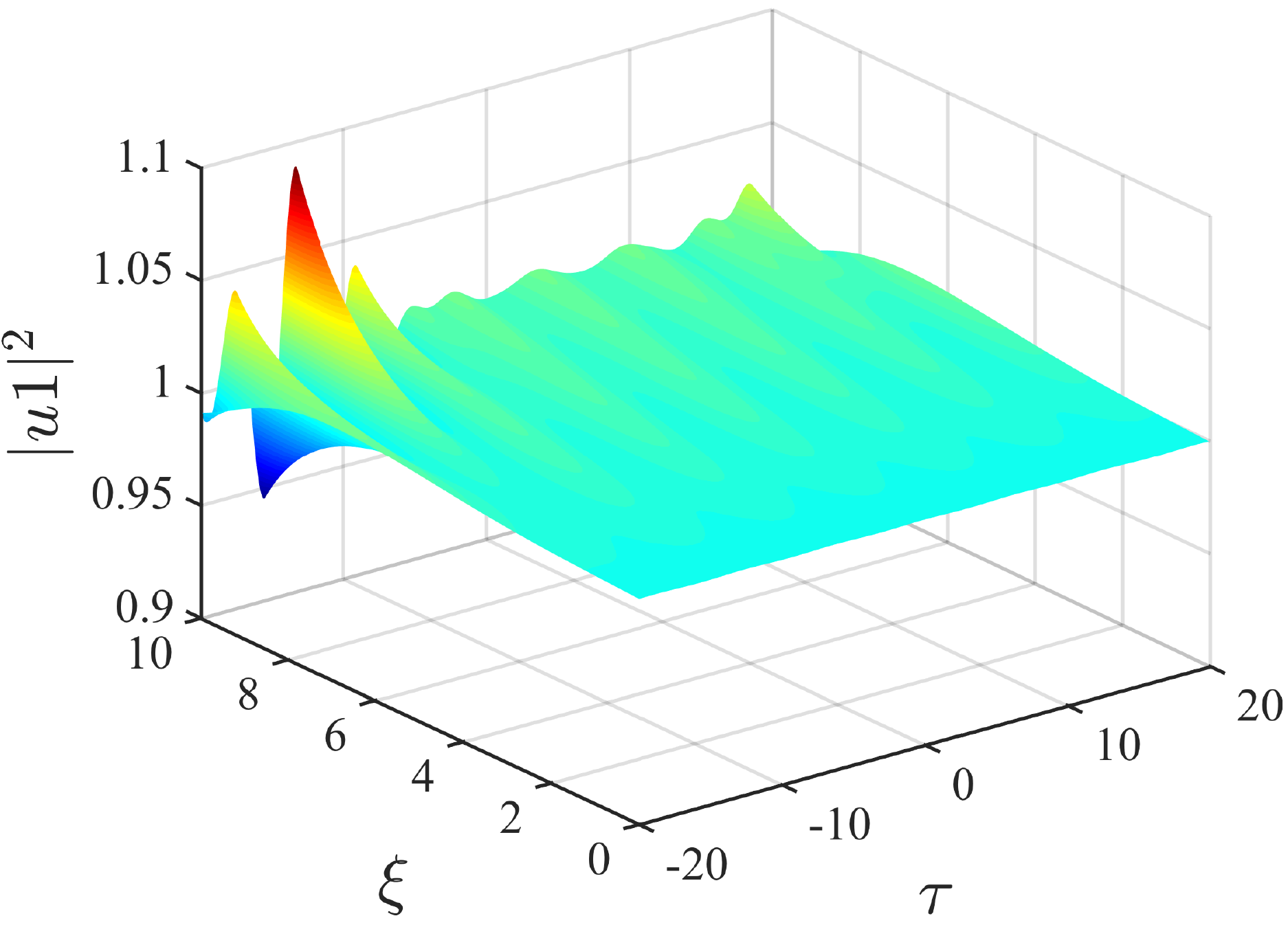}} %
\subfloat[]{\includegraphics[width=0.38\linewidth]{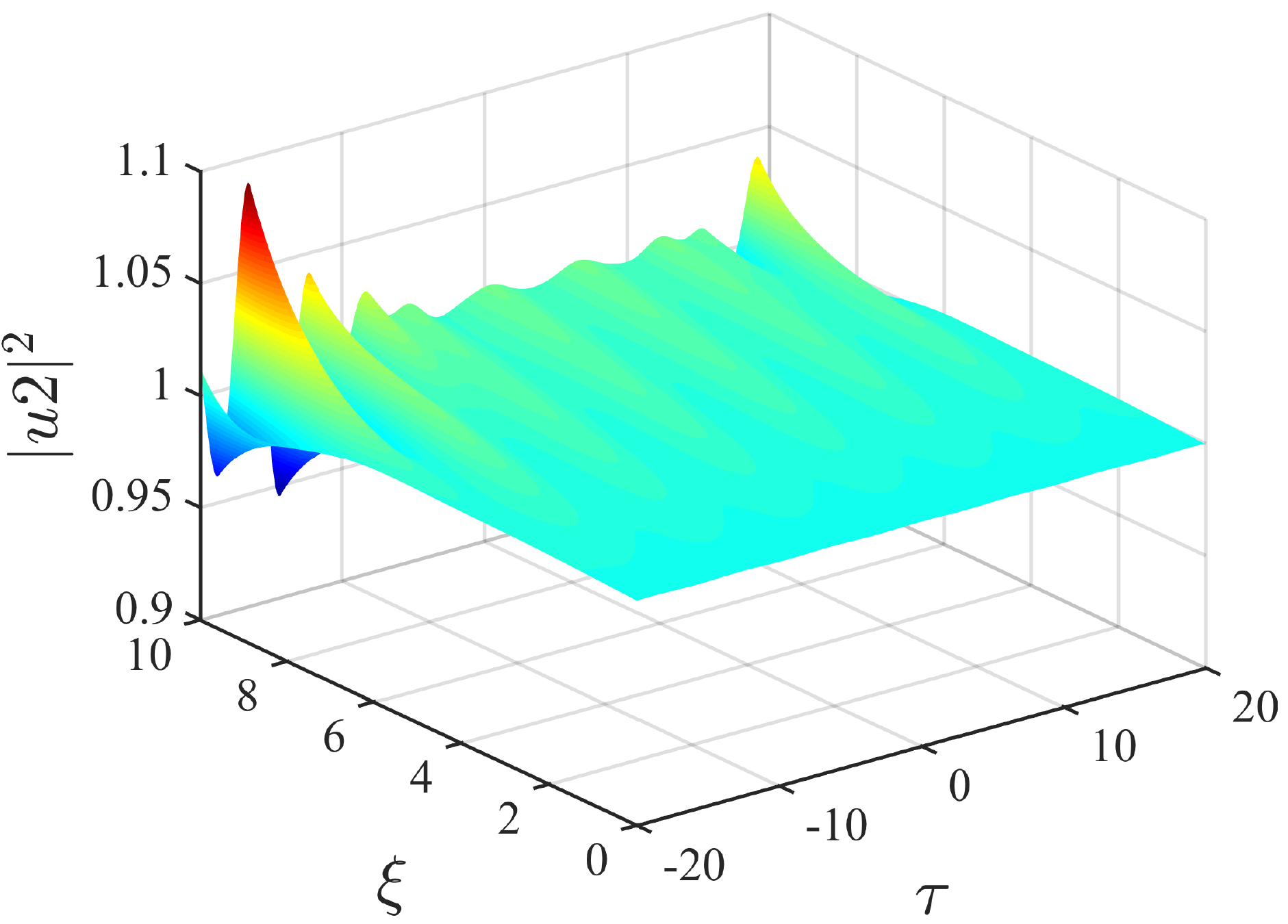}}
\caption{The influence of the group-velocity mismatch, $\protect\rho =1$, on
the evolutions of the MI in the bar (\textbf{a}) and cross (\textbf{b}) channels in the
anomalous-GVD regime. Other system parameters are, $A_{1}=A_{2}=\protect%
\omega _{0}=\protect\alpha =\Gamma =\protect\kappa =1,\protect\chi =0$ and $%
a_{0}=0.0001$.}
\label{fig:S2}
\end{figure}

\begin{figure}[H]
\centering
\subfloat[]{\includegraphics[width=0.45\linewidth]{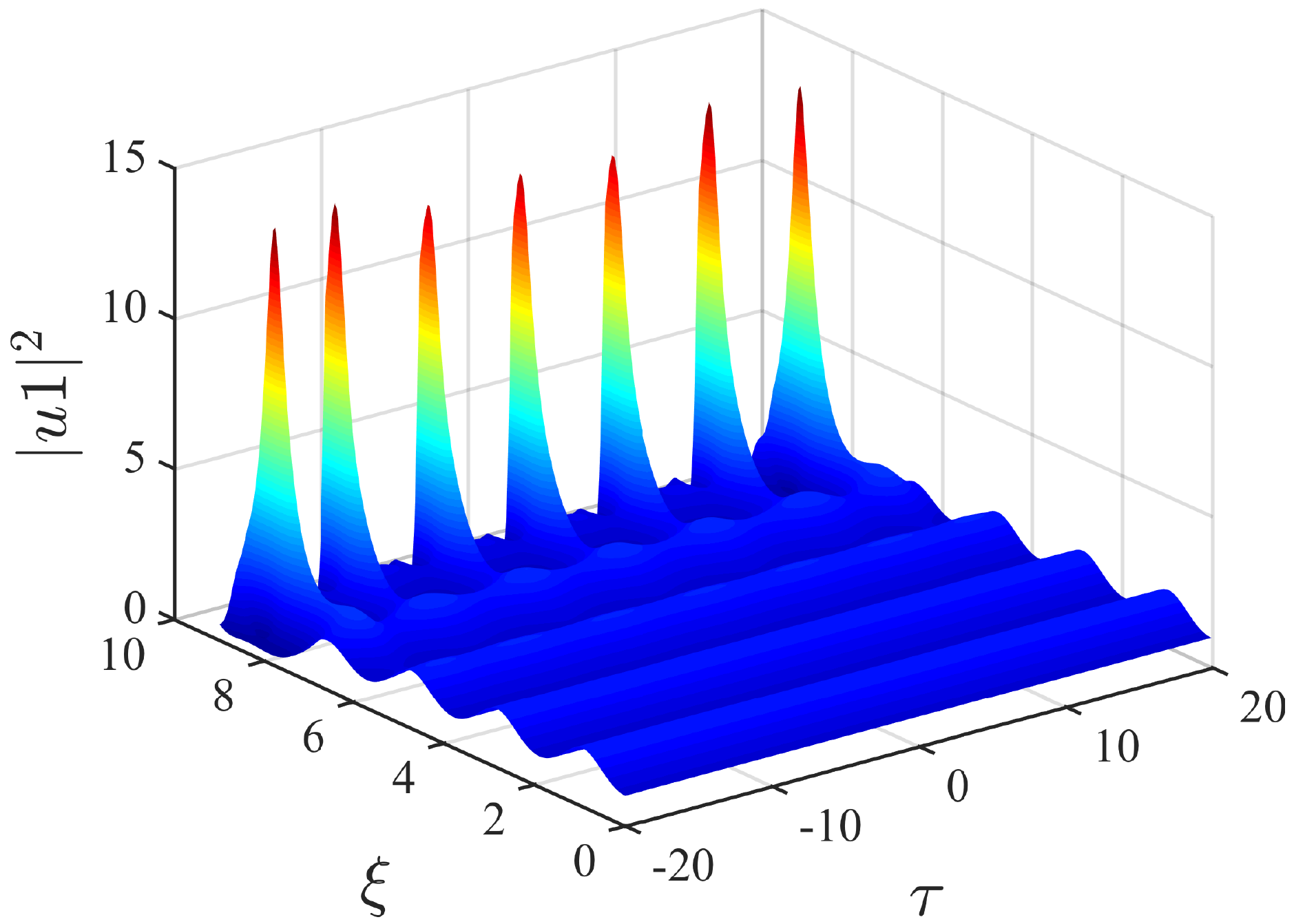}} %
\subfloat[]{\includegraphics[width=0.45\linewidth]{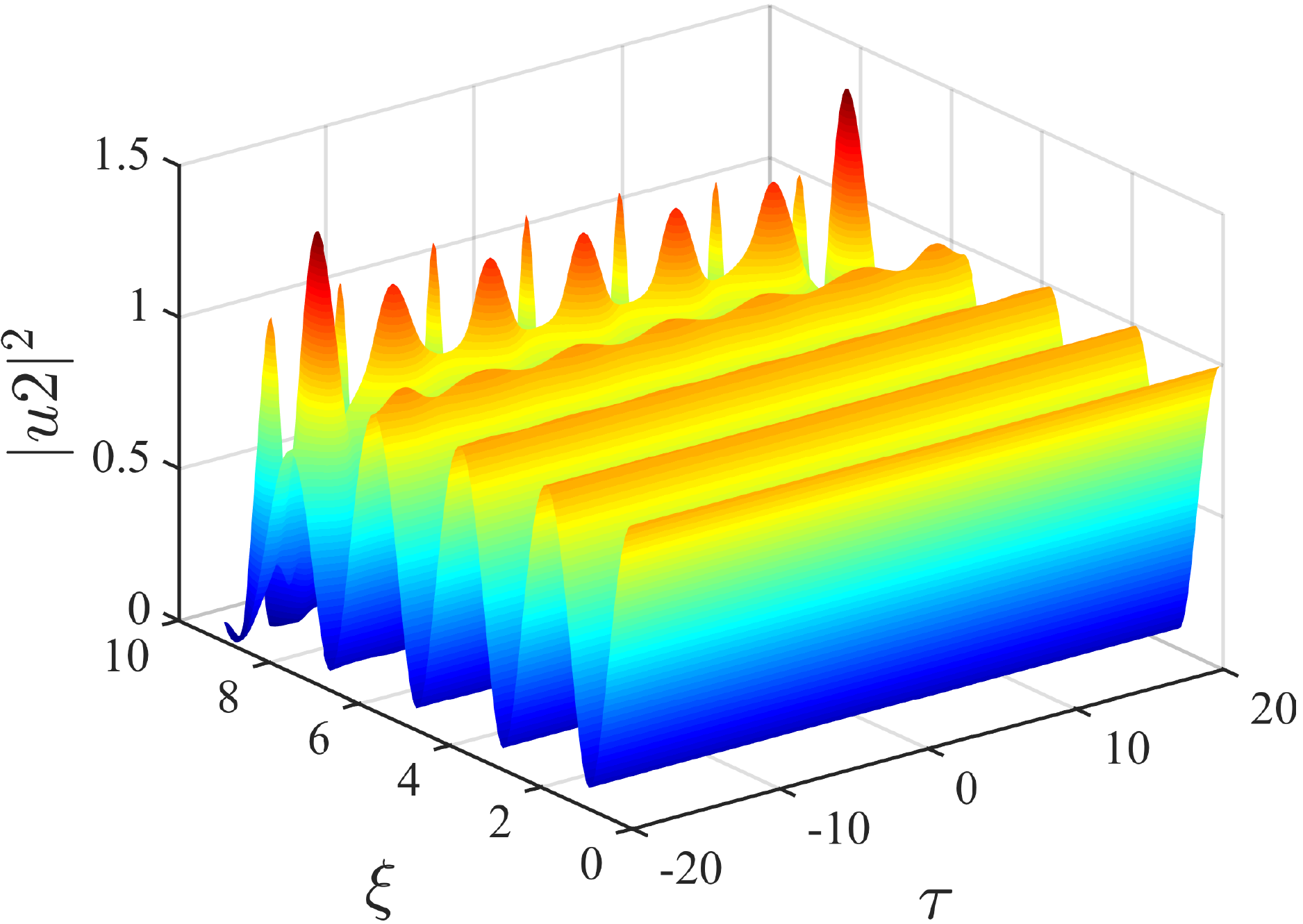}}
\caption{The influence of the phase-velocity mismatch, $\protect\chi =1$, on
the evolution of the MI in the bar~(\textbf{a}) and cross (\textbf{b}) channels in the
anomalous-GVD dispersion regime. Other system parameters are $A_{1}=A_{2}=%
\protect\omega _{0}=\protect\alpha =\Gamma =\protect\kappa =1,\protect\rho %
=0 $ and $a_{0}=0.0001$.}
\label{fig:S3}
\end{figure}

\begin{figure}[H]
\centering
\subfloat[]{\includegraphics[width=0.45\linewidth]{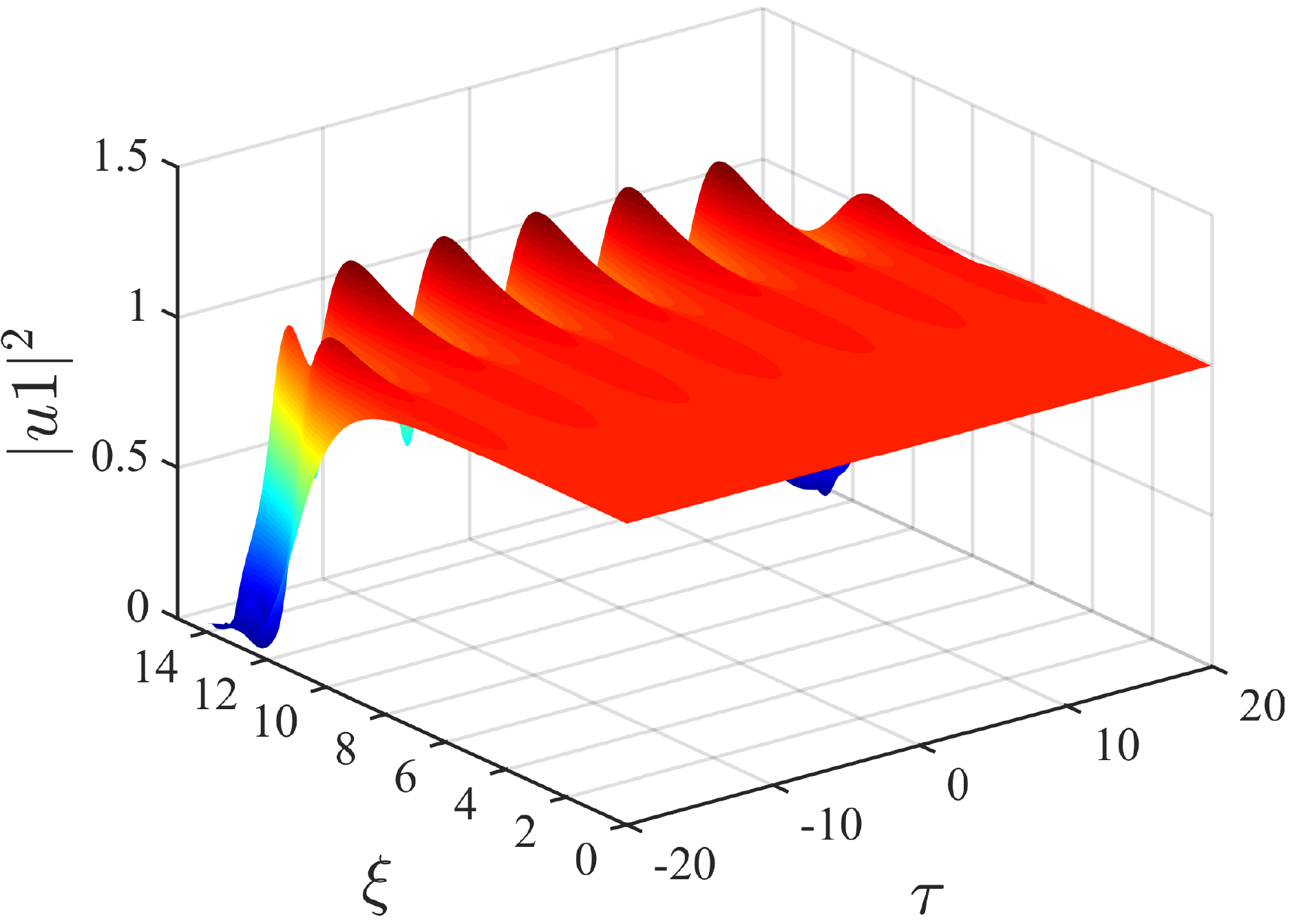}} %
\subfloat[]{\includegraphics[width=0.45\linewidth]{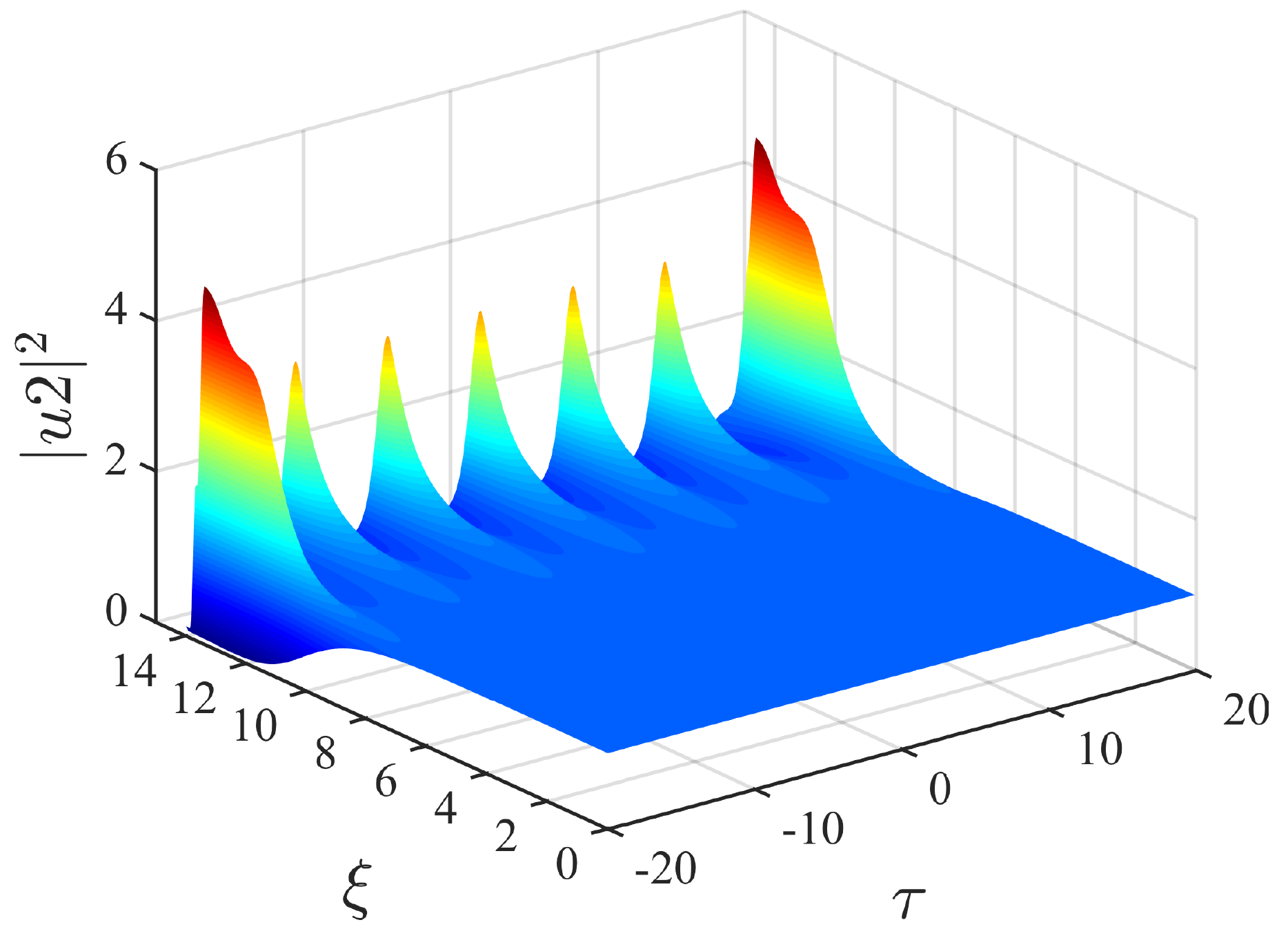}}
\caption{\textls[-18]{The MI evolution in the bar\ (\textbf{a}) and cross (\textbf{b}) cores, in the case
of the anomalous GVD in the bar channel, and zero GVD ($\protect\alpha =0$)
in the cross channel.} Other parameters are $A_{1}=A_{2}=\protect\omega %
_{0}=\Gamma =\protect\kappa =1$ and $a_{0}=0.0001$.}
\label{fig:S4}
\end{figure}

\begin{figure}[H]
\centering
\subfloat[]{\includegraphics[width=0.5\linewidth]{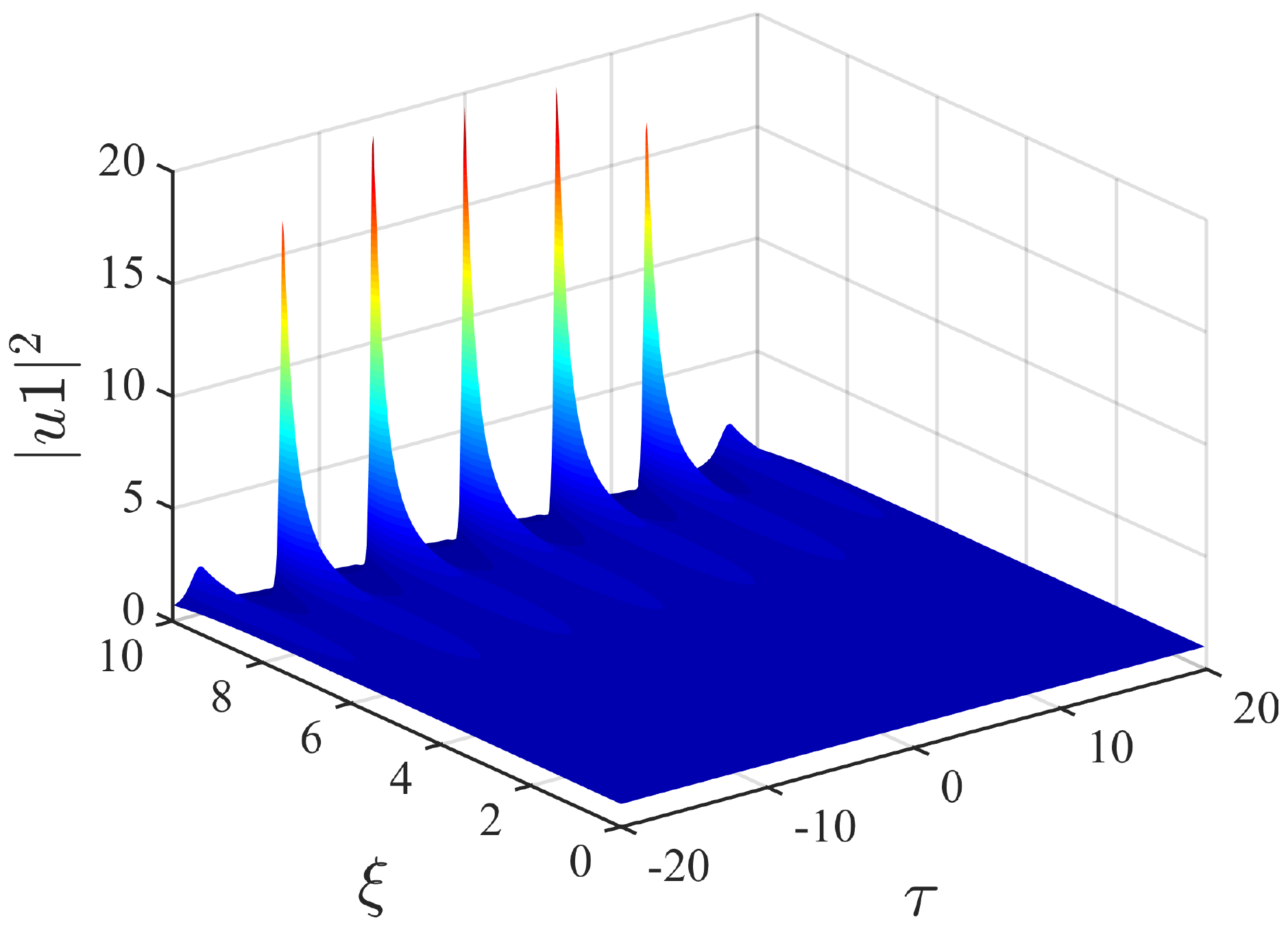}} %
\subfloat[]{\includegraphics[width=0.5\linewidth]{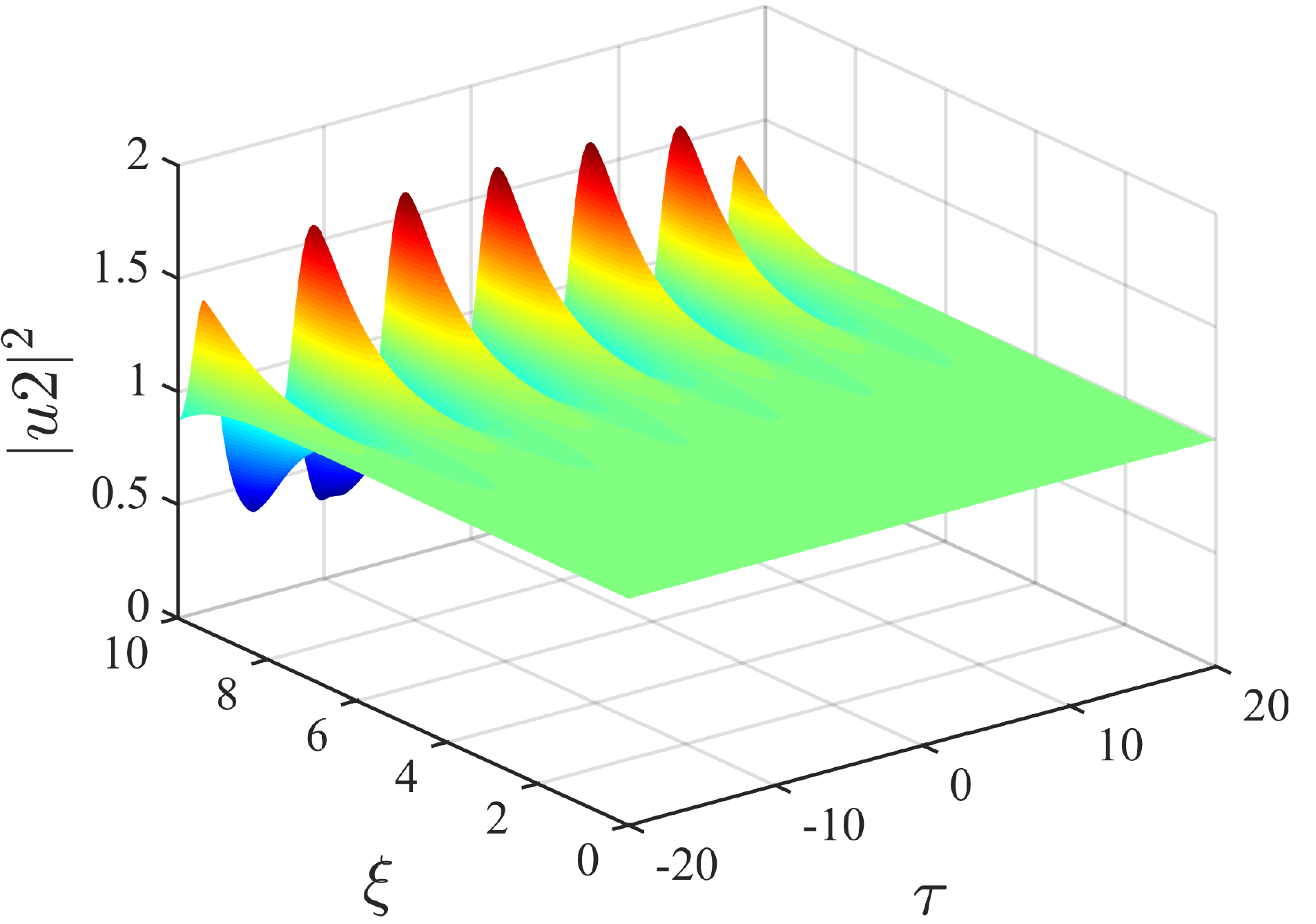}}
\caption{The same as in Figure \protect\ref{fig:S4}, but when the difference
in the GVD coefficients is $\protect\alpha =2$.} % We think it is not necessary to once again write the explanations for the subfigure a and b. And the above sentence is enough.
\label{fig:S5}
\end{figure}

\begin{figure}[H]
\centering
\subfloat[]{\includegraphics[width=0.48\linewidth]{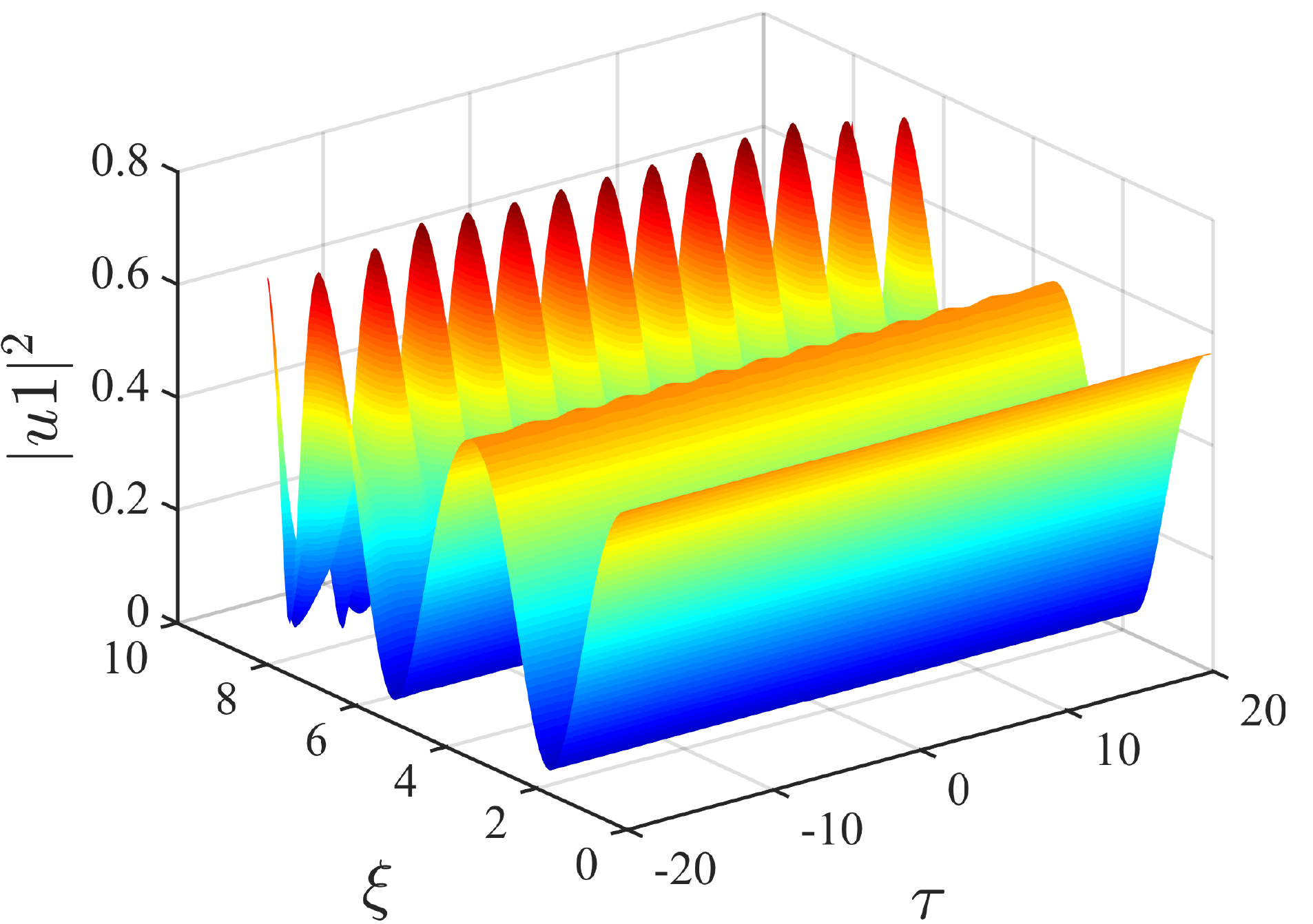}} %
\subfloat[]{\includegraphics[width=0.48\linewidth]{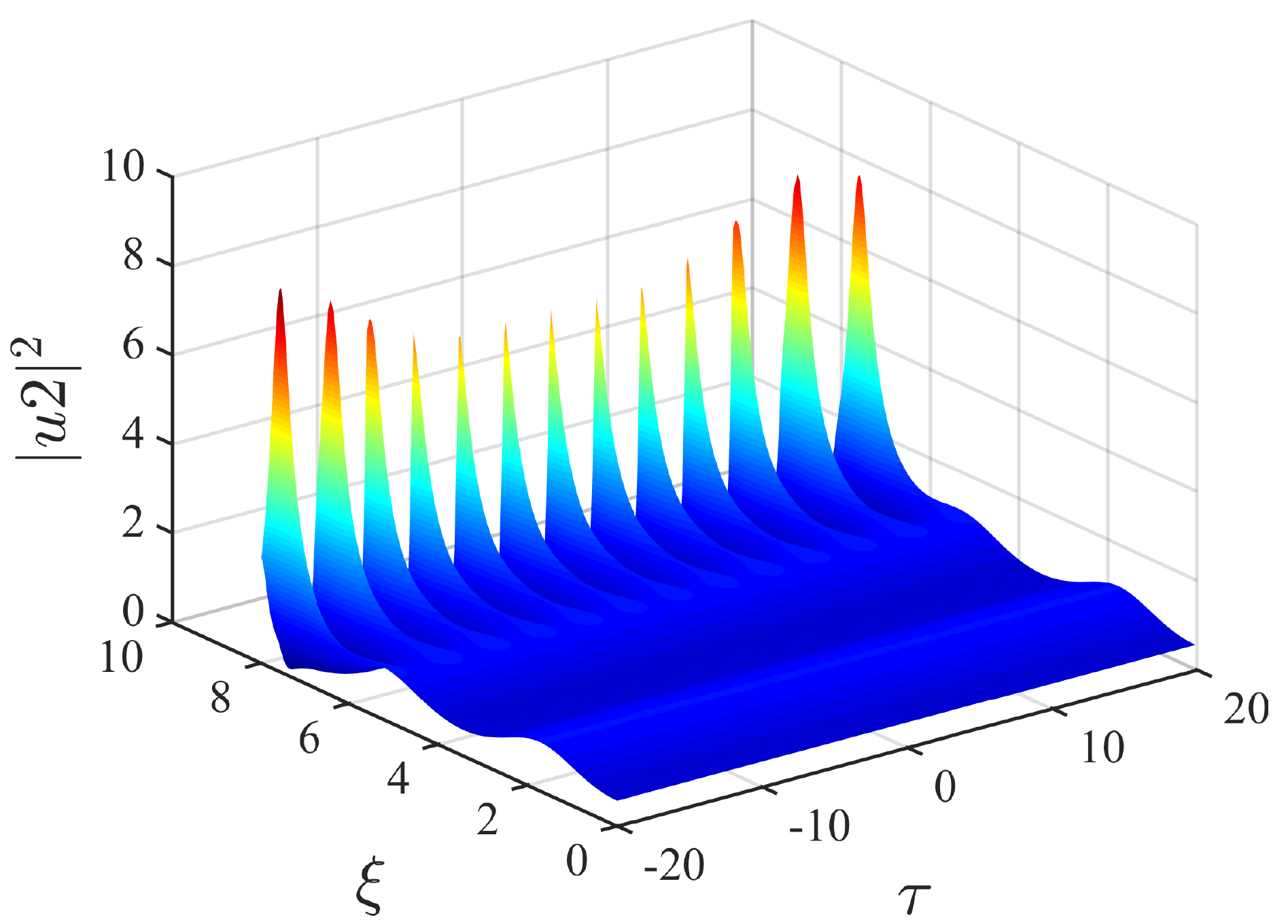}}
\caption{The influence of the ratio of the nonlinearity coefficients in the
two cores, $\Gamma =2$, on the MI evolution in the bar (\textbf{a}) and cross (\textbf{b})
channels in the anomalous-GVD regime. Other parameters are $A_{1}=A_{2}=0.75,%
\protect\omega _{0}=2,\protect\alpha =\protect\kappa =1,\protect\rho =%
\protect\chi =0$ and $a_{0}=0.0001$.}
\label{fig:S6}
\end{figure}

\begin{figure}[H]
\centering
\subfloat[]{\includegraphics[width=0.45\linewidth]{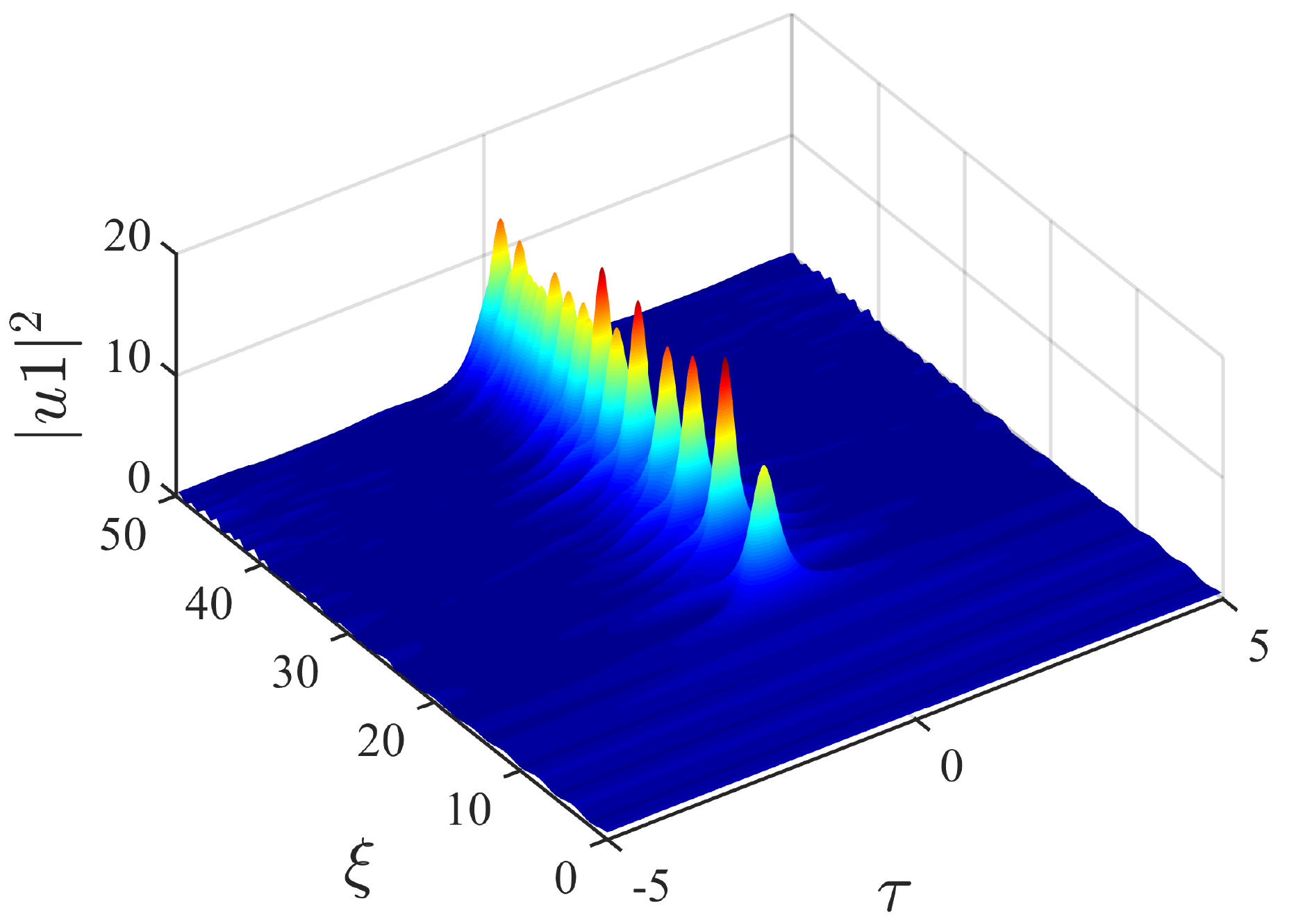}} %
\subfloat[]{\includegraphics[width=0.45\linewidth]{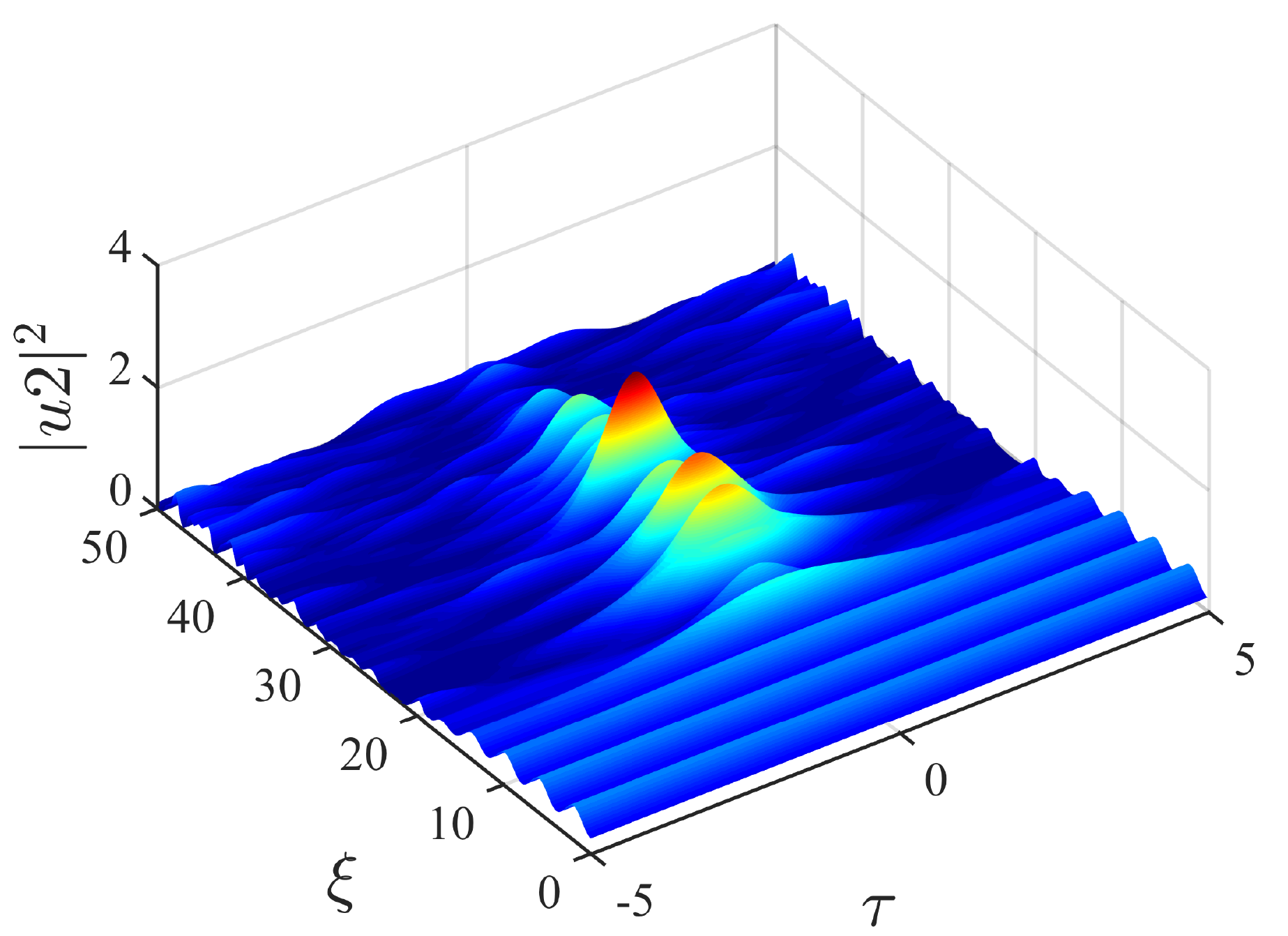}}
\caption{The MI-driven evolution in the bar (\textbf{a}) and cross (\textbf{b}) channels in
the anomalous-GVD regime ($\protect\sigma =1$) for initial CW amplitudes $%
A_{1}=0.75,A_{2}=0.5$ and perturbation parameters $a_{0}=0.0009,\protect%
\omega _{0}=1$. Other parameters are $\protect\alpha =2,\Gamma =1,\protect%
\rho =0.01,\protect\chi =0.001$ and $\protect\kappa =1$.}
\label{fig:S7}
\end{figure}

\begin{figure}[H]
\centering
\subfloat[]{\includegraphics[width=0.45\linewidth]{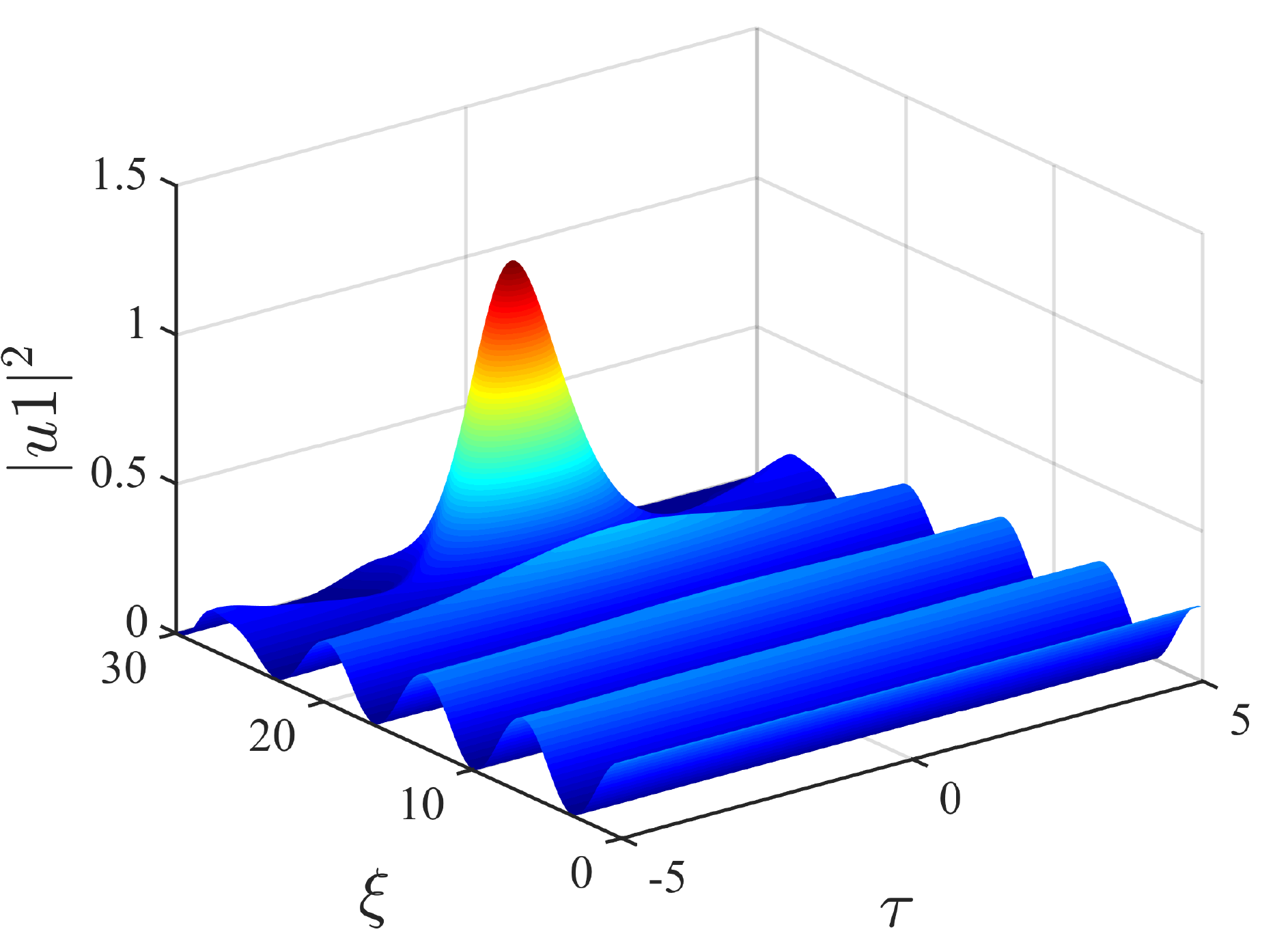}} %
\subfloat[]{\includegraphics[width=0.45\linewidth]{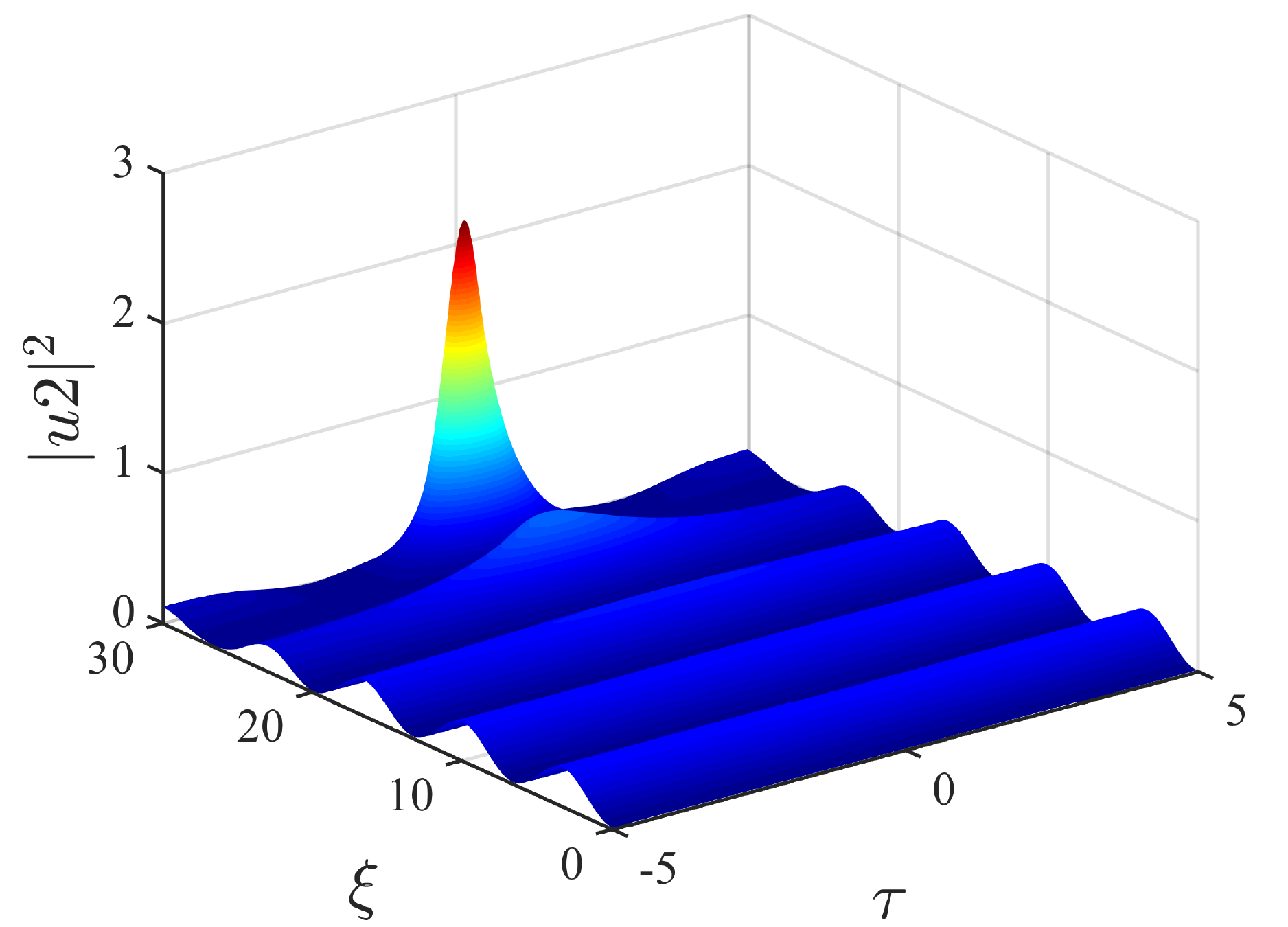}}
\caption{The same as in Figure \protect\ref{fig:S7}, but for 
$A_{1}=0.5,A_{2}=0.1$, $a_{0}=0.0007,$ $\protect\chi =0.01$ and 
$\protect\kappa =0.5$.} %We think it is not necessary to once again write the explanations for the subfigure a and b. And the above sentence is enough.
\label{fig:S8}
\end{figure}

\begin{figure}[H]
\centering
\subfloat[]{\includegraphics[width=0.45\linewidth]{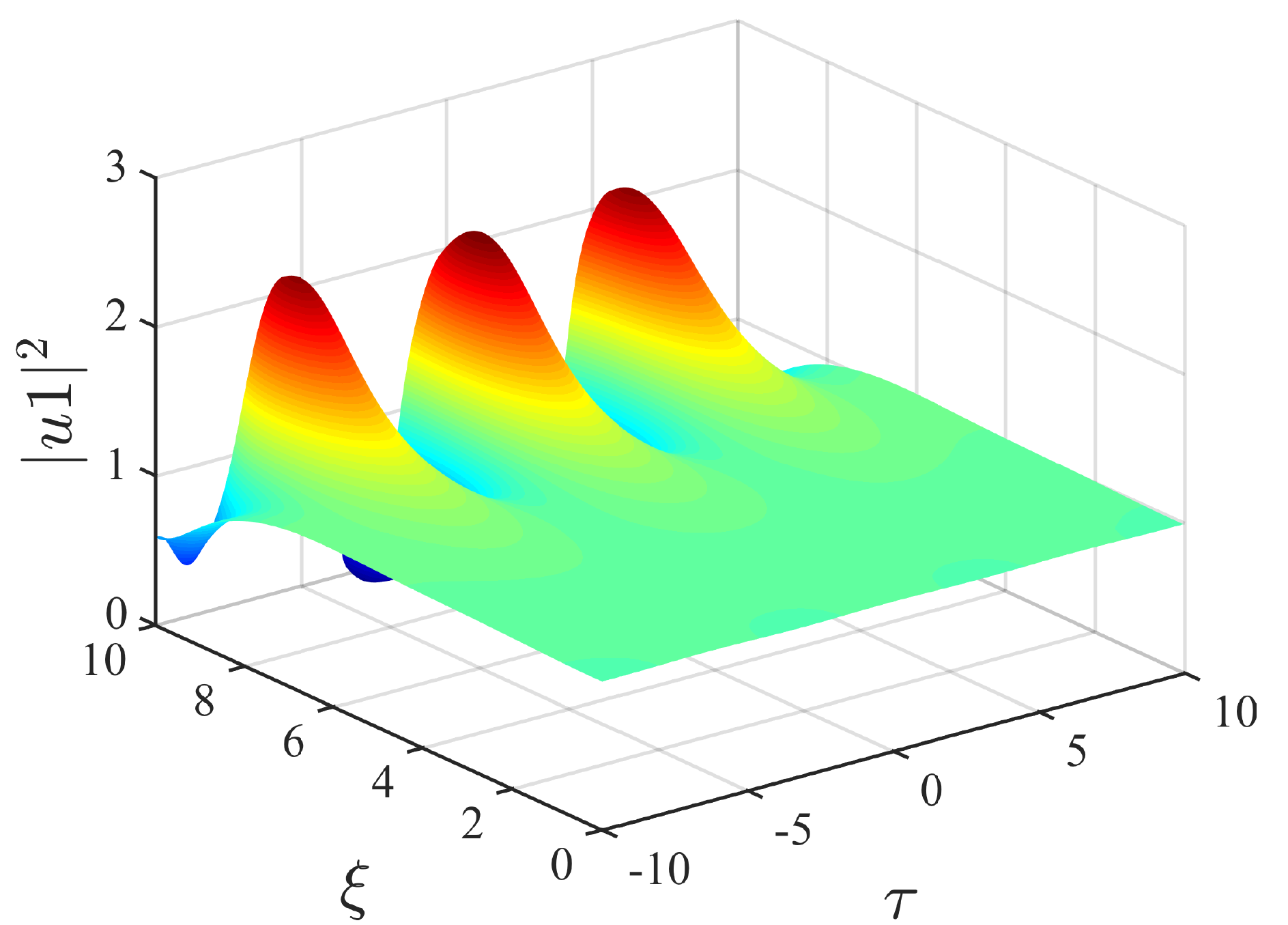}} %
\subfloat[]{\includegraphics[width=0.45\linewidth]{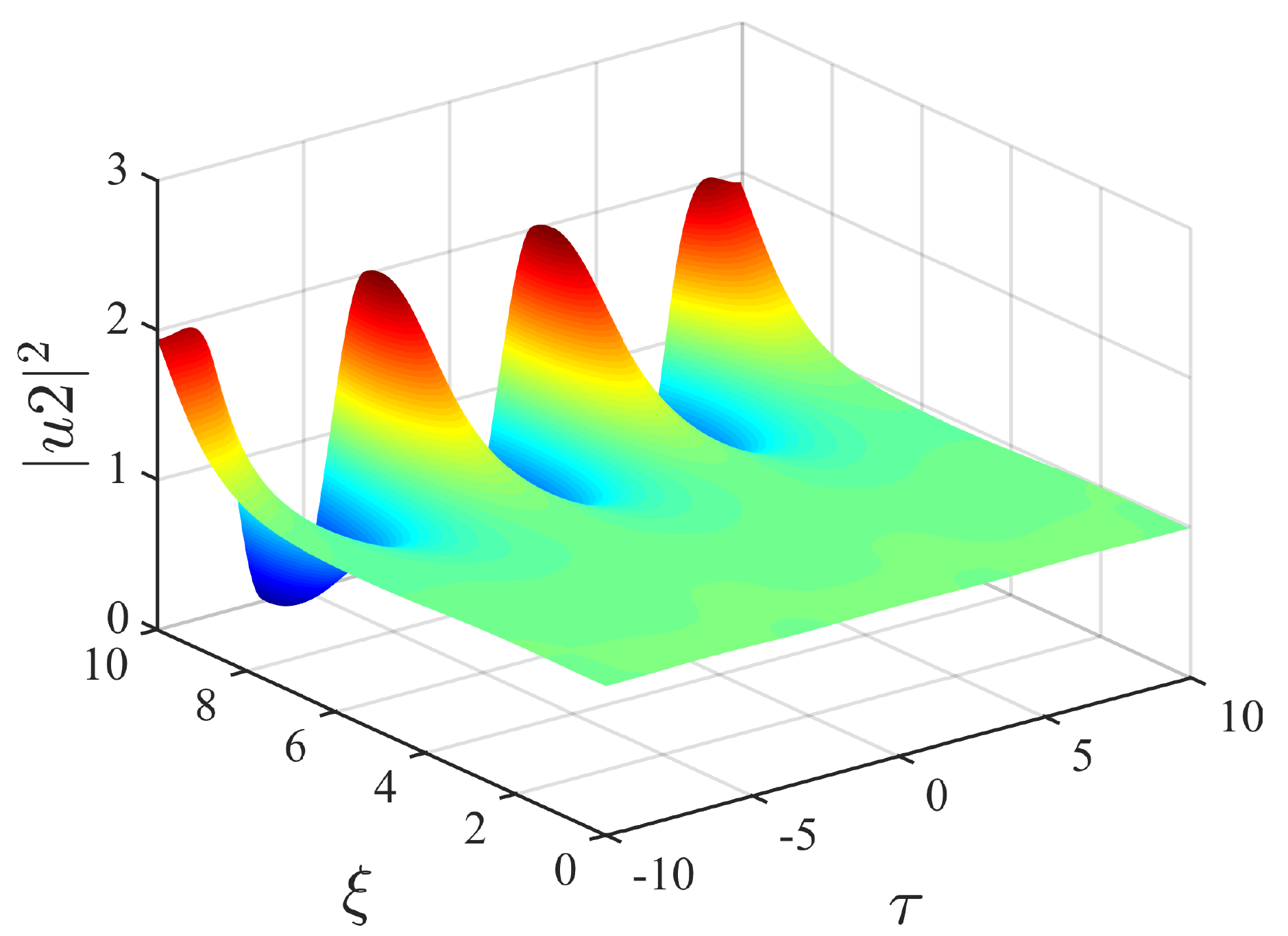}}
\caption{The MI evolution in the bar (\textbf{a}) and cross (\textbf{b}) channels in the
normal-GVD regime ($\protect\sigma =-1$) for the amplitudes of the CW
components $A_{1}=A_{2}=1$ and perturbation parameters $a_{0}=0.002,\protect%
\omega _{0}=1$. Other parameters are $\protect\alpha =2,\Gamma =1,\protect%
\rho =0.01,\protect\chi =0.001$ and $\protect\kappa =1$.}
\label{fig:S9}
\end{figure}

\begin{figure}[H]
\centering
\subfloat[]{\includegraphics[width=0.5\linewidth]{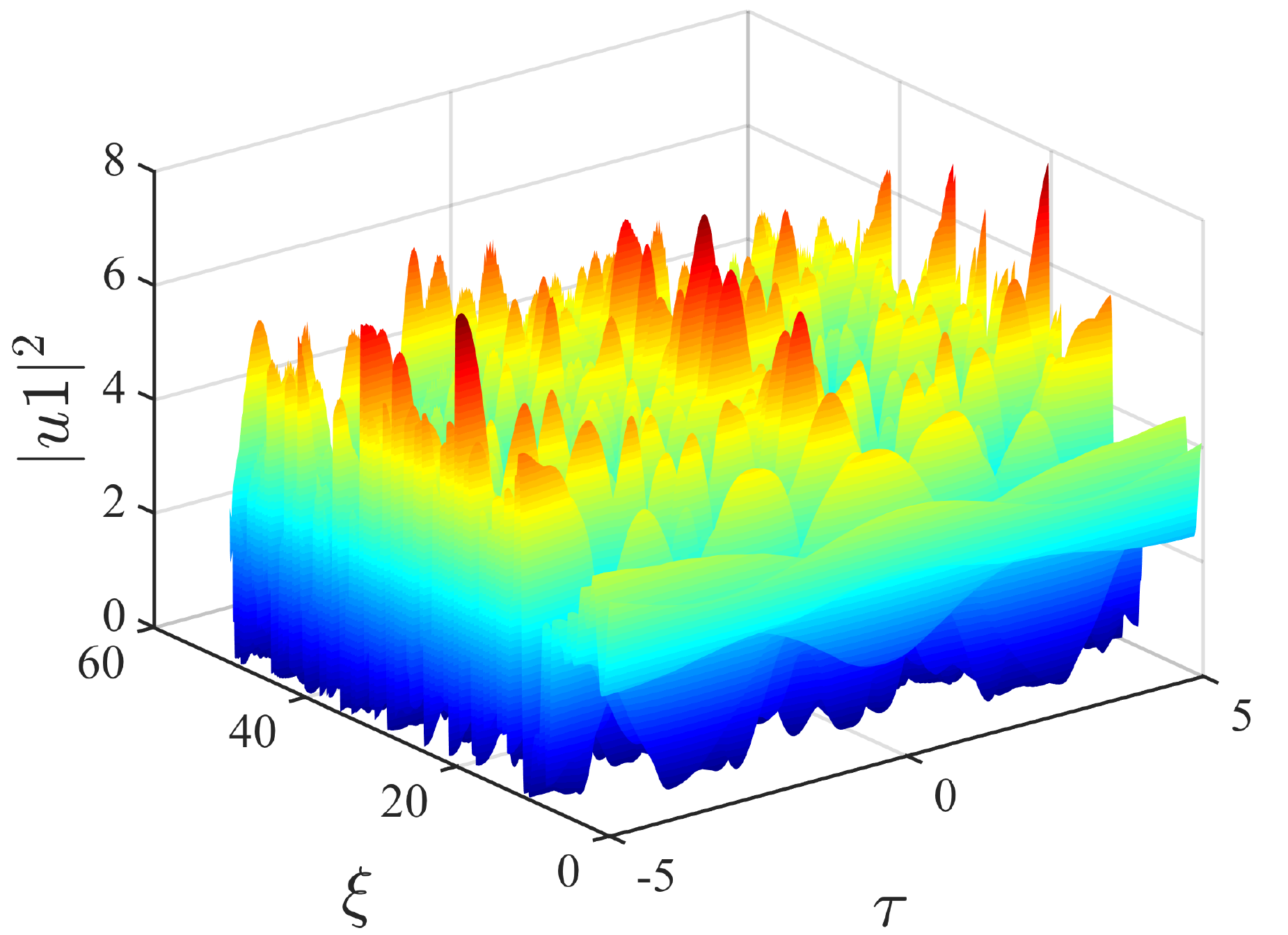}} %
\subfloat[]{\includegraphics[width=0.5\linewidth]{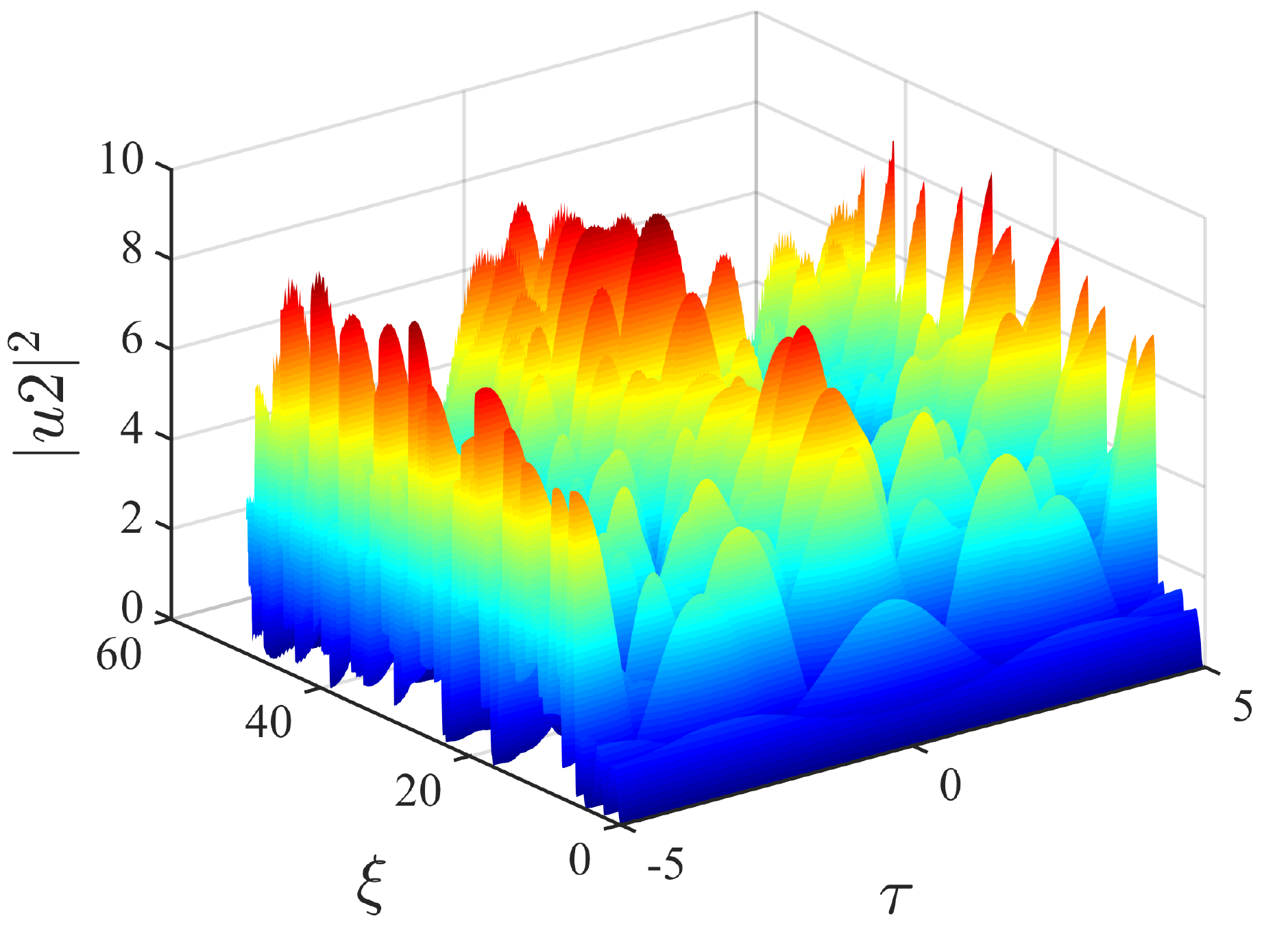}}
\caption{Creation of a chaotic (turbulent) state by the MI in the bar (\textbf{a})
and cross (\textbf{b}) channels in the normal-GVD dispersion regime ($\protect\sigma %
=-1$) for CW amplitudes $A_{1}=2,A_{2}=0.001$, with perturbation parameters $%
a_{0}=0.09,\protect\omega _{0}=1$. Other parameters are $\protect\alpha %
=2,\Gamma =0.5$, $\protect\rho =0.01,\protect\chi =0.02$ and $\protect\kappa %
=1$.}
\label{fig:S10}
\end{figure}

Various outcomes of the MI development for CW states with different
parameters are displayed in Figures \ref{fig:S1}--\ref{fig:S10}. First, %Yes. Now it is revised to Figures 14--23 
in Figure \ref{fig:S1}, we show the results for the symmetric coupler in the
anomalous-GVD regime when the amplitudes of two CW components are equal ($%
A_{1}=A_{2}=1$). {As seen in the figure, a periodic chain of well-shaped soliton-like pulses is
produced on top of the nonzero background in both cores. Longer simulations demonstrate regular dynamics of the quasi-soliton arrays. In this work,
we do not aim to study the latter in detail, as it is not closely related to the initial MI.}
\vspace{-12pt}
\label{sec:5}

We now turn to simulations of the MI in the asymmetric coupler and the analysis
of effects of its different parameters. The impact of the group-velocity
mismatch (walk-off) between the cores in the anomalous-GVD regime is
presented in Figure \ref{fig:S2}. As seen in the figure, pulses generated by
the MI drift away from their original positions, which implies spontaneous
symmetry breaking, as a particular drift direction is selected by the
system. {We have also investigated the spectral evolution of the MI for different values of the group-velocity mismatch. The results (not shown here in detail) corroborate,
in particular, that the group-velocity mismatch has no impact on the instability
spectrum, as predicted by the analytical results in Figure \ref{fig:6}a}. 
Further, Figure \ref{fig:S3} shows the influence of the phase-velocity
mismatch on the MI evolution in the anomalous-GVD regime. In this case, the main
effects are oscillations of the background and retaining of the power
chiefly in the bar channel.
\vspace{-12pt}
\vspace{-20pt}

The role of the ratio of the GVD coefficients in the two cores, $\alpha $,
is shown in \mbox{Figures \ref{fig:S4} and \ref{fig:S5}}, for the case of the
anomalous GVD in both cores. In the case of zero GVD in the cross channel ($%
\alpha =0$) (Figure \ref{fig:S4}) shows that a chain of quasi-solitons with
growing amplitudes is generated on top of a nonzero background in the bar
channel, while narrow growing peaks emerge at edges of the background in the
cross channel. If $\alpha $ increases to $\alpha =2$, the former picture is
essentially reversed, so that a chain of solitons on top of the background
appears in the cross channel, and a chain of very narrow solitons is
generated in the bar channel. In all of these cases, the soliton chains keep
the initial modulation period, $2\pi /\omega _{0}$.

Figure \ref{fig:S6} reveals the impact of the ratio of nonlinearity
coefficients between the two cores. In this case, the MI generates a chain
of very narrow solitons with a higher amplitude, whose peak powers are
growing in the cross channel and growing peaks on an oscillating background
with a relatively low amplitude in the bar channel. Next, we plug in all of the
parameters, to identify their combined effect on the MI evolution in the
anomalous-GVD regime, in Figures \ref{fig:S7} and \ref{fig:S8}. In the former
case, it~is observed that the MI gives rise to a single soliton in the bar
channel, whereas the field in the cross channel decays into radiation. In
the latter case, a single soliton is generated too (which is natural for the
case of the anomalous GVD), but with components in both cores.

Focusing our attention on the asymmetric coupler in the normal-GVD regime,
in Figure \ref{fig:S9}, we address the case when the amplitudes of the two CW
components are equal. In this case as well, a periodic array of peaks with
growing amplitudes is generated in both the bar and cross channels. However,
its shape is essentially different from the soliton chains displayed above
in the anomalous-GVD regime, as in the present case, the array is built of
alternating peaks and wells. \textls[-15]{Lastly, if the amplitudes of the two CW states
are widely different, such as in the case of a large amplitude in the bar
channel and a relatively small one in the cross channel, the MI evolution
leads to a chaotic state, as shown in Figure \ref{fig:S10}.}
\vspace{-12pt}
\vspace{-20pt}
\unskip
\vspace{-20pt}
\vspace{-20pt}
\vspace{-20pt}

\section{Conclusions}

\label{sec:6} In this work, we have investigated the MI (modulational
instability) in the model of asymmetric dual-core NLDCs (nonlinear
directional couplers), based on the system of nonlinear Schr\"{o}dinger
equations, which include differences in the GVD and nonlinearity coefficient
in the two cores, as well as the group- and phase-velocity mismatch between them.
\textls[-18]{The MI of symmetric and asymmetric CW states in the NLDC against small
perturbations was investigated using the linearized equations for the
perturbations. This was followed by direct simulations to investigate the
nonlinear development of the MI.}

First, we have considered the dependence of the MI gain spectra on the total
power of the two-component CW states in the coupler with the anomalous sign
of the GVD in both cores. It was found that the MI bands in the asymmetric
couplers are broader in comparison with their symmetric counterparts. Then,
we focused on the impact of the magnitude of the inter-core coupling
coefficient,~$\kappa $, demonstrating that the increase of $\kappa $ leads
to gradual suppression of the MI. Next, a large GVD coefficient in the bar
channel, in comparison with the cross channel, generates very broad MI
spectra with large values of the instability gain. If the asymmetry between
the cores is introduced only through the difference in the GVD coefficients,
high values of the group-velocity mismatch cause splitting of the single MI
band into two. The effect of asymmetry between two components of the CW
state, $\eta $, was identified as well. It was found that the MI gain and
bandwidth reduce with the increase of $\eta $ from small values to one,
while the further increase of $\eta $ leads to shrinkage of the MI band.

Next, the MI was explored in the normal-GVD regime, in which the MI occurs
in two separated spectral bands. The increase of the coupling coefficient makes
the size of the MI gain in the two bands strongly different. The influence
of the difference in the GVD and nonlinearity coefficients was analyzed, as well.
The increase of these coefficients leads, respectively, to the decrease and
increase of the MI gain in the two bands.

Noteworthy results were produced by the analysis of the MI in the coupler
with opposite signs of the GVD in the cores. While the difference in the
negative values of the GVD coefficient, and in the nonlinearity
coefficients, produce approximately the same effects as in the anomalous-GVD
regime, the response to the increase of the coupling coefficient is similar
to that in the case of the normal GVD, leading to the increase of the MI gain. A
notable effect was observed with the variation of the group-velocity
mismatch, $\rho $, between the cores: the increase of $\rho $ from small
values to one suppresses the MI, which disappears at $\rho =1$. It appears
again and enhances at $\rho >1$. The asymmetry ratio of the two components
of the underlying CW state, $\eta $, also produces a nontrivial effect:
while the MI is absent at small values of $\eta $, it appears at $\eta
\gtrsim 0.5$ in the form of a single spectral band, which grows up to $\eta
=1$ and then splits into two bands.

Finally, we have also performed systematic simulation of the nonlinear
development of the MI in different regimes, which were studied analytically.
\textls[-15]{Typical outcomes feature the generation of periodic chains of growing peaks in
the anomalous-GVD regime. In particular, the group-velocity mismatch
naturally causes a walk-off effect, while the phase-velocity mismatch and
difference in the nonlinearity coefficients produce oscillations on the
background, on top of which soliton arrays emerge. }The difference in the GVD
coefficients facilitates the generation of arrays of very narrow solitary
pulses in the bar channel, whereas arrays of regular pulses appear in the
cross channel. The formation of a single soliton is possible as well. In the
normal-GVD regime, the formation of arrayed peaks with a growing amplitude was
observed. The MI of the CW states with widely different amplitudes of its
two components may produce a turbulent state.

\textls[-15]{These results, especially the generation of regular arrays of solitary
pulses and of a single pulse, can find applications for the design of signal
sources for optical systems. The variations of many parameters that control
the dynamics of the asymmetric couplers may be used to optimize these
applications.}

For further work, it may be relevant to take into regard higher-order
terms, such as those accounting for the third-order dispersion and
self-steepening, and analyze their effects on the MI in the asymmetric
dispersive nonlinear couplers.

\vspace{6pt}

%%%%%%%%%%%%%%%%%%%%%%%%%%%%%%%%%%%%%%%%%%
\acknowledgments{A.G. % They are now corrected and seem right
 is grateful to Professor M. Lakshmanan and National
Academy of Sciences, India (NASI), for providing the Senior Research Fellowship
(Research Associate) under the NASI Platinum Jubilee Senior Scientist Fellowship
project of M.L.}

%%%%%%%%%%%%%%%%%%%%%%%%%%%%%%%%%%%%%%%%%%
\authorcontributions{A.G. conceived of the idea of the present work and has
performed the analytical and numerical computations. B.A.M refined the
analytical calculations. B.A.M., A.G., A.M. and A.U. conducted the
interpretation of the results. A.G. wrote an initial draft. B.A.M. 
finalized the article. } % We have now corrected the main names and so the abbreviations too.

%%%%%%%%%%%%%%%%%%%%%%%%%%%%%%%%%%%%%%%%%%
\conflictsofinterest{The authors declare no conflict of interest.}

\appendix
\appendixtitles{no} 
\appendixsections{multiple} 
\section{\label{Appendix A}}
Elements of matrix $\mathbf{M}$ in Equation (\ref{eqn:15}) are:
\begin{equation}
\begin{split}
m_{11}& =-K-(\sigma _{1}\Omega ^{2}/2)+\eta ^{2}S-\kappa /\eta , \\
m_{12}& =m_{21}=m_{34}=m_{43}=\kappa , \\
m_{13}& =\eta ^{2}S, \\
m_{14}& =m_{23}=m_{32}=m_{41}=0, \\
m_{22}& =-K+\rho \Omega -(\sigma _{1}\alpha \Omega ^{2}/2)+\Gamma S-\kappa
\eta , \\
m_{24}& =\Gamma S, \\
m_{31}& =\eta ^{2}S, \\
m_{33}& =K-(\sigma _{1}\Omega ^{2}/2)+\eta ^{2}S-\kappa /\eta , \\
m_{42}& =\Gamma S, \\
m_{44}& =K-\rho \Omega -(\sigma _{1}\alpha \Omega ^{2}/2)+\Gamma S-\kappa
\eta , \\
\end{split}%
\end{equation}%
where $S=P/(1+\eta ^{2})$. Coefficients of quartic Equation (\ref{eqn:16})\
for $K$, as functions of $\Omega $, are given by:
\begin{gather}
a=2 \rho \Omega \\
\scalebox{.92}[0.92]{$b=2 S \eta \kappa +2 S \Gamma \eta \kappa -2 \kappa ^2-\frac{\kappa ^2}{\eta
^2}-\eta ^2 \kappa ^2+\Omega ^2 \Big(\rho ^2+S \eta ^2 \sigma _1-\frac{%
\kappa \sigma _1}{\eta }+S \alpha \Gamma \sigma _1-\alpha \eta \kappa \sigma
_1\Big)+ \Omega ^4 \Big(-\frac{\sigma _1^2}{4}-\frac{1}{4} \alpha ^2 \sigma
_1^2 \Big)$} \\
c=\Big(-4 S \eta \kappa \rho +2 \kappa ^2 \rho +\frac{2 \kappa ^2 \rho }{%
\eta ^2}\Big) \Omega +\Omega ^3 \Big(-2 S \eta ^2 \rho \sigma _1+\frac{2
\kappa \rho \sigma _1}{\eta }\Big)+\frac{1}{2} \rho \Omega ^5 \sigma _1^2
\end{gather}
%\begin{scriptsize}
\begin{equation}
\scalebox{.95}[0.95]{$\begin{array}{l}
d=\Omega ^{2}\Big(2S\eta \kappa \rho ^{2}-\frac{\kappa ^{2}\rho ^{2}}{\eta
^{2}}+2S^{2}\Gamma \eta ^{3}\kappa \sigma _{1}-S\Gamma \kappa ^{2}\sigma
_{1}-S\eta ^{4}\kappa ^{2}\sigma _{1}+2S^{2}\alpha \Gamma \eta \kappa \sigma
_{1}-\frac{S\alpha \Gamma \kappa ^{2}\sigma _{1}}{\eta ^{2}}-S\alpha \eta
^{2}\kappa ^{2}\sigma _{1}\Big) \\
+\Omega ^{4}\Big(S\eta ^{2}\rho ^{2}\sigma _{1}-\frac{\kappa \rho ^{2}\sigma
_{1}}{\eta }-\frac{1}{2}S\Gamma \eta \kappa \sigma _{1}^{2}+\frac{1}{4}\eta
^{2}\kappa ^{2}\sigma _{1}^{2}+S^{2}\alpha \Gamma \eta ^{2}\sigma _{1}^{2}-%
\frac{S\alpha \Gamma \kappa \sigma _{1}^{2}}{\eta }-S\alpha \eta ^{3}\kappa
\sigma _{1}^{2} \\
+\frac{1}{2}\alpha \kappa ^{2}\sigma _{1}^{2}-\frac{1}{2}S\alpha ^{2}\eta
\kappa \sigma _{1}^{2}+\frac{\alpha ^{2}\kappa ^{2}\sigma _{1}^{2}}{4\eta
^{2}}\Big)+\Omega ^{6}\Big(-\frac{1}{4}\rho ^{2}\sigma _{1}^{2}-\frac{1}{4}%
S\alpha \Gamma \sigma _{1}^{3}+\frac{1}{4}\alpha \eta \kappa \sigma _{1}^{3}-%
\frac{1}{4}S\alpha ^{2}\eta ^{2}\sigma _{1}^{3}+\frac{\alpha ^{2}\kappa
\sigma _{1}^{3}}{4\eta }\Big) \\
+\frac{1}{16}\alpha ^{2}\Omega ^{8}\sigma _{1}^{4}
\end{array}$}
\end{equation}
%\end{scriptsize}


\begin{thebibliography}{999}
	\providecommand{\natexlab}[1]{#1}
	
	\bibitem[Benjamin and Feir(1967)]{benjamin1967}
	Benjamin, T.B.; Feir, J.
	\newblock The disintegration of wave trains on deep water Part 1. Theory.
	\newblock {\em J. Fluid Mech.} {\bf 1967}, {\em 27},~417--430.
	
	\bibitem[Hasegawa(1984)]{hasegawa1984}
	Hasegawa, A.
	\newblock Generation of a train of soliton pulses by induced modulational
	instability in optical fibers.
	\newblock {\em Opt. Lett.} {\bf 1984}, {\em 9},~288--290.
	
	\bibitem[Tai \em{et~al.}(1986)Tai, Hasegawa, and Tomita]{tai1986}
	Tai, K.; Hasegawa, A.; Tomita, A.
	\newblock Observation of modulational instability in optical fibers.
	\newblock {\em Phys. Rev. Lett.} {\bf 1986}, {\em 56},~135.
	
	\bibitem[Agrawal(1987)]{agarwal1}
	Agrawal, G.P.
	\newblock Modulation instability induced by cross-phase modulation.
	\newblock {\em Phys. Rev. Lett.} {\bf 1987}, {\em 59},~880.
	
	\bibitem[Zakharov \em{et~al.}(2006)Zakharov, Dyachenko, and
	Prokofiev]{zakharov2006freak}
	Zakharov, V.E.; Dyachenko, A.; Prokofiev, A.
	\newblock Freak waves as nonlinear stage of Stokes wave modulation instability.
	\newblock {\em Eur. J. Mech. B/Fluids} {\bf 2006}, {\em
		25},~677--692.
	
	\bibitem[Melville(1982)]{melville1982instability}
	Melville, W.
	\newblock The instability and breaking of deep-water waves.
	\newblock {\em J. Fluid Mech.} {\bf 1982}, {\em 115},~165--185.
	
	\bibitem[Konotop and Salerno(2002)]{konotop2002}
	Konotop, V.; Salerno, M.
	\newblock \textls[-15]{Modulational instability in Bose-Einstein condensates in optical
	lattices.}
	\newblock {\em Phys. Rev.~A} {\bf 2002}, {\em 65},~021602.
	
	\bibitem[Li \em{et~al.}(2005)Li, Li, Malomed, Mihalache, and Liu]{li2005}
	Li, L.; Li, Z.; Malomed, B.A.; Mihalache, D.; Liu, W.
	\newblock Exact soliton solutions and nonlinear modulation instability in
	spinor Bose-Einstein condensates.
	\newblock {\em Phys. Rev. A} {\bf 2005}, {\em 72},~033611.
	
	\bibitem[Bhat \em{et~al.}(2015)Bhat, Mithun, Malomed, and Porsezian]{SOC}
	Bhat, I.A.; Mithun, T.; Malomed, B.; Porsezian, K.
	\newblock Modulational instability in binary spin-orbit-coupled Bose-Einstein
	condensates.
	\newblock {\em Phys. Rev. A} {\bf 2015}, {\em 92},~063606.
	
	\bibitem[Taniuti and Washimi(1968)]{taniuti1968self}
	Taniuti, T.; Washimi, H.
	\newblock Self-trapping and instability of hydromagnetic waves along the
	magnetic field in a cold plasma.
	\newblock {\em Phys. Rev. Lett.} {\bf 1968}, {\em 21},~209.
	
	\bibitem[Galeev \em{et~al.}(1975)Galeev, Sagdeev, Sigov, Shapiro, and
	Shevchenko]{galeev1975nonlinear}
	Galeev, A.; Sagdeev, R.; Sigov, Y.S.; Shapiro, V.; Shevchenko, V.
	\newblock Nonlinear theory of the modulation instability of plasma waves.
	\newblock {\em Sov. J. Plasma Phys.} {\bf 1975},
	{\em 1}, 5--10.
	
	\bibitem[Zakharov and Ostrovsky(2009)]{zakharov2009}
	Zakharov, V.; Ostrovsky, L.
	\newblock Modulation instability: The beginning.
	\newblock {\em Phys. D Nonlinear Phenom.} {\bf 2009}, {\em 238}, 540--548.
	
	\bibitem[Boggio \em{et~al.}(2001)Boggio, Tenenbaum, and Fragnito]{boggio2001}
	Boggio, J.; Tenenbaum, S.; Fragnito, H.
	\newblock Amplification of broadband noise pumped by two lasers in optical
	fibers.
	\newblock {\em J. Opt. Soc. Am. B} {\bf 2001}, {\em 18},~1428--1435. % We confirm this also.
	
	\bibitem[Tanemura \em{et~al.}(2004)Tanemura, Ozeki, and Kikuchi]{tanemura2004}
	Tanemura, T.; Ozeki, Y.; Kikuchi, K.
	\newblock Modulational instability and parametric amplification induced by loss
	dispersion in optical fibers.
	\newblock {\em Phys. Rev. Lett.} {\bf 2004}, {\em 93},~163902.
	
	\bibitem[H{\"o}{\"o}k and Karlsson(1993)]{hook1993}
	H{\"o}{\"o}k, A.; Karlsson, M.
	\newblock Ultrashort solitons at the minimum-dispersion wavelength: effects of
	fourth-order dispersion.
	\newblock {\em Opt. Lett.} {\bf 1993}, {\em 18},~1388--1390.
	
	\bibitem[Agrawal(2006)]{agrawal2006nonlinear}
	Agrawal, G.
	\newblock \emph{Nonlinear Fiber Optics}; Optics and Photonics, Academic Press, London:
	2006. %The location information is now provided for the publisher.
	
	\bibitem[Hasegawa and Tappert(1973)]{hasegawa1973transmission}
	Hasegawa, A.; Tappert, F.
	\newblock Transmission of stationary nonlinear optical pulses in dispersive
	dielectric fibers. I. Anomalous dispersion.
	\newblock {\em Appl. Phys. Lett.} {\bf 1973}, {\em 23},~142--144.
	
	\bibitem[Rehberg \em{et~al.}(1988)Rehberg, Rasenat, Fineberg, De~La
	Torre~Juarez, and Steinberg]{rehberg1988}
	Rehberg, I.; Rasenat, S.; Fineberg, J.; De~La Torre~Juarez, M.; Steinberg, V.
	\newblock Temporal modulation of traveling waves.
	\newblock {\em Phys. Rev. Lett.} {\bf 1988}, {\em 61},~2449.
	
	\bibitem[Malendevich \em{et~al.}(2001)Malendevich, Jankovic, Stegeman, and
	Aitchison]{malendevich2001}
	Malendevich, R.; Jankovic, L.; Stegeman, G.; Aitchison, J.S.
	\newblock Spatial modulation instability in a Kerr slab waveguide.
	\newblock {\em Opt. Lett.} {\bf 2001}, {\em 26},~1879--1881.
	
	\bibitem[Liou \em{et~al.}(1992)Liou, Cao, McKinstrie, and Agrawal]{liou1992}
	Liou, L.; Cao, X.; McKinstrie, C.; Agrawal, G.P.
	\newblock Spatiotemporal instabilities in dispersive nonlinear media.
	\newblock {\em Phys. Rev. A} {\bf 1992}, {\em 46},~4202.
	
	\bibitem[Greer \em{et~al.}(1989)Greer, Patrick, Wigley, and Taylor]{greer1989}
	Greer, E.; Patrick, D.; Wigley, P.; Taylor, J.
	\newblock Generation of 2 THz repetition rate pulse trains through induced
	modulational instability.
	\newblock {\em Electron. Lett.} {\bf 1989}, {\em 25},~1246--1248.
	
	\bibitem[Agrawal(2001)]{agrawal2001applications}
	Agrawal, G.
	\newblock {\em Applications of Nonlinear Fiber Optics}; Academic Press, London: 2001;
	Chapter 2. %The location information is now provided for the publisher.
	
	\bibitem[Ellingham \em{et~al.}(2005)Ellingham, Ania-Casta{\~n}{\'o}n, Turitsyn,
	Pustovskikh, Kobtsev, and Fedoruk]{ellingham2005}
	Ellingham, T.; Ania-Casta{\~n}{\'o}n, J.; Turitsyn, S.; Pustovskikh, A.;
	Kobtsev, S.; Fedoruk, M.
	\newblock Dual-pump Raman amplification with increased flatness using
	modulation instability.
	\newblock {\em Opt. Express} {\bf 2005}, {\em 13},~1079--1084.
	
	\bibitem[Dudley \em{et~al.}(2009)Dudley, Genty, Dias, Kibler, and
	Akhmediev]{dudley2009}
	Dudley, J.M.; Genty, G.; Dias, F.; Kibler, B.; Akhmediev, N.
	\newblock Modulation instability, Akhmediev Breathers and continuous wave
	supercontinuum generation.
	\newblock {\em Opt. Express} {\bf 2009}, {\em 17},~21497--21508.
	
	\bibitem[Trillo \em{et~al.}(1988)Trillo, Wabnitz, Wright, and
	Stegeman]{trillo1988soliton}
	Trillo, S.; Wabnitz, S.; Wright, E.; Stegeman, G.
	\newblock Soliton switching in fiber nonlinear directional couplers.
	\newblock {\em Opt.~Lett.} {\bf 1988}, {\em 13},~672--674.
	
	\bibitem[Jensen(1982)]{jensen1982nonlinear}
	Jensen, S.
	\newblock The nonlinear coherent coupler.
	\newblock {\em IEEE Trans. Microw. Theory Tech.} {\bf 1982}, {\em 30},~1568--1571. % Yes it is correct and we confirm this.
	
	\bibitem[Maier(1982)]{Maier1982}
	Maier, A.
	\newblock Optical transistors and bistable elements on the basis of non-linear
	transmission of light by the systems with unidirectional coupled waves.
	\newblock {\em Kvantovaya Elektron.} {\bf 1982}, {\em 9},~2296--2302.
	
	\bibitem[Kivshar(1993)]{kivshar1993}
	Kivshar, Y.S.
	\newblock Switching dynamics of solitons in fiber directional couplers.
	\newblock {\em Opt. Lett.} {\bf 1993}, {\em 18},~7--9.
	
	\bibitem[Friberg \em{et~al.}(1988)Friberg, Weiner, Silberberg, Sfez, and
	Smith]{friberg1988}
	Friberg, S.; Weiner, A.; Silberberg, Y.; Sfez, B.; Smith, P.
	\newblock Femotosecond switching in a dual-core-fiber nonlinear coupler.
	\newblock {\em Opt. Lett.} {\bf 1988}, {\em 13},~904--906.
	
	\bibitem[Malomed \em{et~al.}(1996)Malomed, Skinner, Chu, and
	Peng]{malomed1996symmetric}
	Malomed, B.A.; Skinner, I.; Chu, P.; Peng, G.
	\newblock Symmetric and asymmetric solitons in twin-core nonlinear optical
	fibers.
	\newblock {\em Phys. Rev. E} {\bf 1996}, {\em 53},~4084--4091.
	
	\bibitem[Chiang(1995)]{chiang1995intermodal}
	Chiang, K.S.
	\newblock Intermodal dispersion in two-core optical fibers.
	\newblock {\em Opt. Lett.} {\bf 1995}, {\em 20},~997--999.
	
	\bibitem[Chen \em{et~al.}(1992)Chen, Snyder, and Payne]{Chen1992}
	\textls[-15]{Chen, Y.; Snyder, A.W.; Payne, D.N.
	\newblock Twin core nonlinear couplers with gain and loss.}
	\newblock {\em IEEE J. Quantum Electron.} {\bf 1992}, {\em
		28},~239--245.
	
	\bibitem[Govindaraji \em{et~al.}(2014)Govindaraji, Mahalingam, and
	Uthayakumar]{Govin2}
	Govindaraji, A.; Mahalingam, A.; Uthayakumar, A.
	\newblock Femtosecond pulse switching in a fiber coupler with third order
	dispersion and self-steepening effects.
	\newblock {\em Optik Int. J. Light Electron Opt.}
	{\bf 2014}, {\em 125},~4135--4139.
	
	\bibitem[Snyder \em{et~al.}(1991)Snyder, Mitchell, Poladian, Rowland, and
	Chen]{Snyder}
	Snyder, A.W.; Mitchell, D.; Poladian, L.; Rowland, D.R.; Chen, Y.
	\newblock Physics of nonlinear fiber couplers.
	\newblock {\em J. Opt. Soc. Am. B} {\bf 1991}, {\em 8},~2102--2118.
	
	\bibitem[Yang(1991)]{yang1991}
	Yang, C.C.
	\newblock All-optical ultrafast logic gates that use asymmetric nonlinear
	directional couplers.
	\newblock {\em Opt. Lett.} {\bf 1991}, {\em 16},~1641--1643.
	
	\bibitem[Yang and Wang(1992)]{yang1992}
	Yang, C.C.; Wang, A.
	\newblock Asymmetric nonlinear coupling and its applications to logic
	functions.
	\newblock {\em IEEE J. Quantum~Electron.} {\bf 1992}, {\em
		28},~479--487.
	
	\bibitem[Kitayama and Wang(1983)]{kitayama1983}
	Kitayama, K.I.; Wang, S.
	\newblock Optical pulse compression by nonlinear coupling.
	\newblock {\em Appl. Phys. Lett.} {\bf 1983}, {\em 43},~17--19.
	
	\bibitem[Thirstrup(1995)]{thirstrup1995}
	Thirstrup, C.
	\newblock Optical bistability in a nonlinear directional coupler.
	\newblock {\em IEEE J. Quantum Electron.} {\bf 1995}, {\em
		31}, 2101--2106.
	
	\bibitem[Trillo \em{et~al.}(1989)Trillo, Stegeman, Wright, and
	Wabnitz]{trillo1989parametric}
	Trillo, S.; Stegeman, G.; Wright, E.; Wabnitz, S.
	\newblock Parametric amplification and modulational instabilities in dispersive
	nonlinear directional couplers with relaxing nonlinearity.
	\newblock {\em J. Opt. Soc. Am. B} {\bf 1989}, {\em 6},~889--900.
	
	\bibitem[Tasgal and Malomed(1999)]{tasgal1999}
	Tasgal, R.S.; Malomed, B.A.
	\newblock Modulational instabilities in the dual-core nonlinear optical fiber.
	\newblock {\em Phys. Scr.} {\bf 1999}, {\em 60},~418.
	
	\bibitem[Li \em{et~al.}(2011)Li, Chiang, and Chow]{li1}
	Li, J.H.; Chiang, K.S.; Chow, K.W.
	\newblock Modulation instabilities in two-core optical fibers.
	\newblock {\em J. Opt. Soc. Am. B} {\bf 2011}, {\em 28}, 1693--1701.
	
	\bibitem[Li \em{et~al.}(2012)Li, Chiang, Malomed, and Chow]{li2}
	Li, J.H.; Chiang, K.S.; Malomed, B.A.; Chow, K.W.
	\newblock Modulation instabilities in birefringent two-core optical fibres.
	\newblock {\em J. Phys. B At. Mol. Opt. Phys.}
	{\bf 2012}, {\em 45},~165404.
	
	\bibitem[Nithyanandan \em{et~al.}(2013)Nithyanandan, Raja, and
	Porsezian]{nithya}
	Nithyanandan, K.; Raja, R.V.J.; Porsezian, K.
	\newblock Modulational instability in a twin-core fiber with the effect of
	saturable nonlinear response and coupling coefficient dispersion.
	\newblock {\em Phys. Rev. A} {\bf 2013}, {\em 87},~043805.
	
	\bibitem[Porsezian \em{et~al.}(2009)Porsezian, Murali, Malomed, and
	Ganapathy]{porseziancgl}
	Porsezian, K.; Murali, R.; Malomed, B.A.; Ganapathy, R.
	\newblock Modulational instability in linearly coupled complex cubic--quintic
	Ginzburg--Landau equations.
	\newblock {\em Chaos Solitons Fractals} {\bf 2009}, {\em 40},~1907--1913.
	
	\bibitem[Ganapathy \em{et~al.}(2006)Ganapathy, Malomed, and
	Porsezian]{ganapathy}
	Ganapathy, R.; Malomed, B.A.; Porsezian, K.
	\newblock Modulational instability and generation of pulse trains in asymmetric
	dual-core nonlinear optical fibers.
	\newblock {\em Phys. Lett. A} {\bf 2006}, {\em 354},~366--372.
	
	\bibitem[Malomed \em{et~al.}(1996)Malomed, Peng, and Chu]{Peng}
	Malomed, B.A.; Peng, G.; Chu, P.
	\newblock Nonlinear-optical amplifier based on a dual-core fiber.
	\newblock {\em Opt. Lett.} {\bf 1996}, {\em 21}, 330--332.
	
	\bibitem[Kaup and Malomed(1998)]{Kaup98}
	Kaup, D.J.; Malomed, B.A.
	\newblock Gap solitons in asymmetric dual-core nonlinear optical fibers.
	\newblock {\em J. Opt. Soc. Am. B} {\bf 1998}, {\em 15}, 2838--2846.
	
	\bibitem[Govindaraji \em{et~al.}(2015)Govindaraji, Mahalingam, and
	Uthayakumar]{Govin3}
	Govindaraji, A.; Mahalingam, A.; Uthayakumar, A.
	\newblock Numerical investigation of dark soliton switching in asymmetric
	nonlinear fiber couplers.
	\newblock {\em Appl. Phys. B} {\bf 2015}, {\em 120},~341--348.
	
	\bibitem[Li \em{et~al.}(2012)Li, Zhang, and Hua]{Li2012}
	Li, Q.; Zhang, A.; Hua, X.
	\newblock Numerical simulation of solitons switching and propagating in
	asymmetric directional couplers.
	\newblock {\em Opt. Commun.} {\bf 2012}, {\em 285},~118--123.
	
	\bibitem[Govindaraji \em{et~al.}(2014)Govindaraji, Mahalingam, and
	Uthayakumar]{Govin1}
	Govindaraji, A.; Mahalingam, A.; Uthayakumar, A.
	\newblock Dark soliton switching in nonlinear fiber couplers with gain.
	\newblock {\em Opt. Laser Technol.} {\bf 2014}, {\em 60},~18--21.
	
	\bibitem[He \em{et~al.}(2011)He, Xie, and Xiang]{He2011}
	He, X.; Xie, K.; Xiang, A.
	\newblock Optical solitons switching in asymmetric dual-core nonlinear fiber
	couplers.
	\newblock {\em \mbox{Optik Int. J. Light Electron Opt}.}
	{\bf 2011}, {\em 122},~1222--1224.
	
	\bibitem[Shum and Liu(2002)]{shum2002}
	Shum, P.; Liu, M.
	\newblock Effects of intermodal dispersion on two-nonidentical-core coupler
	with different radii.
	\newblock {\em IEEE Photonics Technol. Lett.} {\bf 2002}, {\em
		14},~1106--1108.
	
	\bibitem[N\'{o}brega \em{et~al.}(2000)N\'{o}brega, da~Silva, and
	Sombra]{Nobrega}
	N\'{o}brega, K.; da~Silva, M.; Sombra, A.
	\newblock Multistable all-optical switching behavior of the asymmetric
	nonlinear directional coupler.
	\newblock {\em Opt. Commun.} {\bf 2000}, {\em 173},~413--421.
	
	\bibitem[Kaup \em{et~al.}(1997)Kaup, Lakoba, and Malomed]{Kaup97}
	Kaup, D.J.; Lakoba, T.I.; Malomed, B.A.
	\newblock Asymmetric solitons in mismatched dual-core optical fibers.
	\newblock {\em J. Opt. Soc. Am. B} {\bf 1997}, {\em 14},~1199--1206.
	
	\bibitem[Atai and Malomed(2003)]{atai200355}
	Atai, J.; Malomed, B.A.
	\newblock Stability and interactions of solitons in asymmetric dual-core
	optical waveguides.
	\newblock {\em Opt. Commun.} {\bf 2003}, {\em 221},~55--62.
	
	\bibitem[Atai and Malomed(2002)]{atai2002spatial}
	Atai, J.; Malomed, B.A.
	\newblock Spatial solitons in a medium composed of self-focusing and
	self-defocusing layers.
	\newblock {\em Phys. Lett. A} {\bf 2002}, {\em 298},~140--148.
	
	\bibitem[Zafrany \em{et~al.}(2005)Zafrany, Malomed, and Merhasin]{Chaoszaf}
	Zafrany, A.; Malomed, B.A.; Merhasin, I.M.
	\newblock Solitons in a linearly coupled system with separated dispersion and
	nonlinearity.
	\newblock {\em Chaos Interdiscip. J. Nonlinear Sci.} {\bf
		2005}, {\em 15},~037108.
	
	\bibitem[Govindaraji \em{et~al.}(2016)Govindaraji, Mahalingam, and
	Uthayakumar]{Govin4}
	Govindaraji, A.; Mahalingam, A.; Uthayakumar, A.
	\newblock Interaction dynamics of bright solitons in linearly coupled
	asymmetric systems.
	\newblock {\em Opt. Quantum Electron.} {\bf 2016}, {\em 48},~563.
	
	\bibitem[Xu \em{et~al.}(2001)Xu, Zhang, Chen, Luo, and Liu]{xu2001}
	Xu, W.C.; Zhang, S.M.; Chen, W.C.; Luo, A.P.; Liu, S.H.
	\newblock Modulation instability of femtosecond pulses in dispersion-decreasing
	fibers.
	\newblock {\em Opt. Commun.} {\bf 2001}, {\em 199},~355--360.
	
\end{thebibliography}
\end{document}